\documentclass[aps,prd,showpacs,superscriptaddress,groupedaddress,nofootinbib]{revtex4-2}  
\usepackage[T1]{fontenc}
\usepackage{lmodern} 
\usepackage{graphicx}  
\usepackage{dcolumn}   
\usepackage{bm}        
\usepackage{amssymb}   
\usepackage{amsmath}
\usepackage{bbold}
\usepackage{array}
\usepackage{makecell}
\usepackage{braket}
\usepackage{longtable}
\usepackage{supertabular,booktabs}

\usepackage{titlesec} 
\usepackage[colorlinks,citecolor=blue,urlcolor=blue,hypertexnames=true]{hyperref}
\usepackage{subfigure}

\usepackage[usenames,dvipsnames,svgnames,table]{xcolor} 

\newcommand{\be}{\begin{equation}}
\newcommand{\ee}{\end{equation}}
\newcommand{\bea}{\begin{eqnarray}}
\newcommand{\eea}{\end{eqnarray}}
\newcommand{\bml}{\begin{subequations}}
\newcommand{\eml}{\end{subequations}}
\newcommand{\bfig}{\begin{figure}}
\newcommand{\efig}{\end{figure}}

\newcommand{\bmat}{\begin{pmatrix}}
\newcommand{\emat}{\end{pmatrix}}
\usepackage{graphicx,slashed,booktabs,xcolor,multirow,float,
amsfonts,bbold,mathtools,sidecap,tikz,bm,enumitem}
\usepackage{multirow}
\usepackage{bbding}
\usepackage{titlesec}
\usepackage{hyperref}
\usepackage{wasysym}
\usepackage{amssymb}
\usepackage{pifont}
\newcommand{\grad}{\nabla}

\renewcommand{\d}{\mathrm{d}}

\renewcommand{\leq}{\leqslant}

\def\be{\begin{equation}}
\def\ee{\end{equation}}
\def\ba{\begin{eqnarray}}
\def\ea{\end{eqnarray}}

\def\d{\mathrm{d}}
\def\p{{\cal P}}

\def\L*{{\cal L}_*}
\def\L{\mathcal{L}}
\def\({\left(}
\def\){\right)}

\def\p{\partial}

\def\p{\partial}

\def\<{\langle}
\def\>{\rangle}

\def\cs2{c_{s}^{2}}

 \def\p{\partial}

 \def\be   {\begin{equation}}   \def\ee   {\end{equation}}
 \def\ba   {\begin{array}}      \def\ea   {\end{array}}
 \def\bea  {\begin{eqnarray}}   \def\eea  {\end{eqnarray}}
 \def\bean {\begin{eqnarray*}}  \def\eean {\end{eqnarray*}}

\titleclass{\subsubsubsection}{straight}[\subsection]

\newcounter{subsubsubsection}[subsubsection]
\renewcommand\thesubsubsubsection{\thesubsubsection.\arabic{subsubsubsection}}

\titleformat{\subsubsubsection}
  {\normalfont\normalsize\bfseries}{\thesubsubsubsection}{1em}{}
\titlespacing*{\subsubsubsection}
{0pt}{3.25ex plus 1ex minus .2ex}{1.5ex plus .2ex}

\makeatletter
\renewcommand\paragraph{\@startsection{paragraph}{5}{\z@}%
  {3.25ex \@plus1ex \@minus.2ex}%
  {-1em}%
  {\normalfont\normalsize\bfseries}}
\renewcommand\subparagraph{\@startsection{subparagraph}{6}{\parindent}%
  {3.25ex \@plus1ex \@minus .2ex}%
  {-1em}%
  {\normalfont\normalsize\bfseries}}
\def\toclevel@subsubsubsection{4}
\def\toclevel@paragraph{5}
\def\toclevel@paragraph{6}
\def\l@subsubsubsection{\@dottedtocline{4}{7em}{4em}}
\def\l@paragraph{\@dottedtocline{5}{10em}{5em}}
\def\l@subparagraph{\@dottedtocline{6}{14em}{6em}}
\makeatother

\begin{document}


\definecolor{lime}{HTML}{A6CE39}
\DeclareRobustCommand{\orcidicon}{\hspace{-2.1mm}
\begin{tikzpicture}
\draw[lime,fill=lime] (0,0.0) circle [radius=0.13] node[white] {{\fontfamily{qag}\selectfont \tiny \,ID}}; \draw[white, fill=white] (-0.0525,0.095) circle [radius=0.007]; 
\end{tikzpicture} \hspace{-3.7mm} }
\foreach \x in {A, ..., Z}{\expandafter\xdef\csname orcid\x\endcsname{\noexpand\href{https://orcid.org/\csname orcidauthor\x\endcsname} {\noexpand\orcidicon}}}
\newcommand{\orcidauthorA}{0000-0002-0459-3873}
\newcommand{\orcidauthorB}{0000-0001-9434-0505}
\newcommand{\orcidauthorC}{0000-0003-1081-0632}


\title{\textcolor{Sepia}{\textbf \huge\Large 
Galileon inflation evades the no-go for PBH formation in the single-field framework}}


\author{{\large  Sayantan Choudhury\orcidA{}${}^{1}$}}
\email{sayantan\_ccsp@sgtuniversity.org,  \\ sayanphysicsisi@gmail.com (Corresponding author)}
\author{\large Sudhakar~Panda\orcidB{}${}^{2,3}$}
\email{panda@niser.ac.in }
\author{ \large M.~Sami\orcidC{}${}^{1,4,5}$}
\email{ sami\_ccsp@sgtuniversity.org,  samijamia@gmail.com}

\affiliation{ ${}^{1}$Centre For Cosmology and Science Popularization (CCSP),\\
        SGT University, Gurugram, Delhi- NCR, Haryana- 122505, India,}
\affiliation{${}^{2}$School of Physical Sciences,  National Institute of Science Education and Research, Bhubaneswar, Odisha - 752050, India,}
\affiliation{${}^{3}$ Homi Bhabha National Institute, Training School Complex, Anushakti Nagar, Mumbai - 400085, India,}
\affiliation{${}^{4}$Center for Theoretical Physics, Eurasian National University, Astana 010008, Kazakhstan.}
	\affiliation{${}^{5}$Chinese Academy of Sciences,52 Sanlihe Rd, Xicheng District, Beijing.}

\begin{abstract}
We consider Galileon inflation in the Effective Field Theory (EFT) framework and examine the possibility for PBH formation during slow roll (SR) to ultra slow roll (USR) transitions. We show that loop corrections to the power spectrum, in this case, do not impose additional constraints on the  masses of PBHs produced.
We indicate that the remarkable non-renormalization property of Galileans due to generalized shift symmetry dubbed as Galilean symmetry is responsible for protecting PBH formation from quantum loop corrections.  

\end{abstract}

\maketitle
\tableofcontents
\newpage

\section{Introduction}
Galileon field theory came into the limelight thanks to massive gravity, known as dRGT in 2010 \cite{deRham:2010ik,deRham:2010kj}. In fact, the first attempt to assign mass to  graviton was made by Pauli-Firz as early as 1939 \cite{Fierz:1939ix}. It was pointed out in 1970 that the Fierz-Pauli theory suffers from the vDVZ discontinuity \cite{vanDam:1970vg,Zakharov:1970cc}, namely, that its predictions do not reduce to those of the Einstein theory when the graviton mass is turned zero. Vainshtein demonstrated in 1972 \cite{Vainshtein:1972sx} that the puzzle could be resolved by promoting the theory to a non-linear background. The latter, however, inevitably leads to the Boulware-Deser ghost \cite{Boulware:1972yco} and massive gravity was almost forgotten until 2010. Indeed, in 2008, based upon generalized shift symmetry dubbed Galileon symmetry, Nicolis et al. \cite{Nicolis:2008in} constructed a higher-order derivative scalar field Lagrangian with a unique structure in Minkowski space-time that gives rise to second order equations of motion free from Ostrogradsky instabilities. The framework was readily generalized to curved space-time in ref. \cite{Deffayet:2009wt}. 
In an attempt to get rid of the Boulware-Deser ghost in dRGT in 2010, the authors constructed the mass term for graviton keeping in mind the covariant Galileon construct such that in the decoupling limit, relevant to local gravity constraints, the longitudinal part of the spin-2 field (scalar field $\phi$) was a Galileon field, which gets successfully screened out by the Vainstein mechanism adhering to local gravity constraints. Let us note that the historic motivation {\it a la} Fierz-Pauli was related to the construction of a consistent relativistic equation for the massive spin-$2$ field, whereas the contemporary reason for assigning a tiny mass of the order of $H_0\sim 10^{-33}eV$ to graviton has been to account for the late-time acceleration. Interestingly, the resulting setup is ghost free and does not suffer from vDVZ discontinuity. Unfortunately, FLRW cosmology is absent from this framework. Efforts were then made to address the issue by invoking the non-trivial fiducial metric both fixed and dynamical (bi-gravity), in place of the flat Minkowski metric used in dRGT. The latter, however, brings in the Higuchi ghost \footnote{If Higuchi bound, $m_g>\sqrt{2} H$, is violated then it is identified to be a ghost. Here $m_g$ represents the graviton mass. } \cite{Higuchi:1986py}, making the scenario, in particular, unsuited to late-time cosmology, apart from several other theoretical issues associated with massive gravity. Perhaps, the graviton mass is strictly zero, or it is too challenging to build a consistent theory of massive gravity.  Some authors have pointed out that the situation might be remedied in Lorentz-violating framework of massive gravity theories, see Refs. \cite{Blas:2009my,Blas:2014ira,Blas:2007zz,Gabadadze:2004iv,Berezhiani:2007zf,deRham:2014zqa} for details.  

Nonetheless, irrespective of massive gravity, the Galileon field is promising in its own right from a phenomenological point of view. However, the lowest non-trivial Galileon Lagrangian that includes $\mathcal{L}_3\sim (\partial \phi)^2(\Box \phi)$, is not relevant to late time acceleration \cite{Chow:2009fm}. It was first demonstrated in ref. \cite{Nicolis:2008in} followed by ref. \cite{Deffayet:2009wt} that the inclusion of higher order terms $\mathcal{L}_4$ and/or $\mathcal{L}_5 $ (built from first and second order differential operators using four and five $\phi$ fields respectively) in the Galileon Lagrangian \footnote{Taken together with $\mathcal{L}_1$ and $\mathcal{L}_2$ represented by a linear term in $\phi$ and standard kinetic term, makes the complete structure of Galileon Lagrangian in $3+1$-space time dimensions.} gives rise to a stable de-Sitter solution relevant to late-time cosmology. This work led to the proliferation of phenomenological applications of Galileon fields to gravity, quantum field theory and cosmology \cite{Kobayashi:2010wa,Jain:2010ka,Gannouji:2010au,Ali:2010gr,deRham:2011by,Tsujikawa:2010sc,Burrage:2010rs,DeFelice:2010jn,DeFelice:2010gb,Babichev:2010jd,DeFelice:2010pv,DeFelice:2010nf,Hinterbichler:2010xn,Kobayashi:2010cm,Deffayet:2010qz,Burrage:2010cu,Mizuno:2010ag,Nesseris:2010pc,Khoury:2010xi,DeFelice:2010as,Kimura:2010di,Zhou:2010di,Hirano:2010yf,Kamada:2010qe,VanAcoleyen:2011mj,Hirano:2011wj,Li:2011sd,Pujolas:2011he,Kobayashi:2011pc,DeFelice:2011zh,Khoury:2011da,Trodden:2011xh,Burrage:2011bt,Liu:2011ns,Kobayashi:2011nu,PerreaultLevasseur:2011wto,deRham:2011by,Clifton:2011jh,Endlich:2011vg,Brax:2011sv,Gao:2011mz,DeFelice:2011uc,Gao:2011qe,Babichev:2011iz,DeFelice:2011hq,Khoury:2011ay,Qiu:2011cy,Renaux-Petel:2011rmu,DeFelice:2011bh,Kimura:2011td,Wang:2011dt,Kimura:2011dc,DeFelice:2011th,Appleby:2011aa,DeFelice:2011aa,Zhou:2011ix,Goon:2012mu,Shirai:2012iw,Goon:2012dy,deRham:2012az,Ali:2012cv,Liu:2012ww,Choudhury:2012yh,Choudhury:2012whm,Barreira:2012kk,Gubitosi:2012hu,Barreira:2013jma,deFromont:2013iwa,Deffayet:2013lga,Arroja:2013dya,Li:2013tda,Sami:2013ssa,Khoury:2013tda,Burrage:2015lla,Koyama:2015vza,Brax:2015dma,Saltas:2016nkg,Ishak:2018his}. Interestingly, as early as 1972, Horndeski constructed a general scalar field Lagrangian that gives rise to second order equations of motion, which is treated to be a generalised Galileon framework \cite{Horndeski:1974wa}. 

It may be noted that there has been a renewed interest in the study of primordial black holes for various reasons  \cite{Zeldovich:1967lct,Hawking:1974rv,Carr:1974nx,Carr:1975qj,Chapline:1975ojl,Carr:1993aq,Kawasaki:1997ju,Yokoyama:1998pt,Kawasaki:1998vx,Rubin:2001yw,Khlopov:2002yi,Khlopov:2004sc,Saito:2008em,Khlopov:2008qy,Carr:2009jm,Choudhury:2011jt,Lyth:2011kj,Drees:2011yz,Drees:2011hb,Ezquiaga:2017fvi,Kannike:2017bxn,Hertzberg:2017dkh,Pi:2017gih,Gao:2018pvq,Dalianis:2018frf,Cicoli:2018asa,Ozsoy:2018flq,Byrnes:2018txb,Ballesteros:2018wlw,Belotsky:2018wph,Martin:2019nuw,Ezquiaga:2019ftu,Motohashi:2019rhu,Fu:2019ttf,Ashoorioon:2019xqc,Auclair:2020csm,Vennin:2020kng,Nanopoulos:2020nnh,Gangopadhyay:2021kmf,Inomata:2021uqj,Stamou:2021qdk,Ng:2021hll,Wang:2021kbh,Kawai:2021edk,Solbi:2021rse,Ballesteros:2021fsp,Rigopoulos:2021nhv,Animali:2022otk,Correa:2022ngq,Frolovsky:2022ewg,Escriva:2022duf,Karam:2022nym,Ozsoy:2023ryl,Ivanov:1994pa,Afshordi:2003zb,Frampton:2010sw,Carr:2016drx,Kawasaki:2016pql,Inomata:2017okj,Espinosa:2017sgp,Ballesteros:2017fsr,Sasaki:2018dmp,Ballesteros:2019hus,Dalianis:2019asr,Cheong:2019vzl,Green:2020jor,Carr:2020xqk,Ballesteros:2020qam,Carr:2020gox,Ozsoy:2020kat,Baumann:2007zm,Saito:2008jc,Saito:2009jt,Choudhury:2013woa,Sasaki:2016jop,Raidal:2017mfl,Papanikolaou:2020qtd,Ali-Haimoud:2017rtz,Di:2017ndc,Raidal:2018bbj,Cheng:2018yyr,Vaskonen:2019jpv,Drees:2019xpp,Hall:2020daa,Ballesteros:2020qam,Ragavendra:2020sop,Carr:2020gox,Ozsoy:2020kat,Ashoorioon:2020hln,Ragavendra:2020vud,Papanikolaou:2020qtd,Ragavendra:2021qdu,Wu:2021zta,Kimura:2021sqz,Solbi:2021wbo,Teimoori:2021pte,Cicoli:2022sih,Ashoorioon:2022raz,Papanikolaou:2022chm,Wang:2022nml,Mishra:2019pzq,ZhengRuiFeng:2021zoz,Cohen:2022clv,Arya:2019wck,Bastero-Gil:2021fac,Correa:2022ngq,Gangopadhyay:2021kmf,Cicoli:2022sih,Brown:2017osf,Palma:2020ejf,Geller:2022nkr,Braglia:2022phb,Kawai:2022emp,Frolovsky:2023xid,Aldabergenov:2023yrk,Aoki:2022bvj,Frolovsky:2022qpg,Aldabergenov:2022rfc,Ishikawa:2021xya,Gundhi:2020kzm,Aldabergenov:2020bpt,Cai:2018dig,Fumagalli:2020adf,Cheng:2021lif,Balaji:2022rsy,Qin:2023lgo,Choudhury:2023kam}. Indeed, apart from being candidates for super-massive black holes in galaxies, PBHs might be hiding secrets about unsettled issues of the early universe, such as those related to dark matter and observed baryon asymmetry in the universe. Leaving aside the late time cosmological relevance, the Galileon field is remarkable from a field theoretic point of view and might play a distinguished role in settling the puzzle related to PBH formation in the framework of single field inflation. Indeed, 
by virtue of Galileon symmetry, the interaction terms in its Lagrangian should include only derivative terms, which however, in general, is plagued with Ostrogradski instabilities. There is a specific structure of terms with fewer than two derivatives per field invariant under the said symmetry that gives rise to equations of motion of second order. It is remarkable that  the galileons are not renormalized by self
loops; even other field loops adhere to the same features provided the couplings respect Galileon symmetry. This is a generic distinguished property of Galileon field theory\footnote{This is true for derivative coupled theories which in general suffer from Ostrogradsky instabilities and lack unitarity at the quantum level.}  with unitarity at the background.  The non-renormalizability of Galileon field might have an important implications for PBH formation for single field inflation.
It may be noted that PBH formation in the framework of single field inflation is under active scrutiny at present \cite{Kristiano:2022maq,Riotto:2023hoz,Choudhury:2023vuj,Choudhury:2023jlt,Kristiano:2023scm,Riotto:2023gpm,Choudhury:2023rks,Firouzjahi:2023aum,Motohashi:2023syh}. It has recently been demonstrated that one loop corrections to power spectrum in $P(X,\phi)$ (where $X\equiv -(\partial_\mu\phi \partial^\mu\phi)/2$) theories severely constrain the mass of PBHs produced during slow roll to ultra-slow role transitions amounting to  a no-go result. Galileon theories due their non-renormalizable property might evade these restrictions and provide an excellent endeavour for PBH formation in the framework of single field inflation.

The organization of this paper is as follows: In section \ref{s2} we have reviewed the Galileon Effective Field Theory.
Section \ref{s3} is devoted to  technical details of  inflationary paradigm using Covariantized Galileon Effective Field Theory (CGEFT). Section \ref{s3} includes explicit discussion of Galileons inflation  in the decoupling limiting. 
Section \ref{s4} is devoted to the computation of the tree level power spectrum from the second order perturbation generated by comoving curvature perturbation in the underlying CGEFT theoretical set up. Section \ref{s5} includes computation of the cut-off regularized one-loop \footnote{For more details regarding the one-loop computation during inflation look at the  refs. \cite{Adshead:2008gk,Senatore:2009cf,Senatore:2012nq,Pimentel:2012tw,Sloth:2006az,Seery:2007we,Seery:2007wf,Bartolo:2007ti,Seery:2010kh,Bartolo:2010bu,Senatore:2012ya,Chen:2016nrs,Markkanen:2017rvi,Higuchi:2017sgj,Syu:2019uwx,Rendell:2019jnn,Cohen:2020php,Green:2022ovz,Premkumar:2022bkm} which we believe will be extremely helpful for the readers. } power spectrum for comoving curvature perturbation from CGEFT set up. Numerical results are given in section \ref{s6}. Our main findings are summarized in section \ref{s7}.

\section{Galileon Effective Field Theory: A old wine in a new glass} \label{s2}

\subsection{Non-Covarinat Galileon Effective Field Theory (NCGEFT)} \label{s2a}

The Galileon action first constructed in \cite{Nicolis:2008in}, is a higher derivative scalar field framework that gives rise to second-order equations of motion in Minkowski space-time.  Later in Ref.\cite{Deffayet:2009wt} the authors constructed a ghost-free and unitarity-preserving version of the Galileon theory in dynamical space-time by allowing a non-minimal coupling to the gravitational background.  Galileon theories are equipped,  at least in planar space,  with the following symmetry on the scalar degree of freedom $\phi$, in addition to the shift symmetry,  which is directly related to the slow-roll feature of inflationary potential \footnote{Galilean symmetry is the extended version of the usual shift symmetry $ \phi \rightarrow \phi+ c$,  which helps to incorporate any type of derivative interactions.  This symmetry in principle cannot commute with the Poincar\'{e} group generated generators, which allows to incorporate mixing of the derivative contributions in the different orders.    Consequently,  within the description of Effective Field Theory (EFT) it must be realized non-linearly in terms of the non-factorizable extended version of the Poincar\'{e} group.}:
\bea \label{GCS}\phi \rightarrow \phi+ c + b_{\mu}x^{\mu}=\phi+ c + b\cdot x\quad\quad\Longrightarrow\quad\quad \partial_{\mu}\phi\rightarrow\partial_{\mu}\phi+b_{\mu},\eea where $c,$ represents a scalar constant,   $b_{\mu}$  represents a vector constant and $x^{\mu}$ describes the corresponding coordinates in $3+1$ space-time dimensions.  Here it is important to note that,  the last term $b_{\mu}x^{\mu}=b\cdot x$ represents the space-time translations.  Its moniker comes from the fact that it imitates the coordinate transformation between non-relativistic inertial frames. Any term $\partial \partial..\, \phi$  with two or more derivatives is obviously inherently Galilean invariant, at least in flat space. The Galilean invariant terms that we'll use in this situation are a particularly unique collection that, among other things, provides a second order equation of motion. These are frequently discussed in literature and continue to garner a lot of interest. 
The number of Galileon terms under discussion is extremely small; there are only five of them in a four-dimensional space-time environment which is appearing along with the Einstein-Hilbert term in the curved gravitation background and described by the following non-covariant version of the representative Galielon Effective Field Theory (NCGEFT) action:
\bea
S= \int d^4 x \sqrt{-g}\Bigg[\frac{M^2_{pl}}{2}R-V_0+ {\cal L}^{\bf NC}_{\phi}\Bigg],
\eea
where we define:
\be {\cal L}^{\bf NC}_{\phi}=\sum^{5}_{i=1}c_i\mathcal{L}^{\bf NC}_i.\ee
The explicit expression for the $\mathcal{L}^{\bf NC}_i\forall i=1,2,\cdots,5$ is given by the following expressions:
\bea
{\cal L}^{\bf NC}_1 & = & \phi, \\
{\cal L}^{\bf NC}_2 & = & -\frac{1}{2} \,  \partial \phi \cdot \p \phi, \\
{\cal L}^{\bf NC}_3 & = & - \frac{1}{2} \,  [\Pi_{\phi}] \, \partial \phi \cdot \partial \phi ,\\
{\cal L}^{\bf NC}_4 & = & - \frac{1}{4} \Big\{ [\Pi_{\phi}]^2 \,  \partial \phi \cdot \partial \phi - 2 \,  [\Pi_{\phi}] \, \partial \phi \cdot \Pi_{\phi} \cdot \partial \phi - [\Pi ^2_{\phi} ] \, \partial \phi \cdot \partial \phi + 2 \, \partial \phi \cdot \Pi ^2_{\phi} \cdot \partial\phi \Big\}, \nonumber \\
{\cal L}^{\bf NC}_5 & = & -\frac{1}{5} \Big\{
[\Pi_{\phi}]^3 \, \partial \phi \cdot \partial \phi- 3 [\Pi_{\phi}]^2_{\phi} \,  \partial \phi \cdot \Pi _{\phi}\cdot \partial \phi
-3 [\Pi_{\phi}] [\Pi^2_{\phi}] \,  \partial \phi \cdot \partial \phi
+6 [\Pi_{\phi}] \, \partial \phi \cdot \Pi^2_{\phi} \cdot \partial\phi \nonumber
 \\
&&
\quad\quad\quad\quad\quad\quad\quad\quad\quad\quad\quad\quad\quad+2 [\Pi ^3_{\phi}] \, \partial \phi \cdot \partial \phi
+3 [\Pi ^2_{\phi}] \, \partial \phi \cdot \Pi_{\phi} \cdot \partial \phi
- 6 \,  \partial \phi \cdot \Pi^3_{\phi} \cdot \partial\phi \Big\},
\eea
where  we use the following short-hand notation for our purpose:
\be (\partial_\mu \partial_\nu \pi)^n \equiv [ \Pi ^n _{\phi}].\ee
In this context,  the brackets $[...]$ stands for the trace operator and $'\dot'$ represents the standard Lorentz invariant contraction of space-time indices.  For an example,  one can write the following:
\be  [\Pi_{\phi}] \, \partial \phi \cdot \partial \phi \equiv \Box\phi \,\partial_{\mu}\phi\partial^{\mu}\phi.\ee
Also,  in this context $c_i$ are the generic coefficients which mimics the role of Wilson coefficients within the framework of EFT.  One may consider more terms in the NCGEFT Lagrangian. But in four space-time dimensions those additional contributions turn out to be trivial and one can recast all of them as total derivatives.

From the above set of NCGEFT Lagrangian one can write the following equation:
\be {\cal E}:\equiv \delta_{\phi}{\cal L}^{\bf NC}_{\phi}=\sum^{5}_{i=1}c_i\delta_{\phi}\mathcal{L}^{\bf NC}_i =\sum^{5}_{i=1}c_i {\cal E}_i=-T^{\mu}_{\mu},\ee
in the present context ${\cal E}_i \forall i=1,2,\cdots,5$ are defined as:
\bea {\cal E}_1 &=& \delta_{\phi}{\cal L}^{\bf NC}_1=1,\\
{\cal E}_2 &=& \delta_{\phi}{\cal L}^{\bf NC}_2=\Box\phi,\\ 
{\cal E}_3 &=& \delta_{\phi}{\cal L}^{\bf NC}_3=(\Box\phi)^2-(\partial_{\mu}\partial_{\nu}\phi)^2,\\ 
{\cal E}_4 &=& \delta_{\phi}{\cal L}^{\bf NC}_4=(\Box\phi)^3-3\Box\phi (\partial_{\mu}\partial_{\nu}\phi)^2+2(\partial_{\mu}\partial_{\nu}\phi)^3,\\
{\cal E}_4 &=& \delta_{\phi}{\cal L}^{\bf NC}_5=(\Box\phi)^4-6(\Box\phi)^2 (\partial_{\mu}\partial_{\nu}\phi)^2 +8\Box\phi (\partial_{\mu}\partial_{\nu}\phi)^3+3[ \Pi ^2 _{\phi}]^2-6(\partial_{\mu}\partial_{\nu}\phi)^4. \eea
Here it is explicitly appearing that the equation of motion are of the second order,  which guarantees that the underlying theory must be free from the Ostrogradski ghost instability.  Also at the quantum mechanical level NCGEFT leads to a set up where unitarity preserves.  Several generalization have been proposed of the mentioned theory using which in refs. \cite{Kobayashi:2010wa,Jain:2010ka,Gannouji:2010au,Ali:2010gr,deRham:2011by,Tsujikawa:2010sc,DeFelice:2010gb,DeFelice:2010pv,DeFelice:2010nf,Kobayashi:2010cm,Deffayet:2010qz,Burrage:2010cu,Mizuno:2010ag,Nesseris:2010pc,Khoury:2010xi,DeFelice:2010as,Kimura:2010di,Hirano:2010yf,Kamada:2010qe,Hirano:2011wj,Li:2011sd,Kobayashi:2011pc,DeFelice:2011zh,Burrage:2011bt,Liu:2011ns,Kobayashi:2011nu,PerreaultLevasseur:2011wto,deRham:2011by,Clifton:2011jh,Gao:2011mz,DeFelice:2011uc,Gao:2011qe,DeFelice:2011hq,Qiu:2011cy,Renaux-Petel:2011rmu,DeFelice:2011bh,Wang:2011dt,Kimura:2011dc,DeFelice:2011th,Appleby:2011aa,DeFelice:2011aa,Zhou:2011ix,Shirai:2012iw,deRham:2012az,Ali:2012cv,Liu:2012ww,Choudhury:2012yh,Choudhury:2012whm,Barreira:2012kk,Gubitosi:2012hu,Arroja:2013dya,Sami:2013ssa,Khoury:2013tda,Burrage:2015lla,Koyama:2015vza,Saltas:2016nkg} studied various cosmological phenomena,  where the set up needs to be embedded in curved de Sitter background to serve the purpose.  However,  to preserve the unitarity it was found that the non-covariant Lagrangians ${\cal L}^{\bf NC}_i \forall i=1,2,\cdots,5$ receive large corrections from renormalization in the mentioned curved background geometry.

\subsection{Covarinat Galileon Effective Field Theory (CGEFT)}
\label{s2b}

Now to get an inflationary solution out of the present NCGEFT set up one needs to break the corresponding exact Galilean symmetry in this context.  Now let us talk about the decoupling limit,  $M_{pl}\rightarrow\infty$ and $3H^2M^2_{pl}=V_0$ having $H$ fixed,  on which the Galilean shift symmetry is still exact and for this reason, any soft breaking which happened later is suppressed by the Planck scale in the previously mentioned coefficients.  Now it is important to note that,  despite softy breaking the Galilean shift symmetry the kinetic term and the linear potential term,  $V(\phi)=V_0-\lambda^3\phi$ with $c_1=\lambda^3$ for the Galileon do not receive any further large contribution from the renormalization procedure.  Hence we don't have any unpredictive outcomes out of the present theoretical setup.  For the rest of the purpose, we  use the fact that due to having soft breaking of the corresponding symmetry the coupling with the gravitational sector is strictly bound to be suppressed by the $\Lambda/M_{pl}$ contribution,  which helps to treat the underlying theory in the realistic theoretical regimes.

Considering the ghost free theory from Ostrogradski instability at least in the classical regime further improved version of GEFT was introduced in the curved space background in ref \cite{Deffayet:2009wt}. This version is commonly referred as the Covariantized Galileon Effective Field Theory (CGEFT).  Starting from a five dimensional covering theory in curved background such CGEFT can be constructed easily and the corresponding action of the theory is given by:
		\bea
S= \int d^4 x \sqrt{-g}\Bigg[\frac{M^2_{pl}}{2}R-V_0+ {\cal L}^{\bf C}_{\phi}\Bigg],
\eea
where we define:
\be {\cal L}^{\bf C}_{\phi}=\sum^{5}_{i=1}c_i\mathcal{L}^{\bf C}_i.\ee
The explicit expression for the $\mathcal{L}^{\bf C}_i\forall i=1,2,\cdots,5$ is given by the following expressions:
\bea
{\cal L}^{\bf C}_1 & = & \phi, \\
 {\cal L}^{\bf C}_2&=&-\frac{1}{2} (\grad \phi)^2 ,\\
	{\cal L}^{\bf C}_3&=&\frac{c_3}{\Lambda^3} (\grad \phi)^2 \Box \phi ,\\
	{\cal L}^{\bf C}_4&=& -\frac{c_4}{\Lambda^6} (\grad \phi)^2 \Big\{
					(\Box \phi)^2 - (\grad_\mu \grad_\nu \phi)
					(\grad^\mu \grad^\nu \phi)
					- \frac{1}{4} R (\grad \phi)^2
				\Big\},\\
	{\cal L}^{\bf C}_5&=& \frac{c_5}{\Lambda^9} (\grad \phi)^2 \Big\{
					(\Box \phi)^3 - 3 (\Box \phi)( \grad_\mu \grad_\nu \phi)
					(\grad^\mu	 \grad^\nu \phi)
					\nonumber\\
					&&\quad\quad\quad\quad\quad+ 2 ( \grad_\mu  \grad_\nu \phi)
					(\grad^\nu	 \grad^\alpha \phi)
					(\grad_\alpha \grad^\mu \phi)
					- 6 G_{\mu \nu} \grad^\mu \grad^\alpha \phi
					\grad^\nu \phi \grad_\alpha \phi
				\Big\}.		
	\eea
	Here $R$ and $G_{\mu\nu}$ represent the Ricci scalar and Einstein tensor for the background gravity.  It is important to note that,  the CGEFT with curved gravitational background softly break the Galilean symmetry in the present context. In the above mentioned Lagrangians the coefficients $c_i$ play the same role as mentioned before covariantization. Only one additional thing is to mention here that we have made the construction in such a way that these coefficinets are always appearing as dimensionless fashion. Also, $\Lambda$ represents the underlying mass scale of CGEFT which physically interpreted as the cutoff scale of the EFT set up. Our usual notion tells us that EFT should not be valid beyond this mentioned cutoff scale. However, it is very important to mention that if the Vainshtein effect is active then the quantum fluctuations can go beyond the cutoff scale $\Lambda$ \cite{deRham:2010eu}. In the above mentioned covariantized Lagragians, ${\cal L}^{\bf C}_4$ and ${\cal L}^{\bf C}_5$, we have incorporated the no-minimal coupling with $G_{\mu\nu}$ and $R$, which are suppressed by powers of $H/\Lambda$ in the corresponding contributions. Although it will turn out that these terms are insignificant in the inflationary regime of interest, where the Galileon self-interactions dominate nonlinearities, we still keep the nonminmal curvature couplings needed for covariantization out of thoroughness.
    Now the careful observation shows that if we fix the coefficients, $c_4 = 0 = c_5 $, then one can recover the covariantized version of the DGP model. However, if the Galileon field $\phi$ is assumed to be only pertinent during inflation, the coefficients $c_i$ are unfixed and ascertained separately from cosmological observations. The ref. \cite{Ali:2010gr} examines various cosmological limits on $c_2$, $c_3$, and $c_4$.

\section{Covariantized Galileon Effective Field Theory (CGEFT) inflation}
\label{s3}	
\subsection{Inflation in the decoupling limiting situation}	
\label{s3a}

In the previously mentioned decoupling limit let us now consider the inflationary solution in quasi de Sitter background geometry. Usually this limit is applicable in the present context when the slow variation in the potential $\Delta V$ in the inflationary effective potential during inflationary epoch satisfy the additional constraint, $|\Delta V/V|\ll 1$. In this decoupling limitnig situation the CGEFT embedded in the quasi de Sitter background is described by the scale factor, $a(t)=\exp(Ht)$, where the Hubble parameter $H$ is not exactly constant and the small deviation from the exact de Sitter solution is characterized by the first slow-roll parameter $\epsilon=-\dot{H}/H^2$.

Let us now consider only the Galileon part of the action on which performing integration by parts and removing the boundary terms during this process we get the following action for the background time dependent homogeneous Galileon field $\bar{\phi}(t)$, which can be written as:
\bea S_0=\int d^4x\,a^3 \,\Bigg\{\frac{c_2}{2}\dot{\bar{\phi}}^2_0+\frac{2c_3H}{\Lambda^3}\dot{\bar{\phi}}^3_0+\frac{9c_4H^2}{2\Lambda^6}\dot{\bar{\phi}}^4_0+\frac{6c_5H^3}{\Lambda^9}\dot{\bar{\phi}}^5_0+\lambda^3\bar{\phi}_0\Bigg\},\eea
which after defining the following new coupling constant:
\bea \label{Z} Z\equiv \frac{H\dot{\bar{\phi}}_0}{\Lambda^3},\eea
can be further recast in the following simplified form:
\bea S_0=\int d^4x\,a^3 \,\Bigg\{\dot{\bar{\phi}}^2_0\Bigg(\frac{c_2}{2}+2c_3Z+\frac{9c_4}{2}Z^2+6c_5Z^3\Bigg)+\lambda^3\bar{\phi}_0\Bigg\}.\eea
This give rise to following solution:
\bea \dot{\bar{\phi}}_0=\frac{\Lambda^3}{12H}\frac{c_2}{c_3}\Bigg[-1+\sqrt{1+\frac{8c_3}{c^2_2}\frac{\lambda^3}{\Lambda^3}}\Bigg]=
\left\{
	\begin{array}{ll}
		\displaystyle \frac{\lambda^3}{3c_2H}\quad\quad\quad & \mbox{when}\quad  Z\ll 1  \;(\rm Weakly-coupled \;solution)  \\ \\
			\displaystyle 
			\displaystyle \sqrt{\frac{\Lambda^3}{18c_3}\frac{\lambda^3}{H^2}}\quad\quad\quad & \mbox{when }  Z\gg 1  \;(\rm Strongly-coupled \;solution)
	\end{array}
\right. \eea
In the weakly coupled regime ($Z\ll 1$) the underlying theory approaches to the usual canonical slow-roll inflation. On the other hand, in the strong coupling regime ($Z\gg 1$) the theory approaches to the DGP model. When $Z\gtrsim 1$, the underlying theory interpolates between the weak and strong coupling regime and in that case the Galileon interactions become significant to serve the purpose of inflation. In such a situation the relative contributions are controlled compared to the higher derivative to the lower derivative terms due to having positive powers of the coupling parameter $Z$ in this construction in the decoupling limit. The best possible explanation of this fact is that is the decoupling limit due to having no interaction with the gravitational sector the non-minimal coupling with the gravity becomes less important, though we need to take into account the non-linear interactions due to having various types of derive terms in the Galileon sector. However, in the weakly coupled regime ($Z\ll 1$) one cannot neglect the mixing contribution with the non-minimal gravitational interactions with Galileon, which it is expect to give rise to significant changes in the features of canonical single-field inflation. We don't bother about this situation because in this paper we restrict our interest to the intermediate regime where the coupling parameter $Z\gtrsim 1$.

\subsection{Underlying connection with the good-old Effective Field Theory of inflation}
\label{s3b}

 \subsubsection{Effective action in the Unitary gauge}
\label{s3b1}
 
In ref. \cite{Cheung:2007st} the authors has explicitly shown that the underlying properties of Effective Field Theory (EFT) set up can able to fix the structure and the behaviour of perturbations for an inflationary paradigm in presence of quasi de Sitter background geometrical construction and for this reason it can be interpreted in a completely model independent fashion. It can be very easily mapped in terms of the large classes of $P(X,\phi)$ theories studied in refs. \cite{Alishahiha:2004eh,Mazumdar:2001mm,Choudhury:2002xu,Panda:2005sg,Chingangbam:2004ng,Armendariz-Picon:1999hyi,Garriga:1999vw,Choudhury:2017glj,Naskar:2017ekm,Choudhury:2015pqa,Choudhury:2014sua,Choudhury:2014kma,Choudhury:2013iaa,Baumann:2022mni,Baumann:2018muz,Baumann:2015nta,Baumann:2014nda,Baumann:2009ds} . Such an important theoretical construction in made by implementing broken time diffeomorphism symmetry and non-linear realization under the Lorentz invariance in this context. In this section our prime objective is to construct the good old EFT generalized version of CGEFT inflationary action valid for small quantum fluctuations. However, due to having an additional Galilean symmetry, it is extremely important to know about the modifications and the corresponding constraints one needs to seriously take care of during such generalization.

In ref. \cite{Cheung:2007st} the authors constructed the EFT setup with the help of a single scalar field inflationary paradigm. Next, the effective action is written in a specific gravitational gauge where the constant time slices overlap with uniform $\phi$ slices. Surprisingly, though in ref. \cite{Cheung:2007st} have not used any specific form of the effective potential and the kinetic interactions in terms of the scalar field $\phi$. But in the end, the underlying theoretical construction performed in ref. \cite{Cheung:2007st} exactly mimics a general $P(X,\phi)$ type of theories, which has been recently shown in refs. \cite{Choudhury:2017glj,Choudhury:2023jlt,Choudhury:2023rks}. To serve our purpose in the present context of the discussion, we consider a unit vector $n^{\mu}$ which is defined in the direction of normal of the constant time slices, and given by the following expression:
\bea n_{\mu}&=&\frac{\partial_{\mu}t}{\sqrt{-g^{\mu \nu}\partial_{\mu}t \partial_{\nu}t}}
=\frac{\delta_{\mu}^0}{\sqrt{-g^{00}}}.\eea
The EFT action for the cosmological perturbations is further constructed in terms of the operators which are invariant under the spatial diffeomorphism symmetry associated with the reparametrization of the dimension spatially induced metric, which is given by the following expression:
\bea h_{\mu \nu}&=&g_{\mu \nu}+n_{\mu} n_{\nu}.\eea
Using the Ricci scalar, time-dependent part of the metric tensor, $g^{00}$ and the extrinsic curvature tensor $K_{\mu\nu}$, which is defined as:
\bea K_{\mu \nu}&=&h^{\sigma}_{\mu}\nabla_{\sigma} n_{\nu}\nonumber\\
&=&\left[\frac{\delta^{0}_{\mu}\partial_{\nu}g^{00}+\delta^{0}_{\nu}\partial_{\mu}g^{00}}{2(-g^{00})^{3/2}}
+\frac{\delta^{0}_{\mu}\delta^{0}_{\nu}g^{0\sigma}\partial_{\sigma}g^{00}}{2(-g^{00})^{5/2}}-\frac{g^{0\rho}\left(\partial_{\mu}g_{\rho\nu}+\partial_{\nu}g_{\rho\mu}-\partial_{\rho}g_{\mu\nu}\right)}{2(-g^{00})^{1/2}}\right],\eea
associated with constant time slices the generalized version of the EFT action can be constructed in the present context and described by the following form of the representative action:
\bea
		S &=&
		\int \d^4x \; \sqrt{-g} \;
		\Bigg[ \
			\frac{M^2_{pl}}{2} R
			- c(t) g^{00}
			- \Lambda(t)
			+ \frac{1}{2} M^4_2(t) (g^{00}+1)^2
			+ \frac{1}{3} M^4_3(t) (g^{00}+1)^3
		\nonumber\\ && \mbox{}
		\quad\quad\quad\quad\quad\quad- \frac{\bar{M}^3_1(t)}{2} (g^{00}+1) \delta {K^\mu}_{\mu}
		- \frac{\bar{M}^2_2(t)}{2} ({\delta K^\mu}_\mu)^2
		- \frac{\bar{M}^2_3(t)}{2} \delta K^{\mu\nu} \delta K_{\mu\nu}
		\nonumber\\ && \mbox{}\quad\quad\quad\quad\quad\quad
		- \frac{\bar{M}^3_4(t)}{2} (g^{00}+1)^2 \delta {K^\mu}_{\mu}
		- \frac{\bar{M}^2_5(t)}{2} (g^{00}+1) ({\delta K^\mu}_\mu)^2
		- \frac{\bar{M}^2_6(t)}{2} (g^{00}+1)
			\delta K^{\mu\nu} \delta K_{\mu\nu}
		\nonumber\\ && \mbox{}\quad\quad\quad\quad\quad\quad
		- \frac{\bar{M}_7(t)}{2} ({\delta K^\mu}_\mu)^3
		- \frac{\bar{M}_8(t)}{2} ({\delta K^\mu}_\mu)
			(\delta K^{\rho \sigma} \delta K_{\rho\sigma})
		- \frac{\bar{M}_9(t)}{2} \delta K^{\mu \nu} \delta K_{\nu \sigma}
			{\delta K^\sigma}_\mu
		+\cdots \Bigg],
	\eea
 In the above mentioned EFT action we have organized the operators in terms of the following fluctuations around an unperturbed FLRW background having quasi de Sitter solution, and given by the following expressions:
\bea \delta g^{00}&=&\left(g^{00}-\bar{g}^{00}\right)=\left(g^{00}+1\right),\\ 
\delta K_{\mu\nu}&=&\left(K_{\mu\nu}-a^2Hh_{\mu\nu}\right).\eea 
Here it is important to note that the time dependent coefficients, $c(t)$ and $\Lambda(t)$, are precisely non-zero on the quasi de Sitter cosmological background that we are considering for our present analysis, which in turn fixes these time dependent coefficients in terms of the Hubble parameter $H(t)$ during inflationary epoch. On the other hand, the background cosmological evolution cannot able to fix the other time dependent coefficients $M_i(t)\forall i=2,3$ and $\bar{M}_4(t)\forall i=1,2,\cdots,9$ and captures the information regarding different models under consideration. This means that for CGEFT set up one should have very specific choices of these parameters. Because of this fact it is expected that for CGEFT one should get clearly distinctive features in the present context of discussion. 

Now to implement the above mentioned action to describe the cosmological perturbation theory order by order for scalar modes generated from perturbation, one need to recast this action in a more tractable form. This can be done by making use of the following unitary gauge transformation, where broken time diffeomorphism is automatically implemented very clearly:
\bea t\longrightarrow \tilde{t}=t-\pi(t,{\bf x}).\eea
As an immediate outcome of the above mentioned gauge transformation the equal time hypersurfaces are deformed by the amount $\pi(t,{\bf x})$, where the deformation parameter is a space-time dependent quantity. This specific parameter is identified as the Goldstone mode and the corresponding trick implemented in this context is commonly known as  St$\ddot{u}$ckelberg trick. Here lies a very common connection with the $SU(N)$ gauge theory. Under the above mentioned gauge transformation each of the operators written in the above mentioned generalized EFT action transforms which contain series of terms containing the spatial, temporal derivatives and space-time mixing contributions of the Goldstone modes $\pi(t,{\bf x})$.
For more details see the refs. \cite{Choudhury:2017glj,Choudhury:2023rks,Choudhury:2021brg}, where all of these transformations are explicitly pointed very clearly. In this fashion, a consistent elegant theoretical description of the underlying physical set up is built using the tools and techniques of EFT method by making use of the lowest dimensional Wilsonian operators that are compatible with the underlying symmetries in this context.

 \subsubsection{Decoupling limiting situation}
\label{s3b2}

In this subsection we discuss about the implementation of the decoupling limit without which the implementation of the EFT tools and techniques becomes extremely difficult in the context of cosmological perturbation theory. Such difficulties arises because of having the mixing contribution of the Goldstone modes with the gravitational sector through the metric. Now in the presence of large non-linear contributions of the self-interacting terms of the Goldstone modes implementation of the decoupling limit helps to treat the perturbation theory in a more trustworthy fashion by neglecting the background gravity. Such limit can be implemented by taking $M_{pl}\rightarrow\infty$ keeping the Hubble parameter $H$ fixed in the quasi de Sitter cosmological background. 
We quickly summarise the underlying argument in this case.
The Ricci scalar will provide the most pertinent kinetic term for the measure variation $\delta g^{00}$. Therefore, by re-scaling the fluctuation:
\bea \delta g^{00} \rightarrow \delta g^{00}_c = M_{pl} \delta g^{00},\eea 
we can proceed to canonical normalisation. The nonlinear operator in the EFT action produce the most significant kinetic term for the Goldstone modes. If it has minimal derivatives, it will take the shape $M^4 \dot{\pi}^2$, where $M$ is some mixture of the coupling parameters $M_i$ or $\bar{M}_i$. In this context the canonically normalized Goldstone modes are given by the following expression:
\bea \pi_c=M^2\pi.\eea
For wave numbers $k$ that fulfil the following constraint:
\be k \gtrsim E_{\mathrm{mix}} = \frac{M^2}{M_{pl}},\ee a mixing term such as $M^4 \dot{\pi} \delta g^{00}$ is insignificant at the quadratic level in contrast to the Goldstone kinetic term. The same is true for cubic terms, where the leading blending term $M^4 \dot{\pi}^2 \delta g^{00}$ is insignificant compared to $M^4 \dot{\pi}^3$ under the same circumstance. Similar justifications can be made if the prime cubic contributions appearing with a physical scale other than $M$ or if the most significant kinetic term for Goldstone mode has higher derivatives. For our forecasts to be accurate to a relative inaccuracy of order $E_{\mathrm{mix}}/ H$, and the corresponding mixing scale $E_{\mathrm{mix}}\ll H$ in the decoupling limit. The measure can be assumed to be unperturbed in the limit where we operate in the sections that follow.Since the unperturbed measure is spatially smooth and can be used as the backdrop of quasi de Sitter geometry, working in the uniform curvature gauge is the most practical option in this context.

\section{Tree level scalar power spectrum from CGEFT}
\label{s4}
Here we discuss the second order perturbation generated for comoving curvature perturbation from the underlying CGEFT theoretical set up. Next we construct the classical equation of motion for the generated comoving curvature perturbation modes in Fourier space. In the corresponding literature such classical equation of motion is known as Mukhanov Sasaki (MS) equation which we have derived from the underlying CGEFT set up. Then we have solved the MS equation in three regions of interest, which are first Slow Roll (SRI) phase, then an Ultra Slow Roll (USR) phase and finally second Slow Roll (SRII) phase followed by the ending of inflation just after that.
To analytically solve the MS equation in the SRI phase we have used the Bunch Davies quantum initial condition provided in terms of the specific choices of the Bogoliubov coefficients which satisfy appropriate normalization condition. Once we provide such information the corresponding mode computed from the SRI phase will going to be automatically fixed in terms of the provided information. Then with the help of continuity of the modes and its corresponding canonically conjugate momenta are continuous at the SRI to USR and USR to SRII transition scales one can compute the explicit expressions for Bogoliubov coefficients in the USR and SRII phases, which in tern fix the structure of the solution of the MS equations computed in these two mentioned phases. The new structure of the computed Bogoliubov coefficients in these mentioned two phases confirms that the quantum initial vacuum is shifted from Bunch Davies and basically we are dealing with non Bunch Davies quantum states in these mentioned two phases. Next, we elaborately discuss the canonical quantization of the of the comoving curvature perturbation whose corresponding Fourier mode solutions are obtained in SRI, USR and SRII phases by solving MS equation. Next, we compute the result for the power spectrum from comoving curvature perturbation at the tree level using the underlying CGEFT set up.
\subsection{Second order perturbation from scalar mode from CGEFT}
\label{s4a}

Following the discussions in the previous section in this section our aim is to study the effect of primordial fluctuations around the quasi de Sitter cosmological background in the decoupling limit. In terms of the Goldstone modes the second oreder perturbed action in the present context can be described by the following equation:
\bea S^{(2)}_{\pi}&=&\int d^4x \, a^3\Bigg[{\cal A}\, \dot{\pi}^2-\frac{{\cal B}}{a^2}\,\left(\partial\pi\right)^2\Bigg]=\int d^4x  
 \,a^3{\cal A}\,\Bigg[ \dot{\pi}^2-\frac{c^2_s}{a^2}\,\left(\partial\pi\right)^2\Bigg].\eea
Here the effective sound speed $c_s$ for the CGEFT set up can be expressed as:
\bea\label{cs1} c_s=\sqrt{\frac{{\cal B}}{{\cal A}}},\eea
where the newly defined quantities ${\cal A}$ and ${\cal B}$ are the time-dependent coefficients, can be expressed in terms of the coupling constants of the original CGEFT action by the following expressions:
\bea {\cal A}:&\equiv& \frac{\dot{\bar{\phi}}^2_0}{2}\Bigg(c_2+12c_3Z+54c_4Z^2+120c_5Z^3\Bigg),\\
    {\cal B}:&\equiv& \frac{\dot{\bar{\phi}}^2_0}{2}\Bigg\{c_2+4c_3\Bigg(2Z+\frac{\ddot{\bar{\phi}}_0}{\Lambda^3}\Bigg)+2c_4\Bigg[13Z^2+\frac{6}{\Lambda^6}\bigg(\dot{H}\dot{\bar{\phi}}^2_0+2H\dot{\bar{\phi}}_0\ddot{\bar{\phi}}_0\bigg)\Bigg]+\frac{24c_5}{\Lambda^9}H\dot{\bar{\phi}}^2_0\Bigg[2\dot{\bar{\phi}}_0\bigg(H^2+\dot{H}\bigg)+3H\ddot{\bar{\phi}}_0\Bigg]\Bigg\}\nonumber\\
    &=& \frac{\dot{\bar{\phi}}^2_0}{2}\Bigg\{c_2+4c_3\Bigg(2Z-\frac{H\dot{\bar{\phi}}_0}{\Lambda^3}\eta\Bigg)+2c_4\Bigg[13Z^2-\frac{6}{\Lambda^6}\dot{\bar{\phi}}^2_0H^2\big(\epsilon+2\eta\big)\Bigg]-\frac{24c_5}{\Lambda^9}H^3\dot{\bar{\phi}}^3_0\big(2\epsilon+1\big)\Bigg\}.\quad\quad\eea
    Here we have introduced the first and second slow-roll parameter by the following expression:
     \bea \epsilon=-\frac{\dot{H}}{H^2},\quad\quad\quad\quad\eta=-\frac{\ddot{\bar{\phi}}_0}{H\dot{\bar{\phi}}_0}.\eea
    which will be going to be extremely useful for the rest of the analysis performed in this paper.
 Now from the good-old Goldstone EFT set up we know that the effective sound speed parameter $c_s$ can be expressed in terms of the EFT coefficient $M_2$, by the following expression:
\bea \label{cs2} c_s=\frac{1}{\displaystyle\sqrt{1-\frac{2M^4_2}{\dot{H}M^2_{pl}}}}.\eea
Further comparing equation (\ref{cs1}) and equation (\ref{cs2}), one can express the EFT coefficient $M_2$ in terms of the coefficients of the CGEFT set up by the following expression:
\bea \frac{M^4_2}{\dot{H}M^2_{pl}}=\frac{1}{2}\Bigg(1-\frac{{\cal A}}{{\cal B}}\Bigg).\eea

 The point to be noted here that, the spatial component of the metric perturbation is described by the following equation:
   \bea g_{ij}\sim a^{2}(t)\left[\left(1+2\zeta(t,{\bf x})\right)\delta_{ij}\right]~~\forall~~~i=1,2,3,\eea
   where the scale factor is $a(t)=\exp(Ht)$, which we have already mentioned before. In this equation the comoving curvature perturbation is described by the symbol $\zeta(t,{\bf x})$, which plays the most significant role in the present computational purpose.
   
Further replacing the Goldstone modes in terms of the comoving curvature perturbation using the following representative equation \footnote{More details on this issue can be found in the refs. \cite{Burrage:2010cu,Cheung:2007st,Choudhury:2017glj,Choudhury:2023jlt,Choudhury:2023rks}.}:
\bea \zeta(t,{\bf x})\approx -H\pi(t,{\bf x}).\eea
Now, since we we clarly know the underlying connecting relationship between the Goldstone mode and the comoving curvature perturbation, we can immediately express the second order goldstone mode action in terms of the curvature perturbation by the following representative equation:
 \bea 
  	S^{(2)}_{\zeta}&=&\displaystyle \int d^{4}x ~a^3\, \frac{{\cal A}}{H^2}\Bigg(\dot{\zeta}^2-c^2_s\frac{\left(\partial_{i}\zeta\right)^2}{a^2}\Bigg)=\displaystyle \int d^{4}x ~a^3\, \frac{{\cal B}}{c^2_sH^2}\Bigg(\dot{\zeta}^2-c^2_s\frac{\left(\partial_{i}\zeta\right)^2}{a^2}\Bigg).~~~~~\quad\quad\eea
Additionally, for the sake of simplicity, we will now use the conformal time coordinate in place of the physical time coordinate for the remainder of the calculation, which allows us to express the second-order perturbed action as follows \footnote{Here it is important to note that, the conformal time coordinate $\tau$ and the physical time coordinate $t$ are related via, $d\tau=dt/a$. Here in quasi de Sitter background the scale factor in terms of the conformal time coordinate can be expressed as, $a(\tau)=-1/H\tau$, where $-\infty<\tau<0$.}:
  \bea S^{(2)}_{\zeta}=\displaystyle \int d\tau\;  d^3x\;  a^2\; \frac{{\cal A}}{H^2}\big(\zeta^{'2}-c^2_s\left(\partial_i\zeta\right)^2\big)=\displaystyle \int d\tau\;  d^3x\;  a^2\; \frac{{\cal B}}{c^2_sH^2}\big(\zeta^{'2}-c^2_s\left(\partial_i\zeta\right)^2\big).~~~~\quad\eea

\subsection{Constructing Mukhanov Sasaki equation from CGEFT}
\label{s4b}

Now, a new variable is defined for the purpose of redefining of the comoving curvature perturbation field, and it is given by:
 \bea v(\tau, {\bf x})=z(\tau)\zeta(\tau, {\bf x}) \quad{\rm with}\quad z(\tau)=\frac{a\sqrt{2{\cal A}}}{H^2}=\frac{a\sqrt{2{\cal B}}}{c_sH^2},\eea
which, in the context of the topic at hand, is usually referred to as the Mukhanov Sasaki (MS) variable. When written in terms of the MS variable, the second order perturbed action described above has the following canonically normalised form:
\bea
S^{(2)}_{\zeta}=\frac{1}{2}\int d\tau\;  d^3x\;  \bigg(v^{'2}(\tau, {\bf x})-c^2_s\left(\partial_iv(\tau, {\bf x})\right)^2+\frac{z^{''}(\tau)}{z(\tau)}v^{2}(\tau, {\bf x})\bigg).
   \eea
   Next, we will formulate the aforementioned action in the Fourier space using the ansatz for the Fourier transformation as follows:
  \bea
   v(\tau,{\bf x})=\int\frac{d^3{\bf k}}{(2\pi)^3}\; e^{i{\bf k}.{\bf x}}\;v_{\bf k}(\tau).
   \eea
   The aforementioned action can also be remade as one of the following scalar modes after being Fourier transformed:
 \bea
S^{(2)}_{\zeta}&=&\frac{1}{2}\int \frac{d^3{\bf k}}{(2\pi)^3}\;d\tau\; e^{i{\bf k}.{\bf x}}\;\bigg(|v^{'}_{\bf k}(\tau)|^{2}-\omega^2(k,c_s,\tau)|v_{\bf k}(\tau)|^{2}\bigg).
   \eea
   The MS equation for the scalar perturbed modes can then be written by varying the previously specified action,
  \bea
 v^{''}_{\bf k}(\tau)+\omega^2(k,c_s,\tau)v_{\bf k}(\tau)=0\,,
   \eea
   where the following equation provides the expression for the effective time-dependent frequency, 
\bea \omega^2(k,c_s,\tau):=\left(c^2_sk^2-\frac{z^{''}(\tau)}{z(\tau)}\right)\quad\displaystyle{\rm where}\quad\frac{z^{''}(\tau)}{z(\tau)}\approx\frac{2}{\tau^2}.\eea
The normalisation condition given below, which is expressed in terms of the Klein Gordon product for the scalar perturbed modes, fixes the formal structure of the general answer:
\be \left(v_{\bf k}(\tau),v^{'}_{\bf k}(\tau)\right)_{\bf KG}:=\bigg(v^{'*}_{\bf k}(\tau)v_{\bf k}(\tau)-v^{'}_{\bf k}(\tau)v^{*}_{\bf k}(\tau)\bigg)=i.\ee
We will solve the MS equation for a given important physical framework in the next subsection. This will be very helpful for the remainder of the subject material of this article.

\subsection{Classical solution of Mukhanov Sasaki equation from CGEFT}
\label{s4c}

In this article, we'll take a look at a physical structure made up of the points-by-point chronological sequence listed below 
:
  \begin{enumerate}
  
      \item \underline{\bf Regime I (SRI):} First, we take into account a Slow Roll (SRI) regime that lasts for the conformal time scale $\tau<\tau_s$. The SRI transitions to an Ultra Slow Roll (USR) region at $\tau=\tau_s$. This entails that SRI terminates at $\tau=\tau_s$ in this design.

      \item \underline{\bf Regime II (USR):} The Ultra Slow Roll (USR) regime is the next thing we take into consideration. It has a starting point at the conformal time scale $\tau=\tau_s$ and an ending point at scale $\tau=\tau_e$. The second transition scale $\tau=\tau_e$, where USR to the second Slow Roll (SRII) transition occurs, is handled in this construction as the structure.

      \item \underline{\bf Regime III (SRII):} Last but not least, we take into account the second Slow Roll (SRII) regime, which begins at $\tau=\tau_e$ and the inflation stops shortly thereafter at $\tau=\tau_{\rm end}$.

  \end{enumerate}
In the following parts of this article, it is our responsibility to specifically examine the classical answer and its quantum consequences from each of the three regimes stated.

\subsubsection{Region I: First Slow Roll (SRI) region}
\label{s4c1}

The general solution of the MS equation for the scalar perturbed mode is given by the following formula, which is expressly true during the first SR (SRI) phase ($\tau<\tau_s$):
\bea
   v_{\bf k}(\tau)&=&\frac{\alpha^{(1)}_{\bf k}}{\sqrt{2c_sk}}\left(1-\frac{i}{kc_s\tau}\right)\; e^{-ikc_s\tau}+\frac{\beta^{(1)}_{\bf k}}{\sqrt{2c_sk}}\left(1+\frac{i}{kc_s\tau}\right)\; e^{ikc_s\tau},
   \eea
   where choosing the correct initial condition sets the answer in terms of the coefficients $\alpha^{(1)}_{\bf k}$ and $\beta^{(1)}_{\bf k}$, which are shown in the above equation. As long as the empirical requirements from the CMB for inflation are met within the SR period, it is theoretically possible to select any starting quantum vacuum state. The most well-known, or Bunch Davies quantum vacuum state, is the one we ultimately opt for. This is essentially a Euclidean vacuum, and it is described by the following equation, which in our case fixes the initial state in the first SR (SRI) period. 
\bea \label{bG1a}&&\alpha^{(1)}_{\bf k}=1, \\
 \label{bG1b}&&\beta^{(1)}_{\bf k}=0.\eea
 Such an initial condition will be very helpful for the subsequent calculations carried out in the remainder of the document. The scalar mode function can be stated in terms of the simplest abbreviation in the first SR (SRI) area after applying the aforementioned starting condition: 
 \bea
 v_{\bf k}(\tau)=\frac{1}{\sqrt{2c_sk}}\left(1-\frac{i}{kc_s\tau}\right)\; e^{-ikc_s\tau}.
 \eea
 One can further write down the formula for the comoving curvature perturbation in terms of the conformal time, effective sound speed parameter $c_s$, and momentum scale by making use of the previously mentioned solution of MS equation in the first SR regime, $\tau<\tau_s$:
 \bea
 \zeta_{\bf k}(\tau)=-H\pi_{\bf k}(\tau)=\frac{v_{\bf k}(\tau)}{z}&=&\left(\frac{iH^2}{2\sqrt{{\cal A}}}\right)\frac{1}{(c_sk)^{3/2}}\left(1+ikc_s\tau\right)\; e^{-ikc_s\tau}\nonumber\\
 &=&\left(\frac{ic_sH^2}{2\sqrt{{\cal B}}}\right)\frac{1}{(c_sk)^{3/2}}\left(1+ikc_s\tau\right)\; e^{-ikc_s\tau}.\quad
 \eea
Although $\epsilon$ changes very slowly with time in the SRI area, it is roughly a constant value.

\subsubsection{Region II: Ultra Slow Roll (USR) region}
\label{s4c2}

The conformal time scale window $\tau_s\leq \tau \leq \tau_e$, which contains the region where the associated scalar modes are valid, can be used to visualise the Ultra Slow Roll (USR) regime. The crossover scale from the first SR (SRI) to USR is denoted in this definition by $\tau_s$. On the other hand, the time scale $\tau_e$ describes the conclusion of the USR regime as well as the inflationary paradigm. The time dependence of the first slow-roll parameter, which can be stated in terms of the first SR component as follows, can be explicitly written down in the USR regime:
\bea \epsilon(\tau)=\epsilon \;\left(\frac{a(\tau_s)}{a(\tau)}\right)^{6}=\epsilon  \;\left(\frac{\tau}{\tau_s}\right)^{6}\quad\quad\quad{\rm where}\quad\quad\tau_s\leq\tau\leq \tau_e.\eea
In this instance, $\epsilon$ is the first slow-roll parameter in the SR region that we specifically specified in the first part of the discussion. The aforementioned mathematical form demonstrates that this parameter is roughly a constant amount at the point where the first SR (SRI) to USR shift occurs, i.e. at $\tau=\tau_s$ where we have $\epsilon(\tau_s)=\epsilon$, which actually corresponds to the first SR (SRI) regime. The first SR (SRI) to USR transition, which occurs at the scale $\tau>\tau_s$, is when the departure from the constant pattern first manifests. The fact that we explicitly took into account an abrupt shift from the first SR (SRI) to the USR region for the current computational purpose is essential to note because it will be highly helpful information for the remainder of the article. The solution of the MS equation in USR period for the comoving curvature perturbation can be stated by the following shortened formula:
\bea
 \zeta_{\bf k}(\tau)=-H\pi_{\bf k}(\tau)=\frac{v_{\bf k}(\tau)}{z}&=&\left(\frac{iH^2}{2\sqrt{{\cal A}}}\right)\left(\frac{\tau_s}{\tau}\right)^{3}\frac{1}{(c_sk)^{3/2}}\times\bigg[\alpha^{(2)}_{\bf k}\left(1+ikc_s\tau\right)\; e^{-ikc_s\tau}-\beta^{(2)}_{\bf k}\left(1-ikc_s\tau\right)\; e^{ikc_s\tau}\bigg]\nonumber\\
 &=&\left(\frac{ic_sH^2}{2\sqrt{{\cal B}}}\right)\left(\frac{\tau_s}{\tau}\right)^{3}\frac{1}{(c_sk)^{3/2}}\times\bigg[\alpha^{(2)}_{\bf k}\left(1+ikc_s\tau\right)\; e^{-ikc_s\tau}-\beta^{(2)}_{\bf k}\left(1-ikc_s\tau\right)\; e^{ikc_s\tau}\bigg],\quad \quad
 \eea
It is crucial to note that, the coefficients $\alpha^{(2)}_{\bf k}$ and $\beta^{(2)}_{\bf k}$ in the aforementioned solution, obtained in the USR region, can all be expressed in terms of the initial condition fixed in terms of Bunch Davies vacuum in the first SR (SRI) region via Bogoliubov transformations. This further suggests that the underlying structure of the vacuum state alters in the USR regime compared to the Bunch Davies initial state. Our goal is to find these Bogoliubov coefficients in the USR regime, $\alpha^{(2)}_{\bf k}$ and $\beta^{(2)}_{\bf k}$. It is possible to achieve this by using the following two boundary conditions, which, in theoretical framework, can be understood as the Israel junction condition that we must apply at the first SR (SRI) to USR transition scale, $\tau=\tau_s$ and are provided by the following expressions:
\begin{enumerate}
    \item $\left[\zeta_{\bf k}(\tau)\right]_{\rm SRI, \tau=\tau_s}= \left[\zeta_{\bf k}(\tau)\right]_{\rm USR, \tau=\tau_s}$, i.e. the continuity of the modes obtained from SRI and USR at the crossover point $\tau=\tau_s$.

    \item $\left[\zeta^{'}_{\bf k}(\tau)\right]_{\rm SRI, \tau=\tau_s}= \left[\zeta^{'}_{\bf k}(\tau)\right]_{\rm USR, \tau=\tau_s}$, i.e. the continuity of the conjugate momenta obtained from SRI and USR at the crossover point $\tau=\tau_s$.
\end{enumerate}
The following closed version of the Bogoliubov coefficients, $\alpha^{(2)}_{\bf k}$ and $\beta^{(2)}_{\bf k}$ is obtained by solving the above two constraints that result from applying the aforementioned junction conditions:
\bea \label{bG2a}\alpha^{(2)}_{\bf k}&=&1-\frac{3}{2ik^{3}c^{3}_s\tau^{3}_s}\left(1+k^{2}c^{2}_s\tau^{2}_s\right),\\
\label{bG2b}\beta^{(2)}_{\bf k}&=&-\frac{3}{2ik^{3}c^{3}_s\tau^{3}_s}\left(1+ikc_s\tau_s\right)^{2}\; e^{-2ikc_s\tau_s}.\eea
The changed structure of the quantum vacuum state is fixed because the structure of the Bogoliubov coefficients is now fixed at the first SR (SRI) to USR transition point $\tau=\tau_s$. For the analysis carried out in the remaining sections of the document, this knowledge will be of utmost value.

\subsubsection{Region III: Second Slow Roll (SRII) region}
\label{s4c3}

The conformal time scale window $\tau_e\leq \tau\leq \tau_{\rm end}$, which can be seen in the second Slow Roll (SRII) regime, will now be the centre of our attention. The temporal scale, $\tau_e$, is used in this definition to identify the scale at which the USR and SRII transition from one another. The time scales, $\tau_e$ and $\tau_{\rm end}$, on the other hand, characterise the conclusion of the USR era as well as the inflationary paradigm. The first slow-roll parameter can be stated directly in terms of the first SR (SRI) parameter in the SRII regime as follows:
\bea \epsilon(\tau)=\epsilon \;\left(\frac{a(\tau_s)}{a(\tau_e)}\right)^{6}=\epsilon  \;\left(\frac{\tau_e}{\tau_s}\right)^{6}\quad\quad\quad{\rm where}\quad\quad\tau_e\leq\tau\leq \tau_{\rm end}.\eea 
The first slow-roll parameter in the SRI region, $\epsilon$, is present here and was clearly stated in the first part of the talk. The mathematical form noted above demonstrates that this parameter is roughly a non-constant number at the transition of the USR to SRII scale, which is at $\tau=\tau_e$, which is actually at the end of the USR regime. Although this number won't change between the time intervals $\tau_e<\tau<\tau_{\rm end}$, the deviation from the constant behaviour persists up to the time scale equivalent to the end of inflation which appears at $\tau=\tau_{\rm end}$. Here, we examine a sudden change at the temporal scale $\tau=\tau_e$. The solution of the MS equation alters appropriately in the current situation when this particular time-dependent behaviour and the non-constancy of the first slow-roll parameter are taken into account.
As previously mentioned, we must take into account a second transition from the USR to the second SR (SRII) region, which is anticipated to take place at the transition point $\tau=\tau_e$. In the region $\tau>\tau_e$, the SR features continue with the non-constant value of the first slow-roll parameter. Therefore, the following equation can be used to describe the solution of the MS equation in terms of the scalar modes in the regime $\tau_e\leq\tau\leq \tau_{\rm end}$:
\bea
 \zeta_{\bf k}(\tau)=-H\pi_{\bf k}(\tau)=\frac{v_{\bf k}(\tau)}{z}&=&\left(\frac{iH^2}{2\sqrt{{\cal A}}}\right)\left(\frac{\tau_s}{\tau_e}\right)^{3}\frac{1}{(c_sk)^{3/2}}\times\bigg[\alpha^{(3)}_{\bf k}\left(1+ikc_s\tau\right)\; e^{-ikc_s\tau}-\beta^{(3)}_{\bf k}\left(1-ikc_s\tau\right)\; e^{ikc_s\tau}\bigg]\nonumber\\
 &=&\left(\frac{ic_sH^2}{2\sqrt{{\cal B}}}\right)\left(\frac{\tau_s}{\tau_e}\right)^{3}\frac{1}{(c_sk)^{3/2}}\times\bigg[\alpha^{(3)}_{\bf k}\left(1+ikc_s\tau\right)\; e^{-ikc_s\tau}-\beta^{(3)}_{\bf k}\left(1-ikc_s\tau\right)\; e^{ikc_s\tau}\bigg],\quad\quad
 \eea
It is important to note that, in the aforementioned solution obtained in the second SR region (SRII), the Bogoliubov coefficients $\alpha^{(3)}_{\bf k}$ and $\beta^{(3)}_{\bf k}$ can be expressed in terms of the boundary condition fixed in terms of new modified vacuum in the USR region via Bogoliubov transformations. This further suggests that the vacuum state's underlying structure in the second SR (SRII) region differs from the vacuum state earlier computed in the USR region. Our goal is to directly find the Bogoliubov coefficients $\alpha^{(3)}_{\bf k}$ and $\beta^{(3)}_{\bf k}$ in the second SR (SRII) regime. Applying the following two boundary conditions, which are theoretically equivalent to the Israel junction condition that we must apply at the USR to the second SR (SRII) transition scale, $\tau=\tau_e$, is one way to achieve this goal:
\begin{enumerate}
    \item $\left[\zeta_{\bf k}(\tau)\right]_{\rm USR, \tau=\tau_s}= \left[\zeta_{\bf k}(\tau)\right]_{\rm SRII, \tau=\tau_s}$, i.e. the continuity of the modes obtained from USR and SRII at the crossover point $\tau=\tau_e$.

    \item $\left[\zeta^{'}_{\bf k}(\tau)\right]_{\rm USR, \tau=\tau_s}= \left[\zeta^{'}_{\bf k}(\tau)\right]_{\rm SRII, \tau=\tau_s}$, i.e. the continuity of the conjugate momenta obtained from USR and SRII at the crossover point $\tau=\tau_e$.
\end{enumerate}
The following closed version of the Bogoliubov coefficients $\alpha^{(3)}_{\bf k}$ and $\beta^{(3)}_{\bf k}$ in the second SR (SRII) regime is obtained by imposing the aforementioned junction conditions and solving two constraints:
\bea \label{bG3a}\alpha^{(3)}_{\bf k}&=&-\frac{1}{4k^6c^6_s\tau^3_s\tau^3_e}\Bigg[9\left(kc_s\tau_s-i\right)^2\left(kc_s\tau_e+i\right)^2 e^{2ikc_s(\tau_e-\tau_s)}\nonumber\\
&&\quad\quad\quad\quad\quad\quad\quad\quad\quad\quad\quad\quad\quad\quad\quad\quad-
\left\{k^2c^2_s\tau^2_e\left(2kc_s\tau_e-3i\right)-3i\right\}\left\{k^2c^2_s\tau^2_s\left(2kc_s\tau_s+3i\right)+3i\right\}\Bigg],\\
\label{bG3b}\beta^{(3)}_{\bf k}&=&\frac{3}{4k^6c^6_s\tau^3_s\tau^3_e}\Bigg[\left(kc_s\tau_s-i\right)^2\left\{k^2c^2_s\tau^2_e\left(3-2ikc_s\tau_e\right)+3\right\}e^{-2ikc_s\tau_s}\nonumber\\
&&\quad\quad\quad\quad\quad\quad\quad\quad\quad\quad\quad\quad\quad\quad\quad\quad+i\left(kc_s\tau_e-i\right)^2\left\{3i+k^2c^2_s\tau^2_s\left(2kc_s\tau_s+3i\right)\right\}e^{-2ikc_s\tau_e}\Bigg].\eea

\subsection{Quantization of comoving curvature perturbation from CGEFT}
\label{s4d}

In order to determine the expression for the two-point correlation function and the associated power spectrum in Fourier space, which are required to compute the cosmological correlations, we must specifically quantize the scalar modes properly. In order to do this, we must first build the creation operator, $\hat{a}^{\dagger}_{\bf k}$, and the annihilation operator, $\hat{a}_{\bf k}$, which will, respectively, create an excited state from the original Bunch Davies state and annihilate it. As Bunch Davies' initial condition, $|0\rangle$ must stick to the following restriction in order for the remainder of the purpose to be served,
\be \hat{a}_{\bf k}|0\rangle=0\quad\forall {\bf k}.\ee
For the quantization purpose the following equal time commutation relations (ETCR) has to be satisfied:
\bea &&\left[\hat{\zeta}_{\bf k}(\tau),\hat{\zeta}^{'}_{{\bf k}^{'}}(\tau)\right]_{\bf ETCR}=i\;\delta^{3}\left({\bf k}+{\bf k}^{'}\right),\quad
\left[\hat{\zeta}_{\bf k}(\tau),\hat{\zeta}_{{\bf k}^{'}}(\tau)\right]_{\bf ETCR}=0,\quad
\left[\hat{\zeta}^{'}_{\bf k}(\tau),\hat{\zeta}^{'}_{{\bf k}^{'}}(\tau)\right]_{\bf ETCR}=0.\eea
Once we promote the previously computed classical results of the comoving curvature perturbation and its canonically conjugate momenta as a quantum mechanical operator to pursue the quantization purpose, can be expressed by the following expressions:
\bea \hat{\zeta}_{\bf k}(\tau)=\bigg[{\zeta}_{\bf k}(\tau)\hat{a}_{\bf k}+{\zeta}^{*}_{\bf k}(\tau)\hat{a}^{\dagger}_{-{\bf k}}\bigg]&=&-H\bigg[{\pi}_{\bf k}(\tau)\hat{a}_{\bf k}+{\pi}^{*}_{\bf k}(\tau)\hat{a}^{\dagger}_{-{\bf k}}\bigg]\nonumber\\
&=&\frac{1}{a\sqrt{2{\cal A}}}\bigg[v_{\bf k}(\tau)\hat{a}_{\bf k}+v^{*}_{\bf k}(\tau)\hat{a}^{\dagger}_{-{\bf k}}\bigg]\nonumber\\
&=&\frac{c_s}{a\sqrt{2{\cal B}}}\bigg[v_{\bf k}(\tau)\hat{a}_{\bf k}+v^{*}_{\bf k}(\tau)\hat{a}^{\dagger}_{-{\bf k}}\bigg],\\
\hat{\zeta}^{'}_{\bf k}(\tau)=\bigg[{\zeta}^{'}_{\bf k}(\tau)\hat{a}_{\bf k}+{\zeta}^{*'}_{\bf k}(\tau)\hat{a}^{\dagger}_{-{\bf k}}\bigg]&=&-H\bigg[{\pi}^{'}_{\bf k}(\tau)\hat{a}_{\bf k}+{\pi}^{*'}_{\bf k}(\tau)\hat{a}^{\dagger}_{-{\bf k}}\bigg]-H^{'}\bigg[{\pi}_{\bf k}(\tau)\hat{a}_{\bf k}+{\pi}^{*}_{\bf k}(\tau)\hat{a}^{\dagger}_{-{\bf k}}\bigg]\nonumber\\
&=&\frac{1}{a\sqrt{2{\cal A}}}\bigg[v^{'}_{\bf k}(\tau)\hat{a}_{\bf k}+v^{*'}_{\bf k}(\tau)\hat{a}^{\dagger}_{-{\bf k}}\bigg]-\frac{1}{2{\cal A} a^{2}}\bigg(a\sqrt{2{\cal A}}\bigg)^{'}\bigg[v_{\bf k}(\tau)\hat{a}_{\bf k}+v^{*}_{\bf k}(\tau)\hat{a}^{\dagger}_{-{\bf k}}\bigg]\nonumber\\
&=&\frac{1}{a\sqrt{2{\cal A}}}\bigg[v^{'}_{\bf k}(\tau)\hat{a}_{\bf k}+v^{*'}_{\bf k}(\tau)\hat{a}^{\dagger}_{-{\bf k}}\bigg]-\frac{1}{2{\cal A} a^{2}}\bigg(a\sqrt{2{\cal A}}\bigg)^{'}\bigg[v_{\bf k}(\tau)\hat{a}_{\bf k}+v^{*}_{\bf k}(\tau)\hat{a}^{\dagger}_{-{\bf k}}\bigg].\quad\quad \eea
which will be very beneficial when we execute the one loop calculation using the in-in formalism in the following part.

 This can also be written as any possible commutation relation between the creation and destruction operators, as shown above:
\bea \left[\hat{a}_{\bf k},\hat{a}^{\dagger}_{{\bf k}^{'}}\right]_{\bf ETCR}=\delta^{3}\left({\bf k}+{\bf k}^{'}\right),\quad\quad
 \left[\hat{a}_{\bf k},\hat{a}_{{\bf k}^{'}}\right]_{\bf ETCR}=0, \quad\quad\left[\hat{a}^{\dagger}_{\bf k},\hat{a}^{\dagger}_{{\bf k}^{'}}\right]_{\bf ETCR}=0.\eea

\subsection{Tree level power spectrum from comoving curvature perturbation from CGEFT}
\label{s4e}

Since we are aware that the co-moving curvature disturbance happens at a late time scale, $\tau\rightarrow 0$, the appropriate tree-level input to the two-point cosmological correlation function can be shown as follows:
\bea \langle \hat{\zeta}_{\bf k}\hat{\zeta}_{{\bf k}^{'}}\rangle_{{\bf Tree}} =H^2\langle \hat{\pi}_{\bf k}\hat{\pi}_{{\bf k}^{'}}\rangle_{{\bf Tree}}=(2\pi)^{3}\;\delta^{3}\left({\bf k}+{\bf k}^{'}\right)\frac{2\pi^2}{k^3}\Delta^{2}_{\zeta,{\bf Tree}}(k),\quad\eea
where the dimensionless form of the tree level power spectrum, which is used in the observational probe is given by the following expression in the Fourier space:
\bea \Delta^{2}_{\zeta,{\bf Tree}}(k)=\frac{k^{3}}{2\pi^{2}}\langle\langle \hat{\zeta}_{\bf k}\hat{\zeta}_{-{\bf k}}\rangle\rangle_{(0,0)}=\frac{k^{3}}{2\pi^{2}}\left[{\zeta}_{\bf k}(\tau){\zeta}_{-{\bf k}}(\tau)\right]_{\tau\rightarrow 0}=\frac{k^{3}}{2\pi^{2}}|{\zeta}_{\bf k}(\tau)|^{2}_{\tau\rightarrow 0}=\frac{k^{3}H^2}{2\pi^{2}}|{\pi}_{\bf k}(\tau)|^{2}_{\tau\rightarrow 0}.\eea

As of right now, we already know from the analysis that we have done that the first SR (SRI), USR, and second SR (SRII) regions, which we have directly estimated in this work, have distinct solutions to the modes for the scalar cosmological perturbations.
With the aid of computed scalar modes from the first SR (SRI), USR, and second SR (SRII) regions, the dimensionless power spectrum can be calculated here at the tree level as follows:  
\bea  \Delta^{2}_{\zeta,{\bf Tree}}(k)
&=& \displaystyle\left(\frac{H^{4}}{8\pi^{2}{\cal A} c^3_s}\right)_* \times
\nonumber\\&&\displaystyle\left\{
	\begin{array}{ll}
		\displaystyle\left(1+k^{2}c^{2}_s\tau^{2}\right)& \mbox{when}\quad  k\leq k_s  \;(\rm SRI)  \\  \\
			\displaystyle 
			\displaystyle\left(\frac{k_e}{k_s }\right)^{6}\times\left|\alpha^{(2)}_{\bf k}\left(1+ikc_s\tau\right)\; e^{-ikc_s\tau}-\beta^{(2)}_{\bf k}\left(1-ikc_s\tau\right)\; e^{ikc_s\tau}\right|^{2} & \mbox{when }  k_s\leq k\leq k_e  \;(\rm USR)\\ \\
   \displaystyle 
			\displaystyle\left(\frac{k_e}{k_s }\right)^{6}\times\left|\alpha^{(3)}_{\bf k}\left(1+ikc_s\tau\right)\; e^{-ikc_s\tau}-\beta^{(3)}_{\bf k}\left(1-ikc_s\tau\right)\; e^{ikc_s\tau}\right|^{2} & \mbox{when }  k_e\leq k\leq k_{\rm end}  \;(\rm SRII) 
	\end{array}
\right. \eea
Here we need to use the following facts at the horizon crossing point:
\bea -k_s c_s\tau_s=1,\quad\quad -k_e c_s\tau_e=1,\quad\quad -k_s c_s\tau_{\rm end}=1,\eea 
which will going to be extremely helpful for the rest of the computation performed in this paper. The Bogoliubov coefficients for the USR $(\alpha^{(2)},\beta^{(2)})$ and the second SR $(\alpha^{(3)},\beta^{(3)})$ regions are shown here. These coefficients are deduced directly in equations (\ref{bG2a}), (\ref{bG2b}), (\ref{bG3a}), and (\ref{bG3b}), respectively. According to equations (\ref{bG1a}) and (\ref{bG1b}), the Bunch Davies initial condition fixes the Bogoliubov coefficients $(\alpha^{(1)},\beta^{(1)})$ for the first SR region (SRI). The current calculation should also take into account the wave numbers $k_e$ and $k_s$ that relate to the time scales $\tau_e$ and $\tau_s$.  The pivot scale where CMB monitoring is conducted is represented by the sign $*$. From the above structure, it is possible to identify the specific contributions made by the first SR region (SRI), the USR region, and the second SR region (SRII) in the current debate context. Further, using the information provided at the horizon-crossing point the total contribution in the tree-level primordial power spectrum can be expressed as:
\bea \Bigg[\Delta^{2}_{\zeta,{\bf Tree}}(k)\Bigg]_{\bf Total}&=&\Bigg[\Delta^{2}_{\zeta,{\bf Tree}}(k\leq k_s)\Bigg]_{\bf SRI}+\Bigg[\Delta^{2}_{\zeta,{\bf Tree}}(k_s\leq k\leq k_e)\Bigg]_{\bf USR}+\Bigg[\Delta^{2}_{\zeta,{\bf Tree}}(k_e\leq k\leq k_{\rm end})\Bigg]_{\bf SRII}\nonumber\\
&=&\Bigg[\Delta^{2}_{\zeta,{\bf Tree}}(k)\Bigg]_{\bf SRI}+\Bigg[\Delta^{2}_{\zeta,{\bf Tree}}(k)\Bigg]_{\bf USR}\Theta(k-k_s)+\Bigg[\Delta^{2}_{\zeta,{\bf Tree}}(k)\Bigg]_{\bf SRII}\Theta(k-k_e)\nonumber\\
&\approx&\Bigg[\Delta^{2}_{\zeta,{\bf Tree}}(k)\Bigg]_{\bf SRI}\Bigg\{1+\left(\frac{k_e}{k_s }\right)^{6}\Bigg[\Sigma(k)\Theta(k-k_s)+\Upsilon(k)\Theta(k-k_e)\Bigg]\Bigg\}.\eea
where the power spectrum in the SRI region can be recast in the following tractable form:
\bea \Bigg[\Delta^{2}_{\zeta,{\bf Tree}}(k)\Bigg]_{\bf SRI}&=&\left(\frac{H^{4}}{8\pi^{2}{\cal A} c^3_s}\right)_*\Bigg\{1+\Bigg(\frac{k}{k_s}\Bigg)^2\Bigg\}=\left(\frac{H^{4}}{8\pi^{2}{\cal B} c_s}\right)_*\Bigg\{1+\Bigg(\frac{k}{k_s}\Bigg)^2\Bigg\}.\eea
Additionally, we have introduced a momentum dependent factor $\Upsilon(k)$ in the SRII region, which is defined as:
\bea \Sigma(k):=\left|\alpha^{(2)}_{\bf k}-\beta^{(2)}_{\bf k}\right|^2,\quad\quad\quad\Upsilon(k):=\left|\alpha^{(3)}_{\bf k}-\beta^{(3)}_{\bf k}\right|^2,\eea
where the individual Bogoliubov coefficients in the SRII region at the horizon crossing point is further simplified as:
\bea \label{bG22a}\alpha^{(2)}_{\bf k}&=&1+\frac{3}{2i}\left(\frac{k_s}{k}\right)^3\left(1+\left(\frac{k}{k_s}\right)^{2}\right),\\
\label{bG22b}\beta^{(2)}_{\bf k}&=&\frac{3}{2i}\left(\frac{k_s}{k}\right)^3\left(1-i\left(\frac{k}{k_s}\right)\right)^{2}\; e^{-2i\left(\frac{k}{k_s}\right)},\\
\label{bGG3a}\alpha^{(3)}_{\bf k}&=&-\frac{1}{4}\left(\frac{k_s}{k}\right)^3\left(\frac{k_e}{k}\right)^3\Bigg[9\left(\left(\frac{k}{k_s}\right)+i\right)^2\left(\left(\frac{k}{k_e}\right)+i\right)^2 e^{-2i\left(\frac{k}{k_e}-\frac{k}{k_s}\right)}\nonumber\\
&&\quad\quad\quad\quad\quad\quad\quad\quad\quad\quad\quad\quad\quad+
\left\{\left(\frac{k}{k_e}\right)^2\left(2\left(\frac{k}{k_e}\right)+3i\right)+3i\right\}\left\{\left(\frac{k}{k_s}\right)^2\left(-2\left(\frac{k}{k_s}\right)+3i\right)+3i\right\}\Bigg],\\
\label{bGG3b}\beta^{(3)}_{\bf k}&=&\frac{3}{4}\left(\frac{k_s}{k}\right)^3\left(\frac{k_e}{k}\right)^3\Bigg[\left(\left(\frac{k}{k_s}\right)+i\right)^2\left\{\left(\frac{k}{k_e}\right)^2\left(3+2i\left(\frac{k}{k_e}\right)\right)+3\right\}e^{2i\left(\frac{k}{k_s}\right)}\nonumber\\
&&\quad\quad\quad\quad\quad\quad\quad\quad\quad\quad\quad\quad\quad\quad\quad\quad+i\left(\left(\frac{k}{k_e}\right)+i\right)^2\left\{3i+\left(\frac{k}{k_s}\right)^2\left(-2\left(\frac{k}{k_s}\right)+3i\right)\right\}e^{2i\left(\frac{k}{k_e}\right)}\Bigg].\eea
Here we have introduced two distinctive Heaviside Theta functions to join the individual contributions in the overall amplitude of the tree-level primordial power spectrum from different regions at the transition points $\tau=\tau_s$ (SRI to USR) and $\tau=\tau_e$ (USR to SRII) smoothly, and these functions are defined as:
\bea  \Theta(k-k_s)
&=& \left\{
	\begin{array}{ll}
		0\quad\quad\quad\quad\quad\quad\quad\quad & \mbox{when }  k<k_s  \;(\rm SRI)\\ \\
   \displaystyle 
			\displaystyle 1 & \mbox{when }  k_s\leq k< k_e  \;(\rm USR) 
	\end{array}
\right. \\
\Theta(k-k_e)
&=& \left\{
	\begin{array}{ll}
		0\quad\quad\quad\quad\quad\quad\quad\quad & \mbox{when }  k_s<k<k_e  \;(\rm USR)\\ \\
   \displaystyle 
			\displaystyle 1 & \mbox{when }  k_e\leq k\leq k_{\rm end}  \;(\rm SRII) 
	\end{array}
\right.\eea

\section{ Cut-off regularized one-loop power spectrum for comoving curvature perturbation from CGEFT}
\label{s5}

In this section, we discuss the crucial role of the very mild breaking of the Galilean symmetry to remove the harmful contributions from the third-order action, due to the presence of which quantum loop corrections are large and comparable to the tree-level counterpart as appearing in the case of good-old EFT of inflation and $P(X,\phi)$ models of inflation. In the present Galileon, inflation due to the absence of such terms quantum loop effects become suppressed and can be treated as a subdominant correction to the tree-level counterpart of the corresponding spectrum. Next, we discuss the in-in formalism ad its explicit role in the computation of the one-loop contributions in the SRI, USR, and SRII phases respectively. Further, we explicitly compute the cut-off regularized one-loop corrections from SRI, USR, and SRII phases. After that, we have written the full expression for the one-loop corrected power spectrum for comoving curvature perturbation adding all the tree level as well as one loop corrections computed from SRI, USR and SRII phases respectively. 

\subsection{Third order perturbation from comoving curvature perturbation}
\label{s5a}

The next phase of our study will involve calculating the one-loop contributions to the power spectrum directly using the comoving curvature perturbation as input. Now before going to the technical details of the computations of the one-loop effects in the present context let us first mention how exactly Galilean symmetry is going to affect the comoving curvature perturbation, which will be going to provide extremely important information in the present context in terms of the fact that which terms are allowed and which terms are not allowed by such symmetry. 

It should be noted that the non-renormalization\footnote{Non-renormalization $\hat{\rm a}$ {\it la} non-renormalization theorem should not be confused with the non-renormalizability of a theory. The former implies that couplings in the underlying theory are stable under radiative corrections, they do not run, whereas, in a theory that is not renormalizable, the number of counter terms required in the process of renormalization is not finite; four Fermi-theory serves as an example. On the other hand, in a renormalizable theory, the structure of divergences that occurs in lower orders of perturbation theory keeps repeating in higher orders, such that a finite number of counter terms suffices.} theorem in a general field theoretic setup implies that the couplings of the underlying theory do not receive any radiative corrections. Interestingly, this property linked to Galileon symmetry is stronger in this case compared to the other theories, for example, mild breaking of the symmetry preserves the said property. A mild breaking of Galileon symmetry is accomplished by going from the de Sitter to the quasi-de Sitter regime. Secondly, the 
Galileon operators are not renormalized even if 
coupled to
heavy external fields provided  the Galilean symmetry is respected (See refs.\cite{Burrage:2010cu,Goon:2016ihr}for details). 

Let us firs mention the underlying reason for the absence of radiative corrections to the power spectrum; the technical supporting details would then follow. In case of standard single field inflation \footnote{The third order action for standard single field inflation can be expressed as \cite{Choudhury:2023vuj,Choudhury:2023jlt,Choudhury:2023rks}:
\bea &&S^{(3)}_{\zeta}=\int d\tau\;  d^3x\;  M^2_{ pl}a^2\; \bigg(\left(3\left(c^2_s-1\right)\epsilon+\epsilon^2\right)\zeta^{'2}\zeta+\frac{\epsilon}{c^2_s}\bigg(\epsilon+1-c^2_s\bigg)\left(\partial_i\zeta\right)^2\zeta-\frac{2\epsilon}{c^2_s}\zeta^{'}\left(\partial_i\zeta\right)\left(\partial_i\partial^{-2}\left(\frac{\epsilon\zeta^{'}}{c^2_s}\right)\right)\nonumber\\
&&\quad\quad\quad\quad\quad\quad\quad\quad\quad-\frac{1}{aH}\left(1-\frac{1}{c^2_{s}}\right)\epsilon \bigg(\zeta^{'3}+\zeta^{'}(\partial_{i}\zeta)^2\bigg)
 +\frac{1}{2}\epsilon\zeta\left(\partial_i\partial_j\partial^{-2}\left(\frac{\epsilon\zeta^{'}}{c^2_s}\right)\right)^2+\underbrace{\frac{1}{2c^2_s}\epsilon\partial_{\tau}\left(\frac{\eta}{c^2_s}\right)\zeta^{'}\zeta^{2}}+\cdots\bigg),\quad\quad\eea
 where the last highlighted term in the above expression is solely responsible for significant one loop contribution to the power spectrum for the scalar modes. In the case of Galileon due to having softly broken Galilean symmetry, such a contribution is strictly absent, and hence the power spectrum is protected from the radiative corrections. In this article and the rest of the paper, we have discussed this issue in detail.
}, the third-order action for comoving curvature perturbation contains $\zeta^{'}\zeta^{2}$ operator(with coefficient proportional to $\eta'$ which is large during a sharp transition) which is responsible for significant one-loop contribution to the power spectrum (other terms in the action give insignificant contribution\cite{Choudhury:2023vuj,Choudhury:2023jlt,Choudhury:2023rks}). The mentioned operator is absent in single-field Galileon inflation with soft Galilean symmetry breaking for which the non-renormalization theorem is respected. Thus loop corrections, in this case, are insignificant.  Indeed, under Galilean symmetry as written in terms of the field in equation (\ref{GCS}), the comoving curvature perturbation is transformed by the following expression:
\bea \zeta\rightarrow\zeta-\frac{H}{\dot{\bar{\phi}}_0}\left(b\cdot \delta x\right),\eea
using which the time and spatial derivatives of the curvature perturbation can be transformed as:
\bea \zeta^{'}\rightarrow\zeta^{'}-\frac{H}{\dot{\bar{\phi}}_0}b_0, \quad\quad\quad\quad\quad \partial_i\zeta\rightarrow \partial_i\zeta-\frac{H}{\dot{\bar{\phi}}_0}b_i,\quad\quad\quad\quad\quad \partial^2\zeta\rightarrow \partial^2\zeta.\eea
which appears in all contributions in the third-order action. Here one can clearly observe that in $\zeta$, $\zeta^{'}$ and in $\partial_{i}\zeta$ the Galilean symmetry is softly broken which is necessarily needed to implement the inflationary paradigm in the present context of discussion. On the other hand, we can also observe that the term $\partial^2\zeta$ is fully Galilean symmetry protected. For this reason, such contribution has to couple with some other contributions which break the Galilean symmetry softly. Due to having this transformation properties at the level of curvature perturbation, its spatial and temporal derivatives it is quite obvious that to allow very small breaking of the Galilean symmetry to perform inflation some of the terms can be absorbed using field redefinition, some of them can be expressed as the total derivative terms and vanish at the boundary. These contributions are, $\zeta^{'2}\zeta$, $\left(\partial_i\zeta\right)^2\zeta$, $\zeta^{'}\left(\partial_i\zeta\right)\left(\partial_i\partial^{-2}\left(\epsilon\zeta^{'}\right)\right)$, $\zeta\left(\partial_i\partial_j\partial^{-2}\left(\epsilon\zeta^{'}\right)\right)^2$, $\zeta\partial_{\tau}\left(\partial_{i}\zeta\right)^2$ and $\zeta^{'}\zeta^{2}$, which are absent in the third order action due to having very soft breaking of the Galilean symmetry. Out of all of these contributions, the last term, $\zeta^{'}\zeta^{2}$ is most significant in the USR period as its coefficient $\partial_{\tau}\left(\eta/c^2_s\right)$ contributes huge amount in the SRI to USR and USR to SRII transition points. Though the contribution in SRI to USR transition is extremely larger compared to the contribution appearing in USR to SRII transition. Such contributions are extremely harmful for PBH production due to having large one-loop effects that recently have been shown in refs. \cite{Choudhury:2023vuj,Choudhury:2023jlt,Choudhury:2023rks}. Now since this contribution is absent in the presence of soft breaking of Galilean symmetry it is important to emphasize  for the same.
Under the aforementioned transformation the term $\zeta^{'}\zeta^{2}$ transforms as follows:
\bea \zeta^{'}\zeta^{2}\rightarrow \bigg(\zeta^{'}-\frac{H}{\dot{\bar{\phi}}_0}b_0\bigg)\bigg(\zeta-\frac{H}{\dot{\bar{\phi}}_0}\left(b\cdot \delta x\right)\bigg)^2\sim \frac{1}{3}\left(b\cdot \delta x\right)\partial_{\tau}\left(\zeta^2\right)\sim 0 \quad ({\rm at \; boundary}).\eea
Here the contributions, $\left(\frac{H}{\dot{\bar{\phi}}_0}\right)^2\left(b\cdot \delta x\right)\zeta^{'}$, $\left(\frac{H}{\dot{\bar{\phi}}_0}\right)^3 b_0\left(b\cdot \delta x\right)^2$, $\left(\frac{H}{\dot{\bar{\phi}}_0}\right)^2 b_0\left(b\cdot \delta x\right)$, $\left(\frac{H}{\dot{\bar{\phi}}_0}\right)\left(b\cdot \delta x\right)\zeta^{'}$ are not going to contribute in the third order action. Now using field redefinition as well as dumping some of the contributions at the boundary one can immediately either absorb them in the coefficients of the second-order perturbed action or make them completely zero due to having a total derivative structure at the boundary. So the contributions, that will survive due to having the small amount of soft Galilean symmetry break are given by, $\zeta^{'3}$, $\zeta^{'2}\left(\partial^2\zeta\right)$, $\zeta^{'}\left(\partial_i\zeta\right)^2$ and $\left(\partial_i\zeta\right)^2\left(\partial^2\zeta\right)$, which are physically interpreted as the bulk self-interactions of the Galileon in terms of the scalar curvature perturbation. For all of these mentioned surviving contributions, some additional terms appear after performing the aforementioned transformation, out of which some of them can be thrown due to having a total derivative structure at the boundary, and some of them can be absorbed by making use of field redefinition and rest of the contributions which are proportional to the quadratic or cubic in the amount of small breaking of Galilean symmetry can easily be neglected in this construction. Using such contributions one needs to construct the third-order action by performing the cosmological perturbation theory in detail. For more on these aspects see ref. \cite{Burrage:2010cu}, where the authors have constructed the third-order action in detail by allowing a small amount of soft breaking of Galilean symmetry.
Assuming that the curvature perturbation expands the CGEFT action in the third order, the following computation will be done in the present context. For the same, the third-order action can be represented by the following expression:
\bea &&S^{(3)}_{\zeta}=\int d\tau\;  d^3x\;  \frac{a^2}{H^3}\; \bigg[\frac{{\cal G}_1}{a}\zeta^{'3}+\frac{{\cal G}_2}{a^2}\zeta^{'2}\left(\partial^2\zeta\right)+\frac{{\cal G}_3}{a}\zeta^{'}\left(\partial_i\zeta\right)^2+\frac{{\cal G}_4}{a^2}\left(\partial_i\zeta\right)^2\left(\partial^2\zeta\right)\bigg],\eea
where the term $\zeta' \zeta^2$ is absent  if the Galileon symmetry is softly broken; its presence would signal the absence of the underlying Galileon symmetry itself and the violation of non-renormalization theorem.

Here the coupling parameters ${\cal G}_i\forall i=1,2,3,4$ as appearing in the third order perturbed action are given by the following expressions:
\bea {\cal G}_1:&\equiv& \frac{2H\dot{\bar{\phi}}^3_0}{\Lambda^3}\Bigg(c_3+9c_4Z+30c_5Z^2\Bigg),\\
      {\cal G}_2:&\equiv& -\frac{2\dot{\bar{\phi}}^3_0}{\Lambda^3}\Bigg(c_3+6c_4Z+18c_5Z^2\Bigg),\\ 
       {\cal G}_3:&\equiv& -\frac{2H\dot{\bar{\phi}}^3_0}{\Lambda^3}\Bigg(c_3+7c_4Z+18c_5Z^2\Bigg)+\frac{2\dot{\bar{\phi}}^2_0\ddot{\bar{\phi}}_0}{\Lambda^3}\Bigg(c_3+6c_4Z+18c_5Z^2\Bigg)\nonumber\\
       &=&-\frac{2H\dot{\bar{\phi}}^3_0}{\Lambda^3}\Bigg(c_3+7c_4Z+18c_5Z^2\Bigg)-\frac{2\dot{\bar{\phi}}^3_0H\eta}{\Lambda^3}\Bigg(c_3+6c_4Z+18c_5Z^2\Bigg),\\
        {\cal G}_4:&\equiv& \frac{\dot{\bar{\phi}}^3_0}{\Lambda^3}\bigg\{c_3+3c_4Z+6c_5\bigg[Z^2+\frac{\dot{H}\dot{\bar{\phi}}^2_0}{\Lambda^6}\bigg]\bigg\}+\frac{3\dot{\bar{\phi}}^3_0\ddot{\bar{\phi}}_0}{\Lambda^6}\bigg\{c_4+4c_5Z\bigg\}\nonumber\\
        &=&\frac{\dot{\bar{\phi}}^3_0}{\Lambda^3}\bigg\{c_3+3c_4Z+6c_5\bigg[Z^2+\frac{\dot{H}\dot{\bar{\phi}}^2_0}{\Lambda^6}\bigg]\bigg\}-\frac{3\dot{\bar{\phi}}^4_0H\eta}{\Lambda^6}\bigg\{c_4+4c_5Z\bigg\},\eea
        where the factor $Z$ is already defined earlier in equation (\ref{Z}). To implement the fact that the Galileon self couplings dominates over all other contributions, and also to neglect the mixing contributions with the gravitational background in the decoupling limit one should look at the region where $Z\gtrsim 1$. In this work we restrict our analysis by considering $Z\sim 1$, where the nonlinear Galileon interactions are present but not harmful for the rest of the computations performed in this paper. This means in the regime $Z\sim 1$ one can able to control such contributions at the perturbative level of computation. Though we are interested to compute the effect one-loop contribution from the third order action for the comoving curvature perturbation in the rest part of the paper, using the same action it might be really interesting to perform the computations for Bispectrum as well as Trispectrum from SRI, USR and SRII phases and comment on the primordial non-Gaussian \cite{Maldacena:2002vr,Seery:2005wm,Senatore:2009gt,Chen:2006nt,Chen:2010xka,Chen:2009zp,Chen:2009we,Chen:2008wn,Chen:2006xjb,Choudhury:2012whm,Agarwal:2012mq,Holman:2007na,Creminelli:2005hu,Behbahani:2011it,Smith:2009jr,Cheung:2007sv,Creminelli:2006rz,Creminelli:2006gc,Kalaja:2020mkq,Meerburg:2019qqi,Lee:2016vti,Maldacena:2011nz,Werth:2023pfl} features in the present context. 

Now, in terms of the good-old EFT construction the coefficients appearing in front of the operators can be written in terms of CGEFT coefficients:
\bea &&\frac{M^4_3}{H^2M^2_{pl}}=\frac{3}{4}\frac{1}{a^2H^4}\left({\cal G}_3-{\cal G}_1\right),\\
&&\frac{\bar{M}^3_1}{HM^2_{pl}}=\frac{2}{3}\bigg[\left(1-\frac{1}{c^2_s}\right)\epsilon+\frac{1}{a^2H^4}{\cal G}_3\bigg].
\eea
  One can similarly fix the other two coefficients, ${\cal G}_2$ and ${\cal G}_4$ in terms of the EFT coefficients by comparing term by term in the third order action. 

Now for the purpose of specifically extracting the correction from the one-loop quantum effect, we will make use of all of the inputs stated above. In the mentioned contributions as appearing in the third order action we have three possible situation appears on the first and second slow-roll parameters $\epsilon$ and $\eta$, which need to take care of very crucially during the computation:
\begin{enumerate}
    \item \underline{\bf Region I (SRI):} In the first slow-roll regime (SRI), the behaviour of the mentioned two slow roll parameters are given by:
    \bea \epsilon\sim {\rm Constant},\quad\quad\quad \eta\sim 0\quad\quad\quad {\rm where}\quad\quad\tau<\tau_s.\eea
    Here in the conformal time scale $\tau<\tau_s$ represents the SRI region, where at $\tau=\tau_s$ scale SR to USR transition occurs.

    \item \underline{\bf Region II (USR):} In the Ultra slow-roll regime (USR), the behaviour of the mentioned two slow roll parameters are given by:
    \bea \epsilon(\tau)=\epsilon  \;\left(\frac{\tau}{\tau_s}\right)^{6},\quad\quad\quad \eta\sim -6\quad\quad\quad {\rm where}\quad\quad\tau_s\leq\tau\leq \tau_e.\eea
    Here $\epsilon$ is the slow-roll parameter as appearing in the phase SRI. Here at the SRI to USR transition point $\tau=\tau_s$ the behaviour is taken care of.

    \item \underline{\bf Region III (SRII):} In the second slow-roll regime (SRII), the behaviour of the mentioned two slow roll parameters are given by:
    \bea \epsilon(\tau)=\epsilon  \;\left(\frac{\tau_e}{\tau_s}\right)^{6},\quad\quad\quad \eta\sim 0\quad\quad\quad {\rm where}\quad\quad\tau_e\leq\tau\leq \tau_{\rm end}.\eea
    Here $\epsilon$ is the slow-roll parameter as appearing in the phase SRI as already mentioned earlier. Here at the USR to SRII transition point $\tau=\tau_e$ the behaviour is taken care of.
\end{enumerate}
From this construction of three consecutive phases it is very clear that the first slow-roll parameter $\epsilon$ behave smoothly at the SRI to USR and USR to SRII transition points, at $\tau=\tau_s$ and $\tau=\tau_e$ respectively. On the other hand, the second slow-roll parameter $\eta$ around the mentioned transition points can be parametrized by the following functional form:
\bea \eta(\tau)=-6-\Delta\eta\left[\Theta(\tau-\tau_s)-\Theta(\tau-\tau_e)\right],\eea
which implies in the SRII region, $\eta\sim -6-\Delta\eta=0$, because we have to take $\Delta\eta\sim -6$. In the SRI region, where we have have $\tau<\tau_s$, this parameter gives $\eta\sim 0$, just like SRII region. This parametrization is engineered in such a way that if we take the time derivative of the second slow-roll parameter $\eta$ at the transition points then the following contribution appears:
\bea \eta^{'}(\tau)=-\Delta\eta\left[\delta(\tau-\tau_s)-\delta(\tau-\tau_e)\right].\eea
However, in the previously mentioned third order action such terms are forbidden by the small amount of softly breaking of Galilean symmetry. This particular term appears in refs. \cite{Kristiano:2022maq,Riotto:2023hoz,Choudhury:2023vuj,Choudhury:2023jlt,Kristiano:2023scm,Riotto:2023gpm,Choudhury:2023rks,Firouzjahi:2023aum,Motohashi:2023syh}, where such underlying symmetry was absent. So in our paper we don't need take care of the derivative of the second slow-roll parameter $\eta$ at the transition points explicitly. But it clear from the construction that the behaviour of the second slow-roll parameter $\eta$ is not smooth at the mentioned transition points. For this reason to make our further analysis consistent we have used the mentioned parametrization of the second slow-roll parameter $\eta$ at the transition points. 

Before going to the further technical details of the one-loop contributions in the SRI, USR and SRII region in the next subsection it is important to mention the strengths of the four types of the cubic self interactions in the aforementioned three regions, which will going to be extremely helpful to keep track on the final result and its overall contribution added with the tree-level result. All of these four interactions are suppressed in the SRI and SRII period due to having vanishing contribution from the second slow-roll parameter $\eta$ in both of these phases. Though we will compute these contributions for the completeness. On the other hand, due to the presence of the second slow-roll parameter $\eta$ in the two coupling parameters, ${\cal G}_3$ and ${\cal G}_4$, of the last two terms of the representative third order action in the USR phase thse two terms become dominant compared to the other two terms having the coupling coefficients,  ${\cal G}_1$ and ${\cal G}_2$. In the next subsection we will going to explicitly show that whatever amount of enhancement we have from the one-loop contribution will be dominated by the last two terms of the third order action appearing in the USR period. Since we do not have any term involving the time derivative of second slow-roll parameter $\eta$, at the SRI to USR and USR to SRII transition points, the corresponding amount of enhancement of the power spectrum in the USR period in the one-loop contribution will be sufficiently suppressed compared to the results obtained in the refs. \cite{} where such type of contribution is present. Though the suppressed contribution in the USR period is much higher than the one-loop contributions obtained from the SRI and SRII regions. In the next subsection we are going to investigate all of these possibilities in great detail.

\subsection{The direct In-In formalism for the one-loop corrected two-point function}
\label{s5b}

The second part of the article will be devoted to a detailed analysis of every term that arises as a result of the CGEFT framework that we used in our research. We make use of the well-known in-in formula to accomplish this.  This leads to the following two-point function at $\tau\rightarrow 0$, which is given by:
 \bea \label{tpt} \langle\hat{\zeta}_{\bf p}\hat{\zeta}_{-{\bf p}}\rangle:&=&\left\langle\bigg[\overline{T}\exp\bigg(i\int^{\tau}_{-\infty(1-i\epsilon)}d\tau^{'}\;H_{\rm int}(\tau^{'})\bigg)\bigg]\;\;\hat{\zeta}_{\bf p}(\tau)\hat{\zeta}_{-{\bf p}}(\tau)
\;\;\bigg[{T}\exp\bigg(-i\int^{\tau}_{-\infty(1+i\epsilon)}d\tau^{''}\;H_{\rm int}(\tau^{''})\bigg)\bigg]\right\rangle_{\tau\rightarrow 0}.\quad\quad \eea
In the above expression $T$ and $\bar{T}$ represent time ordering and anti-time ordering of the unitary operators which is made up of the time integral of the interaction Hamiltonian, which is described by the following expression:
\bea && H_{\rm int}(\tau)=-\int d^3x\; \frac{a^2}{H^3}\; \bigg[\frac{{\cal G}_1}{a}\zeta^{'3}+\frac{{\cal G}_2}{a^2}\zeta^{'2}\left(\partial^2\zeta\right)+\frac{{\cal G}_3}{a}\zeta^{'}\left(\partial_i\zeta\right)^2+\frac{{\cal G}_4}{a^2}\left(\partial_i\zeta\right)^2\left(\partial^2\zeta\right)\bigg],\eea
where all of these coefficients ${\cal G}_i\forall i=1,2,3,4$ are defined in the previous subsection.

Next expanding  equation (\ref{tpt}) order by order and collecting the non-trival terms which will contribute to the one-loop correction of the two point correlation function of comoving curvature perturbation we get the following simplified expression:
\bea  &&\label{g}\langle\hat{\zeta}_{\bf p}\hat{\zeta}_{-{\bf p}}\rangle=\langle\hat{\zeta}_{\bf p}\hat{\zeta}_{-{\bf p}}\rangle_{(0,0)}+\langle\hat{\zeta}_{\bf p}\hat{\zeta}_{-{\bf p}}\rangle_{(0,2)}+\langle\hat{\zeta}_{\bf p}\hat{\zeta}_{-{\bf p}}\rangle^{\dagger}_{(0,2)}+\langle\hat{\zeta}_{\bf p}\hat{\zeta}_{-{\bf p}}\rangle_{(1,1)},
\eea
where the first terms the tree-level contribution and remaining terms physically represent the one-loop correction to the tree-level contribution of the primordial two-point correlation function. The first term we have already computed in the earlier section of this paper. Now our prime objective is to compute the remaining three terms in the above expression and for the future computational purpose it is good to mention the expressions for these contributions in terms of the interaction Hamiltonian, which is given by the following expressions:
\bea &&\label{A1}\langle\hat{\zeta}_{\bf p}\hat{\zeta}_{-{\bf p}}\rangle_{(0,2)}=\lim_{\tau\rightarrow 0}\left[\int^{\tau}_{-\infty}d\tau_1\;\int^{\tau}_{-\infty}d\tau_1\;\langle \hat{\zeta}_{\bf p}(\tau)\hat{\zeta}_{-{\bf p}}(\tau)H_{\rm int}(\tau_1)H_{\rm int}(\tau_2)\rangle\right],\\
 &&\label{A2}\langle\hat{\zeta}_{\bf p}\hat{\zeta}_{-{\bf p}}\rangle^{\dagger}_{(0,2)}=\lim_{\tau\rightarrow 0}\left[\int^{\tau}_{-\infty}d\tau_1\;\int^{\tau}_{-\infty}d\tau_1\;\langle \hat{\zeta}_{\bf p}(\tau)\hat{\zeta}_{-{\bf p}}(\tau)H_{\rm int}(\tau_1)H_{\rm int}(\tau_2)\rangle^{\dagger}\right],\\
  &&\label{A3}\langle\hat{\zeta}_{\bf p}\hat{\zeta}_{-{\bf p}}\rangle_{(1,1)}=\lim_{\tau\rightarrow 0}\left[\int^{\tau}_{-\infty}d\tau_1\;\int^{\tau}_{-\infty}d\tau_1\;\langle H_{\rm int}(\tau_1)\hat{\zeta}_{\bf p}(\tau)\hat{\zeta}_{-{\bf p}}(\tau)H_{\rm int}(\tau_2)\rangle\right].\eea
  Further, in terms of the individual cubic self interactions the one-loop contributions to the two-point primordial cosmological correlation function can be quantified by the following expressions:
\bea \label{A11}\langle\hat{\zeta}_{\bf p}\hat{\zeta}_{-{\bf p}}\rangle_{(0,2)}&=&\sum^{4}_{i=1}{\bf Z}^{(1)}_i,\\ 
\label{A22}\langle\hat{\zeta}_{\bf p}\hat{\zeta}_{-{\bf p}}\rangle^{\dagger}_{(0,2)}&=&\sum^{4}_{i=1}{\bf Z}^{(2)}_i,\\
\label{A33}\langle\hat{\zeta}_{\bf p}\hat{\zeta}_{-{\bf p}}\rangle^{\dagger}_{(1,1)}&=&\sum^{4}_{i=1}{\bf Z}^{(3)}_i,\eea
where the factors ${\bf Z}^{(1)}_i\forall i=1,2,3,4$, ${\bf Z}^{(2)}_i\forall i=1,2,3,4$ and ${\bf Z}^{(3)}_i\forall i=1,2,3,4$ are described by the following expressions:
\bea \label{A11a}{\bf Z}^{(1)}_1:&=&\lim_{\tau\rightarrow 0}\Bigg[\int^{0}_{-\infty}d\tau_1\frac{a^2(\tau_1)}{H^3(\tau_1)}\frac{{\cal G}_1(\tau_1)}{a(\tau_1)}\;\int^{0}_{-\infty}d\tau_2\;\frac{a^2(\tau_2)}{H^3(\tau_2)}\frac{{\cal G}_1(\tau_2)}{a(\tau_2)}\nonumber\\
  &&\quad\quad\quad\quad\times\int \frac{d^{3}{\bf k}_1}{(2\pi)^3} \int \frac{d^{3}{\bf k}_2}{(2\pi)^3} \int \frac{d^{3}{\bf k}_3}{(2\pi)^3} \int \frac{d^{3}{\bf k}_4}{(2\pi)^3} \int \frac{d^{3}{\bf k}_5}{(2\pi)^3} \int \frac{d^{3}{\bf k}_6}{(2\pi)^3}\nonumber\\
  &&\quad\quad\quad\quad\times \delta^3\bigg({\bf k}_1+{\bf k}_2+{\bf k}_3\bigg) \delta^3\bigg({\bf k}_4+{\bf k}_5+{\bf k}_6\bigg)\nonumber\\
  &&\quad\quad\quad\quad\times \langle \hat{\zeta}_{\bf p}(\tau)\hat{\zeta}_{-{\bf p}}(\tau)\hat{\zeta}^{'}_{{\bf k}_1}(\tau_1)\hat{\zeta}^{'}_{{\bf k}_2}(\tau_1)\hat{\zeta}^{'}_{{\bf k}_3}(\tau_1)\hat{\zeta}^{'}_{{\bf k}_4}(\tau_2)\hat{\zeta}^{'}_{{\bf k}_5}(\tau_2)\hat{\zeta}^{'}_{{\bf k}_6}(\tau_2)\rangle\Bigg],\\ 
  \label{A11b}{\bf Z}^{(1)}_2:&=&\lim_{\tau\rightarrow 0}\Bigg[\int^{0}_{-\infty}d\tau_1\frac{a^2(\tau_1)}{H^3(\tau_1)}\frac{{\cal G}_2(\tau_1)}{a^2(\tau_1)}\;\int^{0}_{-\infty}d\tau_2\;\frac{a^2(\tau_2)}{H^3(\tau_2)}\frac{{\cal G}_2(\tau_2)}{a^2(\tau_2)}\nonumber\\
  &&\quad\quad\quad\quad\times\int \frac{d^{3}{\bf k}_1}{(2\pi)^3} \int \frac{d^{3}{\bf k}_2}{(2\pi)^3} \int \frac{d^{3}{\bf k}_3}{(2\pi)^3} \int \frac{d^{3}{\bf k}_4}{(2\pi)^3} \int \frac{d^{3}{\bf k}_5}{(2\pi)^3} \int \frac{d^{3}{\bf k}_6}{(2\pi)^3}\nonumber\\
  &&\quad\quad\quad\quad\times \delta^3\bigg({\bf k}_1+{\bf k}_2+{\bf k}_3\bigg) \delta^3\bigg({\bf k}_4+{\bf k}_5+{\bf k}_6\bigg)|{\bf K}|^4\nonumber\\
  &&\quad\quad\quad\quad\times \langle \hat{\zeta}_{\bf p}(\tau)\hat{\zeta}_{-{\bf p}}(\tau)\hat{\zeta}^{'}_{{\bf k}_1}(\tau_1)\hat{\zeta}^{'}_{{\bf k}_2}(\tau_1)\hat{\zeta}_{{\bf k}_3}(\tau_1)\hat{\zeta}^{'}_{{\bf k}_4}(\tau_2)\hat{\zeta}^{'}_{{\bf k}_5}(\tau_2)\hat{\zeta}_{{\bf k}_6}(\tau_2)\rangle\Bigg],\\
 \label{A11c}{\bf Z}^{(1)}_3:&=&\lim_{\tau\rightarrow 0}\Bigg[\int^{0}_{-\infty}d\tau_1\frac{a^2(\tau_1)}{H^3(\tau_1)}\frac{{\cal G}_3(\tau_1)}{a(\tau_1)}\;\int^{0}_{-\infty}d\tau_2\;\frac{a^2(\tau_2)}{H^3(\tau_2)}\frac{{\cal G}_3(\tau_2)}{a(\tau_2)}\nonumber\\
  &&\quad\quad\quad\quad\times\int \frac{d^{3}{\bf k}_1}{(2\pi)^3} \int \frac{d^{3}{\bf k}_2}{(2\pi)^3} \int \frac{d^{3}{\bf k}_3}{(2\pi)^3} \int \frac{d^{3}{\bf k}_4}{(2\pi)^3} \int \frac{d^{3}{\bf k}_5}{(2\pi)^3} \int \frac{d^{3}{\bf k}_6}{(2\pi)^3}\nonumber\\
  &&\quad\quad\quad\quad\times \delta^3\bigg({\bf k}_1+{\bf k}_2+{\bf k}_3\bigg) \delta^3\bigg({\bf k}_4+{\bf k}_5+{\bf k}_6\bigg)\left({\bf k}_2\cdot{\bf k}_3\right)\left({\bf k}_5\cdot{\bf k}_6\right)\nonumber\\
  &&\quad\quad\quad\quad\times \langle \hat{\zeta}_{\bf p}(\tau)\hat{\zeta}_{-{\bf p}}(\tau)\hat{\zeta}^{'}_{{\bf k}_1}(\tau_1)\hat{\zeta}_{{\bf k}_2}(\tau_1)\hat{\zeta}_{{\bf k}_3}(\tau_1)\hat{\zeta}^{'}_{{\bf k}_4}(\tau_2)\hat{\zeta}_{{\bf k}_5}(\tau_2)\hat{\zeta}_{{\bf k}_6}(\tau_2)\rangle\Bigg],\\
   \label{A11d}{\bf Z}^{(1)}_4:&=&\lim_{\tau\rightarrow 0}\Bigg[\int^{0}_{-\infty}d\tau_1\frac{a^2(\tau_1)}{H^3(\tau_1)}\frac{{\cal G}_4(\tau_1)}{a^2(\tau_1)}\;\int^{0}_{-\infty}d\tau_2\;\frac{a^2(\tau_2)}{H^3(\tau_2)}\frac{{\cal G}_4(\tau_2)}{a^2(\tau_2)}\nonumber\\
  &&\quad\quad\quad\quad\times\int \frac{d^{3}{\bf k}_1}{(2\pi)^3} \int \frac{d^{3}{\bf k}_2}{(2\pi)^3} \int \frac{d^{3}{\bf k}_3}{(2\pi)^3} \int \frac{d^{3}{\bf k}_4}{(2\pi)^3} \int \frac{d^{3}{\bf k}_5}{(2\pi)^3} \int \frac{d^{3}{\bf k}_6}{(2\pi)^3}\nonumber\\
  &&\quad\quad\quad\quad\times \delta^3\bigg({\bf k}_1+{\bf k}_2+{\bf k}_3\bigg) \delta^3\bigg({\bf k}_4+{\bf k}_5+{\bf k}_6\bigg)\left({\bf k}_1\cdot{\bf k}_2\right)\left({\bf k}_4\cdot{\bf k}_5\right)|{\bf K}|^4\nonumber\\
  &&\quad\quad\quad\quad\times \langle \hat{\zeta}_{\bf p}(\tau)\hat{\zeta}_{-{\bf p}}(\tau)\hat{\zeta}_{{\bf k}_1}(\tau_1)\hat{\zeta}_{{\bf k}_2}(\tau_1)\hat{\zeta}_{{\bf k}_3}(\tau_1)\hat{\zeta}_{{\bf k}_4}(\tau_2)\hat{\zeta}_{{\bf k}_5}(\tau_2)\hat{\zeta}_{{\bf k}_6}(\tau_2)\rangle\Bigg],\eea
  and
\bea \label{A22a}{\bf Z}^{(2)}_1:&=&\lim_{\tau\rightarrow 0}\Bigg[\int^{0}_{-\infty}d\tau_1\frac{a^2(\tau_1)}{H^3(\tau_1)}\frac{{\cal G}_1(\tau_1)}{a(\tau_1)}\;\int^{0}_{-\infty}d\tau_2\;\frac{a^2(\tau_2)}{H^3(\tau_2)}\frac{{\cal G}_1(\tau_2)}{a(\tau_2)}\nonumber\\
  &&\quad\quad\quad\quad\times\int \frac{d^{3}{\bf k}_1}{(2\pi)^3} \int \frac{d^{3}{\bf k}_2}{(2\pi)^3} \int \frac{d^{3}{\bf k}_3}{(2\pi)^3} \int \frac{d^{3}{\bf k}_4}{(2\pi)^3} \int \frac{d^{3}{\bf k}_5}{(2\pi)^3} \int \frac{d^{3}{\bf k}_6}{(2\pi)^3}\nonumber\\
  &&\quad\quad\quad\quad\times \delta^3\bigg({\bf k}_1+{\bf k}_2+{\bf k}_3\bigg) \delta^3\bigg({\bf k}_4+{\bf k}_5+{\bf k}_6\bigg)\nonumber\\
  &&\quad\quad\quad\quad\times \langle \hat{\zeta}_{\bf p}(\tau)\hat{\zeta}_{-{\bf p}}(\tau)\hat{\zeta}^{'}_{{\bf k}_1}(\tau_1)\hat{\zeta}^{'}_{{\bf k}_2}(\tau_1)\hat{\zeta}^{'}_{{\bf k}_3}(\tau_1)\hat{\zeta}^{'}_{{\bf k}_4}(\tau_2)\hat{\zeta}^{'}_{{\bf k}_5}(\tau_2)\hat{\zeta}^{'}_{{\bf k}_6}(\tau_2)\rangle^{\dagger}\Bigg]=\big[{\bf Z}^{(1)}_1\big]^{\dagger},\eea\bea
  \label{A22b}{\bf Z}^{(2)}_2:&=&\lim_{\tau\rightarrow 0}\Bigg[\int^{0}_{-\infty}d\tau_1\frac{a^2(\tau_1)}{H^3(\tau_1)}\frac{{\cal G}_2(\tau_1)}{a^2(\tau_1)}\;\int^{0}_{-\infty}d\tau_2\;\frac{a^2(\tau_2)}{H^3(\tau_2)}\frac{{\cal G}_2(\tau_2)}{a^2(\tau_2)}\nonumber\\
  &&\quad\quad\quad\quad\times\int \frac{d^{3}{\bf k}_1}{(2\pi)^3} \int \frac{d^{3}{\bf k}_2}{(2\pi)^3} \int \frac{d^{3}{\bf k}_3}{(2\pi)^3} \int \frac{d^{3}{\bf k}_4}{(2\pi)^3} \int \frac{d^{3}{\bf k}_5}{(2\pi)^3} \int \frac{d^{3}{\bf k}_6}{(2\pi)^3}\nonumber\\
  &&\quad\quad\quad\quad\times \delta^3\bigg({\bf k}_1+{\bf k}_2+{\bf k}_3\bigg) \delta^3\bigg({\bf k}_4+{\bf k}_5+{\bf k}_6\bigg)|{\bf K}|^4\nonumber\\
  &&\quad\quad\quad\quad\times \langle \hat{\zeta}_{\bf p}(\tau)\hat{\zeta}_{-{\bf p}}(\tau)\hat{\zeta}^{'}_{{\bf k}_1}(\tau_1)\hat{\zeta}^{'}_{{\bf k}_2}(\tau_1)\hat{\zeta}_{{\bf k}_3}(\tau_1)\hat{\zeta}^{'}_{{\bf k}_4}(\tau_2)\hat{\zeta}^{'}_{{\bf k}_5}(\tau_2)\hat{\zeta}_{{\bf k}_6}(\tau_2)\rangle^{\dagger}\Bigg]=\big[{\bf Z}^{(1)}_2\big]^{\dagger},\\
 \label{A22c}{\bf Z}^{(2)}_3:&=&\lim_{\tau\rightarrow 0}\Bigg[\int^{0}_{-\infty}d\tau_1\frac{a^2(\tau_1)}{H^3(\tau_1)}\frac{{\cal G}_3(\tau_1)}{a(\tau_1)}\;\int^{0}_{-\infty}d\tau_2\;\frac{a^2(\tau_2)}{H^3(\tau_2)}\frac{{\cal G}_3(\tau_2)}{a(\tau_2)}\nonumber\\
  &&\quad\quad\quad\quad\times\int \frac{d^{3}{\bf k}_1}{(2\pi)^3} \int \frac{d^{3}{\bf k}_2}{(2\pi)^3} \int \frac{d^{3}{\bf k}_3}{(2\pi)^3} \int \frac{d^{3}{\bf k}_4}{(2\pi)^3} \int \frac{d^{3}{\bf k}_5}{(2\pi)^3} \int \frac{d^{3}{\bf k}_6}{(2\pi)^3}\nonumber\\
  &&\quad\quad\quad\quad\times \delta^3\bigg({\bf k}_1+{\bf k}_2+{\bf k}_3\bigg) \delta^3\bigg({\bf k}_4+{\bf k}_5+{\bf k}_6\bigg)\left({\bf k}_2\cdot{\bf k}_3\right)\left({\bf k}_5\cdot{\bf k}_6\right)\nonumber\\
  &&\quad\quad\quad\quad\times \langle \hat{\zeta}_{\bf p}(\tau)\hat{\zeta}_{-{\bf p}}(\tau)\hat{\zeta}^{'}_{{\bf k}_1}(\tau_1)\hat{\zeta}_{{\bf k}_2}(\tau_1)\hat{\zeta}_{{\bf k}_3}(\tau_1)\hat{\zeta}^{'}_{{\bf k}_4}(\tau_2)\hat{\zeta}_{{\bf k}_5}(\tau_2)\hat{\zeta}_{{\bf k}_6}(\tau_2)\rangle^{\dagger}\Bigg]=\big[{\bf Z}^{(1)}_3\big]^{\dagger},\\
   \label{A22d}{\bf Z}^{(2)}_4:&=&\lim_{\tau\rightarrow 0}\Bigg[\int^{0}_{-\infty}d\tau_1\frac{a^2(\tau_1)}{H^3(\tau_1)}\frac{{\cal G}_4(\tau_1)}{a^2(\tau_1)}\;\int^{0}_{-\infty}d\tau_2\;\frac{a^2(\tau_2)}{H^3(\tau_2)}\frac{{\cal G}_4(\tau_2)}{a^2(\tau_2)}\nonumber\\
  &&\quad\quad\quad\quad\times\int \frac{d^{3}{\bf k}_1}{(2\pi)^3} \int \frac{d^{3}{\bf k}_2}{(2\pi)^3} \int \frac{d^{3}{\bf k}_3}{(2\pi)^3} \int \frac{d^{3}{\bf k}_4}{(2\pi)^3} \int \frac{d^{3}{\bf k}_5}{(2\pi)^3} \int \frac{d^{3}{\bf k}_6}{(2\pi)^3}\nonumber\\
  &&\quad\quad\quad\quad\times \delta^3\bigg({\bf k}_1+{\bf k}_2+{\bf k}_3\bigg) \delta^3\bigg({\bf k}_4+{\bf k}_5+{\bf k}_6\bigg)\left({\bf k}_1\cdot{\bf k}_2\right)\left({\bf k}_4\cdot{\bf k}_5\right)|{\bf K}|^4\nonumber\\
  &&\quad\quad\quad\quad\times \langle \hat{\zeta}_{\bf p}(\tau)\hat{\zeta}_{-{\bf p}}(\tau)\hat{\zeta}_{{\bf k}_1}(\tau_1)\hat{\zeta}_{{\bf k}_2}(\tau_1)\hat{\zeta}_{{\bf k}_3}(\tau_1)\hat{\zeta}_{{\bf k}_4}(\tau_2)\hat{\zeta}_{{\bf k}_5}(\tau_2)\hat{\zeta}_{{\bf k}_6}(\tau_2)\rangle^{\dagger}\Bigg]=\big[{\bf Z}^{(1)}_4\big]^{\dagger},\eea
  and

\bea \label{A33a}{\bf Z}^{(3)}_1:&=&\lim_{\tau\rightarrow 0}\Bigg[\int^{0}_{-\infty}d\tau_1\frac{a^2(\tau_1)}{H^3(\tau_1)}\frac{{\cal G}_1(\tau_1)}{a(\tau_1)}\;\int^{0}_{-\infty}d\tau_2\;\frac{a^2(\tau_2)}{H^3(\tau_2)}\frac{{\cal G}_1(\tau_2)}{a(\tau_2)}\nonumber\\
  &&\quad\quad\quad\quad\times\int \frac{d^{3}{\bf k}_1}{(2\pi)^3} \int \frac{d^{3}{\bf k}_2}{(2\pi)^3} \int \frac{d^{3}{\bf k}_3}{(2\pi)^3} \int \frac{d^{3}{\bf k}_4}{(2\pi)^3} \int \frac{d^{3}{\bf k}_5}{(2\pi)^3} \int \frac{d^{3}{\bf k}_6}{(2\pi)^3}\nonumber\\
  &&\quad\quad\quad\quad\times \delta^3\bigg({\bf k}_1+{\bf k}_2+{\bf k}_3\bigg) \delta^3\bigg({\bf k}_4+{\bf k}_5+{\bf k}_6\bigg)\nonumber\\
  &&\quad\quad\quad\quad\times \langle  \hat{\zeta}^{'}_{{\bf k}_1}(\tau_1)\hat{\zeta}^{'}_{{\bf k}_2}(\tau_1)\hat{\zeta}^{'}_{{\bf k}_3}(\tau_1)
 \hat{\zeta}_{\bf p}(\tau)\hat{\zeta}_{-{\bf p}}(\tau)\hat{\zeta}^{'}_{{\bf k}_4}(\tau_2)\hat{\zeta}^{'}_{{\bf k}_5}(\tau_2)\hat{\zeta}^{'}_{{\bf k}_6}(\tau_2)\rangle\Bigg],\eea\bea
  \label{A33b}{\bf Z}^{(3)}_2:&=&\lim_{\tau\rightarrow 0}\Bigg[\int^{0}_{-\infty}d\tau_1\frac{a^2(\tau_1)}{H^3(\tau_1)}\frac{{\cal G}_2(\tau_1)}{a^2(\tau_1)}\;\int^{0}_{-\infty}d\tau_2\;\frac{a^2(\tau_2)}{H^3(\tau_2)}\frac{{\cal G}_2(\tau_2)}{a^2(\tau_2)}\nonumber\\
  &&\quad\quad\quad\quad\times\int \frac{d^{3}{\bf k}_1}{(2\pi)^3} \int \frac{d^{3}{\bf k}_2}{(2\pi)^3} \int \frac{d^{3}{\bf k}_3}{(2\pi)^3} \int \frac{d^{3}{\bf k}_4}{(2\pi)^3} \int \frac{d^{3}{\bf k}_5}{(2\pi)^3} \int \frac{d^{3}{\bf k}_6}{(2\pi)^3}\nonumber\\
  &&\quad\quad\quad\quad\times \delta^3\bigg({\bf k}_1+{\bf k}_2+{\bf k}_3\bigg) \delta^3\bigg({\bf k}_4+{\bf k}_5+{\bf k}_6\bigg)|{\bf K}|^4\nonumber\\
  &&\quad\quad\quad\quad\times \langle \hat{\zeta}^{'}_{{\bf k}_1}(\tau_1)\hat{\zeta}^{'}_{{\bf k}_2}(\tau_1)\hat{\zeta}_{{\bf k}_3}(\tau_1)\hat{\zeta}_{\bf p}(\tau)\hat{\zeta}_{-{\bf p}}(\tau)\hat{\zeta}^{'}_{{\bf k}_4}(\tau_2)\hat{\zeta}^{'}_{{\bf k}_5}(\tau_2)\hat{\zeta}_{{\bf k}_6}(\tau_2)\rangle\Bigg],\eea\bea
 \label{A33c}{\bf Z}^{(3)}_3:&=&\lim_{\tau\rightarrow 0}\Bigg[\int^{0}_{-\infty}d\tau_1\frac{a^2(\tau_1)}{H^3(\tau_1)}\frac{{\cal G}_3(\tau_1)}{a(\tau_1)}\;\int^{0}_{-\infty}d\tau_2\;\frac{a^2(\tau_2)}{H^3(\tau_2)}\frac{{\cal G}_3(\tau_2)}{a(\tau_2)}\nonumber\\
  &&\quad\quad\quad\quad\times\int \frac{d^{3}{\bf k}_1}{(2\pi)^3} \int \frac{d^{3}{\bf k}_2}{(2\pi)^3} \int \frac{d^{3}{\bf k}_3}{(2\pi)^3} \int \frac{d^{3}{\bf k}_4}{(2\pi)^3} \int \frac{d^{3}{\bf k}_5}{(2\pi)^3} \int \frac{d^{3}{\bf k}_6}{(2\pi)^3}\nonumber\\
  &&\quad\quad\quad\quad\times \delta^3\bigg({\bf k}_1+{\bf k}_2+{\bf k}_3\bigg) \delta^3\bigg({\bf k}_4+{\bf k}_5+{\bf k}_6\bigg)\left({\bf k}_2\cdot{\bf k}_3\right)\left({\bf k}_5\cdot{\bf k}_6\right)\nonumber\\
  &&\quad\quad\quad\quad\times \langle \hat{\zeta}^{'}_{{\bf k}_1}(\tau_1)\hat{\zeta}_{{\bf k}_2}(\tau_1)\hat{\zeta}_{{\bf k}_3}(\tau_1)\hat{\zeta}_{\bf p}(\tau)\hat{\zeta}_{-{\bf p}}(\tau)\hat{\zeta}^{'}_{{\bf k}_4}(\tau_2)\hat{\zeta}_{{\bf k}_5}(\tau_2)\hat{\zeta}_{{\bf k}_6}(\tau_2)\rangle\Bigg],\\ 
   \label{A33d}{\bf Z}^{(3)}_4:&=&\lim_{\tau\rightarrow 0}\Bigg[\int^{0}_{-\infty}d\tau_1\frac{a^2(\tau_1)}{H^3(\tau_1)}\frac{{\cal G}_4(\tau_1)}{a^2(\tau_1)}\;\int^{0}_{-\infty}d\tau_2\;\frac{a^2(\tau_2)}{H^3(\tau_2)}\frac{{\cal G}_4(\tau_2)}{a^2(\tau_2)}\nonumber\\
  &&\quad\quad\quad\quad\times\int \frac{d^{3}{\bf k}_1}{(2\pi)^3} \int \frac{d^{3}{\bf k}_2}{(2\pi)^3} \int \frac{d^{3}{\bf k}_3}{(2\pi)^3} \int \frac{d^{3}{\bf k}_4}{(2\pi)^3} \int \frac{d^{3}{\bf k}_5}{(2\pi)^3} \int \frac{d^{3}{\bf k}_6}{(2\pi)^3}\nonumber\\
  &&\quad\quad\quad\quad\times \delta^3\bigg({\bf k}_1+{\bf k}_2+{\bf k}_3\bigg) \delta^3\bigg({\bf k}_4+{\bf k}_5+{\bf k}_6\bigg)\left({\bf k}_1\cdot{\bf k}_2\right)\left({\bf k}_4\cdot{\bf k}_5\right)|{\bf K}|^4\nonumber\\
  &&\quad\quad\quad\quad\times \langle \hat{\zeta}_{{\bf k}_1}(\tau_1)\hat{\zeta}_{{\bf k}_2}(\tau_1)\hat{\zeta}_{{\bf k}_3}(\tau_1)\hat{\zeta}_{\bf p}(\tau)\hat{\zeta}_{-{\bf p}}(\tau)\hat{\zeta}_{{\bf k}_4}(\tau_2)\hat{\zeta}_{{\bf k}_5}(\tau_2)\hat{\zeta}_{{\bf k}_6}(\tau_2)\rangle\Bigg],\eea
where it is important to note that:
\bea |{\bf K}|=|{\bf k}_1+{\bf k}_2+{\bf k}_3|=|{\bf k}_4+{\bf k}_5+{\bf k}_6|=\sqrt{k^2_1+k^2_2+k^3_3}=\sqrt{k^2_4+k^2_5+k^2_6}.\eea
Also the integral over conformal time as well as the momentum scales has to be tackled by dividing them in the previously mentioned three regions:
\bea {\bf Conformal\;time\;integral:}\quad\quad\quad\quad\lim_{\tau\rightarrow 0}\int^{\tau}_{-\infty}:\equiv \underbrace{\Bigg(\int^{\tau_s}_{-\infty}\Bigg)}_{\bf SRI}+\underbrace{\Bigg(\int^{\tau_e}_{\tau_s}\Bigg)}_{\bf USR}+\underbrace{\Bigg(\int^{\tau_{\rm end}\rightarrow 0}_{\tau_e}\Bigg)}_{\bf SRII},\eea
and 
\bea {\bf Momentum\;integral:}\quad\quad\quad\quad\int^{\infty}_{0}:\equiv \underbrace{\Bigg(\int^{k_s}_{k_*}\Bigg)}_{\bf SRI}+\underbrace{\Bigg(\int^{k_e}_{k_s}\Bigg)}_{\bf USR}+\underbrace{\Bigg(\int^{k_{\rm end}\rightarrow 0}_{k_e}\Bigg)}_{\bf SRII},\eea
where in the above description finite limits of the integration plays the role of IR and UV cut-offs, which play a very crucial role to extract the finite contribution out of the present computation. However, inclusion of such cut-off regulators allow some divergent effects in the final result which one needs to remove by performing renormalization as well as resummation, which we will going to address in detail in the later half of this paper. At present, our prime objective is to explicitly compute the cut-off regulated expressions for the two-point primordial cosmological correlators from curvature perturbation in the three consecutive regions, SRI, USR and SRII respectively. In the next subsection we are going to study these facts in detail.

\subsection{Computation of the cut-off regularized one-loop correction to the tree-level power spectrum}
\label{s5c}

In the following subsection we compute the explicit contributions from the one-loop correction by implementing the cut-off regularization technique in the subsequent aforementioned regions. Before going to the technical details in this section, let us first mention that using all possible types of Wick contractions the one-loop result can be simplified by the following expression:
 \bea \label{OLP} \langle\langle\hat{\zeta}_{\bf p}\hat{\zeta}_{-{\bf p}}\rangle\rangle_{\bf One-loop}&=&\langle\langle\hat{\zeta}_{\bf p}\hat{\zeta}_{-{\bf p}}\rangle\rangle_{(1,1)}+2{\rm Re}\bigg[\langle\langle\hat{\zeta}_{\bf p}\hat{\zeta}_{-{\bf p}}\rangle\rangle_{(0,2)}\bigg]\nonumber\\
 &=&\sum^{4}_{i=1}{\bf Z}^{(3)}_i+2{\rm Re}\Bigg[\sum^{4}_{i=1}{\bf Z}^{(1)}_i\Bigg], \eea 
which is easier to evaluate in the three regions of discussion.

\subsubsection{Result for the Region I (SRI)}
\label{s5c1}

The one-loop contribution to the primordial power spectrum computed from the comoving curvature perturbation in the SRI region is computed as:
 \bea \label{OLP1} \Bigg[\Delta^{2}_{\zeta,{\bf One-loop}}(p)\Bigg]_{\bf SRI}&=&\Bigg[\Delta^{2}_{\zeta,{\bf Tree}}(p)\Bigg]^2_{\bf SRI}\Bigg\{c_{\bf SRI}-\frac{1}{8{\cal A}^2_*\pi^4}\sum^{4}_{i=1}\widetilde{\cal G}_{i,{\bf SRI}} {\bf F}_{i,{\bf SRI}}(k_s,k_*)\Bigg\}\nonumber\\
 &=&\Bigg[\Delta^{2}_{\zeta,{\bf Tree}}(p)\Bigg]^2_{\bf SRI}\Bigg\{c_{\bf SRI}-\frac{c^4_s}{8{\cal B}^2_*\pi^6}\sum^{4}_{i=1}\widetilde{\cal G}_{i,{\bf SRI}} {\bf F}_{i,{\bf SRI}}(k_s,k_*)\Bigg\}.\eea 
where the power spectrum in the SRI region is given by:
\bea \Bigg[\Delta^{2}_{\zeta,{\bf Tree}}(p)\Bigg]_{\bf SRI}&=&\left(\frac{H^{4}}{8\pi^{2}{\cal A} c^3_s}\right)_*\Bigg\{1+\Bigg(\frac{p}{k_s}\Bigg)^2\Bigg\}=\left(\frac{H^{4}}{8\pi^{2}{\cal B} c_s}\right)_*\Bigg\{1+\Bigg(\frac{p}{k_s}\Bigg)^2\Bigg\}\quad\quad{\rm where}\quad p<k_S.\eea
Here $c_{\bf SRI}$ represents the regularization scheme dependent parameter for the present computational purpose in the SRI phase.
Also the coefficients in terms CGEFT couplings can be expressed at the pivot scale $\tau=\tau_*$ by the following expressions:
\bea \widetilde{\cal G}_{1,{\bf SRI}}&=&{\cal G}^2_1(\tau_*)\nonumber\\
    &=&\frac{4H^2(\tau_*)\dot{\bar{\phi}}^6_0(\tau_*)}{\Lambda^6}\Bigg(c_3+9c_4Z_*+30c_5Z^2_*\Bigg)^2,\\ 
    \widetilde{\cal G}_{2,{\bf SRI}}&=&-{\cal G}^2_2(\tau_*)H^2(\tau_*)c^2_s\nonumber\\
    &=&-\frac{4H^2(\tau_*)\dot{\bar{\phi}}^6_0(\tau_*)c^2_s}{\Lambda^6}\Bigg(c_3+6c_4Z_*+18c_5Z^2_*\Bigg)^2,\\
    \widetilde{\cal G}_{3,{\bf SRI}}&=&-\frac{{\cal G}^2_3(\tau_*)}{c^2_s}\nonumber\\
    &=&-\Bigg(\frac{2H(\tau_*)\dot{\bar{\phi}}^3_0(\tau_*)}{\Lambda^3c_s}\Bigg(c_3+7c_4Z_*+18c_5Z^2_*\Bigg)+\frac{2\dot{\bar{\phi}}^3_0(\tau_*)H(\tau_*)\eta_*}{\Lambda^3c_s}\Bigg(c_3+6c_4Z_*+18c_5Z^2_*\Bigg)\Bigg)^2,\\
    \widetilde{\cal G}_{4,{\bf SRI}}&=&\frac{{\cal G}^2_4(\tau_*)H^2(\tau_*)}{c^6_s}\nonumber\\
    &=&\Bigg(\frac{\dot{\bar{\phi}}^3_0(\tau_*)H(\tau_*)}{\Lambda^3c^3_s}\bigg\{c_3+3c_4Z_*+6c_5\bigg[Z^2_*+\frac{\dot{H}(\tau_*)\dot{\bar{\phi}}^2_0(\tau_*)}{\Lambda^6}\bigg]\bigg\}-\frac{3\dot{\bar{\phi}}^4_0(\tau_*)H^2(\tau_*)\eta_*}{\Lambda^6c^3_s}\bigg\{c_4+4c_5Z_*\bigg\}\Bigg)^2.\eea

    Also, the momentum and effective sound speed dependent functions are described by the following expressions:
    \bea {\bf F}_{1,{\bf SRI}}(k_s,k_*)&=&\frac{1}{2}\Bigg[3+\left(\frac{k_*}{k_s}\right)^2\Bigg],\\ 
         {\bf F}_{2,{\bf SRI}}(k_s,k_*)&=&\Bigg[\frac{17}{42}-\frac{2}{3}\left(\frac{k_*}{k_s}\right)^6+\frac{24}{7}\left(\frac{k_*}{k_s}\right)^7-\frac{9}{2}\left(\frac{k_*}{k_s}\right)^8\Bigg],\\ 
         {\bf F}_{3,{\bf SRI}}(k_s,k_*)&=&-\frac{2}{3}\Bigg[1-\left(\frac{k_*}{k_s}\right)^6\Bigg],\\ 
         {\bf F}_{4,{\bf SRI}}(k_s,k_*)&=&-\frac{1}{2}\Bigg[1-\left(\frac{k_*}{k_s}\right)^2\Bigg].\eea
         At the pivot scale $\tau=\tau_*$, the parameters ${\cal A}_*$ and ${\cal B}_*$ are defined by the following expressions:
\bea {\cal A}_*&\equiv& {\cal A}(\tau_*)=\frac{\dot{\bar{\phi}}^2_0(\tau_*)}{2}\Bigg(c_2+12c_3Z_*+54c_4Z^2_*+120c_5Z^3_*\Bigg),\\
    {\cal B}_*&\equiv& {\cal B}(\tau_*)\nonumber\\&=& \frac{\dot{\bar{\phi}}^2_0(\tau_*)}{2}\Bigg\{c_2+4c_3\Bigg(2Z_*-\frac{H(\tau_*)\dot{\bar{\phi}}_0(\tau_*)}{\Lambda^3}\eta_*\Bigg)+2c_4\Bigg[13Z^2_*-\frac{6}{\Lambda^6}\dot{\bar{\phi}}^2_0(\tau_*)H^2(\tau_*)\big(\epsilon(\tau_*)+2\eta(\tau_*)\big)\Bigg]\nonumber\\
    &&\quad\quad\quad\quad\quad\quad\quad\quad\quad\quad\quad\quad-\frac{24c_5}{\Lambda^9}H^3(\tau_*)\dot{\bar{\phi}}^3_0(\tau_*)\big(2\epsilon(\tau_*)+1\big)\Bigg\}.\quad\quad\eea
         which are extremely useful to fix the effective sound speed at that scale by making use of the following relationship:
         \bea c_s=c_s(\tau_*)=\sqrt{\frac{{\cal B}_*}{{\cal A}_*}}.\eea
        Also, it is important to mention the explicit expression for the coupling parameter $Z_*$ at the pivot scale $\tau=\tau_*$ as appearing in the SRI phase is given by the following expression:
         \bea Z_*\equiv Z(\tau_*)= \frac{H(\tau_*)\dot{\bar{\phi}}_0(\tau_*)}{\Lambda^3},\eea
         where $\dot{\bar{\phi}}_0(\tau_*)$ can be expressed as:
         \bea \dot{\bar{\phi}}_0(\tau_*)=\frac{\Lambda^3}{12H(\tau_*)}\frac{c_2}{c_3}\Bigg[-1+\sqrt{1+\frac{8c_3}{c^2_2}\frac{\lambda^3}{\Lambda^3}}\Bigg].\eea
         
\subsubsection{Result for the Region II (USR)}
\label{s5c2}

The one-loop contribution to the primordial power spectrum computed from the comoving curvature perturbation in the USR region is computed as:
 \bea \label{OLP2} \Bigg[\Delta^{2}_{\zeta,{\bf One-loop}}(p)\Bigg]_{\bf USR}&=&\Bigg[\Delta^{2}_{\zeta,{\bf Tree}}(p)\Bigg]^2_{\bf SRI}\Bigg\{c_{\bf USR}+\frac{1}{8{\cal A}^2_*\pi^4}\sum^{4}_{i=1}\widetilde{\cal G}_{i,{\bf USR}} {\bf F}_{i,{\bf USR}}(k_e,k_s)\Bigg\}\Theta(p-k_s)\nonumber\\
 &=&\Bigg[\Delta^{2}_{\zeta,{\bf Tree}}(p)\Bigg]^2_{\bf SRI}\Bigg\{c_{\bf USR}+\frac{c^4_s}{8{\cal B}^2_*\pi^6}\sum^{4}_{i=1}\widetilde{\cal G}_{i,{\bf USR}} {\bf F}_{i,{\bf USR}}(k_e,k_s)\Bigg\}\Theta(p-k_s).\eea 
where the power spectrum in the SRI region is already defined earlier. Here $c_{\bf USR}$ represents the regularization scheme dependent parameter for the present computational purpose in the USR phase. Also  the coefficients in terms CGEFT couplings can be expressed within the interval $\tau_s\leq \tau \leq \tau_e$ by the following expressions:
\bea \widetilde{\cal G}_{1,{\bf USR}}&=&\Bigg(\frac{{\cal G}^2_1(\tau_e)}{c^3_s}\left(\frac{k_e}{k_s}\right)^6-\frac{{\cal G}^2_1(\tau_s)}{c^3_s}\Bigg)\nonumber\\
&=& \Bigg\{\frac{4H^2(\tau_e)\dot{\bar{\phi}}^6_0(\tau_e)}{\Lambda^6c^3_s}\Bigg(c_3+9c_4Z_e+30c_5Z^2_e\Bigg)^2\left(\frac{k_e}{k_s}\right)^6-\frac{4H^2(\tau_s)\dot{\bar{\phi}}^6_0(\tau_s)}{\Lambda^6c^3_s}\Bigg(c_3+9c_4Z_s+30c_5Z^2_s\Bigg)^2\Bigg\},\\
    \widetilde{\cal G}_{2,{\bf USR}}&=&\Bigg(\frac{{\cal G}^2_2(\tau_e)H^2(\tau_e)}{c^5_s}\left(\frac{k_e}{k_s}\right)^4-\frac{{\cal G}^2_2(\tau_s)H^2(\tau_s)}{c^5_s}\Bigg) \nonumber\\
    &=&\Bigg\{\frac{4H^2(\tau_e)\dot{\bar{\phi}}^6_0(\tau_e)}{\Lambda^6c^5_s}\Bigg(c_3+6c_4Z_e+18c_5Z^2_e\Bigg)^2\left(\frac{k_e}{k_s}\right)^4-\frac{4H^2(\tau_s)\dot{\bar{\phi}}^6_0(\tau_s)}{\Lambda^6c^5_s}\Bigg(c_3+6c_4Z_s+18c_5Z^2_s\Bigg)^2\Bigg\},\eea\bea
    \widetilde{\cal G}_{3,{\bf USR}}&=&\Bigg(\frac{{\cal G}^2_3(\tau_e)}{c^6_s}\left(\frac{k_e}{k_s}\right)^3-\frac{{\cal G}^2_3(\tau_s)}{c^6_s}\Bigg)\nonumber\\
    &=&\Bigg(\frac{2H(\tau_e)\dot{\bar{\phi}}^3_0(\tau_e)}{\Lambda^3c^3_s}\Bigg(c_3+7c_4Z_e+18c_5Z^2_e\Bigg)+\frac{2\dot{\bar{\phi}}^3_0(\tau_e)H(\tau_e)\eta(\tau_e)}{\Lambda^3c^3_s}\Bigg(c_3+6c_4Z_e+18c_5Z^2_e\Bigg)\Bigg)^2\left(\frac{k_e}{k_s}\right)^3\nonumber\\
    && \quad\quad-\Bigg(\frac{2H(\tau_s)\dot{\bar{\phi}}^3_0(\tau_s)}{\Lambda^3c^3_s}\Bigg(c_3+7c_4Z_s+18c_5Z^2_s\Bigg)+\frac{2\dot{\bar{\phi}}^3_0(\tau_s)H(\tau_s)\eta(\tau_s)}{\Lambda^3c^3_s}\Bigg(c_3+6c_4Z_s+18c_5Z^2_s\Bigg)\Bigg)^2,\\
    \widetilde{\cal G}_{4,{\bf USR}}&=&\Bigg(\frac{{\cal G}^2_2(\tau_e)H^2(\tau_e)}{c^7_s}\left(\frac{k_e}{k_s}\right)^2-\frac{{\cal G}^2_2(\tau_s)H^2(\tau_s)}{c^7_s}\Bigg)\nonumber\\
    &=&\Bigg(\frac{\dot{\bar{\phi}}^3_0(\tau_e)}{\Lambda^3c^{7/2}_s}\bigg\{c_3+3c_4Z_e+6c_5\bigg[Z^2_e+\frac{\dot{H}(\tau_e)\dot{\bar{\phi}}^2_0(\tau_e)}{\Lambda^6}\bigg]\bigg\}-\frac{3\dot{\bar{\phi}}^4_0(\tau_e)H(\tau_e)\eta(\tau_e)}{\Lambda^6c^{7/2}_s}\bigg\{c_4+4c_5Z_e\bigg\}\Bigg)^2\left(\frac{k_e}{k_s}\right)^2\nonumber\\
    &&\quad\quad-\Bigg(\frac{\dot{\bar{\phi}}^3_0(\tau_s)}{\Lambda^3c^{7/2}_s}\bigg\{c_3+3c_4Z_s+6c_5\bigg[Z^2_s+\frac{\dot{H}(\tau_s)\dot{\bar{\phi}}^2_0(\tau_s)}{\Lambda^6}\bigg]\bigg\}-\frac{3\dot{\bar{\phi}}^4_0(\tau_s)H(\tau_s)\eta(\tau_s)}{\Lambda^6c^{7/2}_s}\bigg\{c_4+4c_5Z_s\bigg\}\Bigg)^2.\eea

    Also, the momentum and effective sound speed dependent functions are described by the following expressions:
    \bea {\bf F}_{1,{\bf USR}}(k_e,k_s)&=&\Bigg[\frac{1}{4}+\frac{9}{4}\left(\frac{k_s}{k_e}\right)^2-\frac{1}{4}\left(\frac{k_s}{k_e}\right)^4-9\left(\frac{k_s}{k_e}\right)^4\ln\left(\frac{k_s}{k_e}\right)-6\sin\left(1+\left(\frac{k_e}{k_s}\right)\right)\sin\left(1-\left(\frac{k_e}{k_s}\right)\right)\nonumber\\
    &&\quad\quad-\frac{9}{8}\left(\frac{k_s}{k_e}\right)^2\Bigg\{2\left(\frac{k_e}{k_s}\right)\sin\left(2\left(\frac{k_e}{k_s}\right)\right)-\cos\left(2\left(\frac{k_e}{k_s}\right)\right)\Bigg\}-6\left(\frac{k_s}{k_e}\right)\sin\left(2\left(\frac{k_e}{k_s}\right)\right)\nonumber\\
    &&\quad\quad-\frac{69}{16}\cos\left(2\left(\frac{k_e}{k_s}\right)\right)-\frac{3}{4}\left(\frac{k_s}{k_e}\right)^2\Bigg\{2\left(\frac{k_e}{k_s}\right)\cos\left(2\left(\frac{k_e}{k_s}\right)\right)-\sin\left(2\left(\frac{k_e}{k_s}\right)\right)\Bigg\}\Bigg],\\
         {\bf F}_{2,{\bf USR}}(k_e,k_s)&=&\Bigg[\frac{1}{8}+\frac{9}{12}\left(\frac{k_s}{k_e}\right)^2+\frac{9}{4}\left(\frac{k_s}{k_e}\right)^4+\frac{9}{4}\left(\frac{k_s}{k_e}\right)^6-\frac{43}{8}\left(\frac{k_s}{k_e}\right)^8-\frac{9}{8}\cos\left(2\left(\frac{k_e}{k_s}\right)\right)\Bigg],\\
         {\bf F}_{3,{\bf USR}}(k_e,k_s)&=&\Bigg[\frac{1}{8}+\frac{9}{12}\left(\frac{k_s}{k_e}\right)^2+\frac{9}{4}\left(\frac{k_s}{k_e}\right)^4+\frac{9}{4}\left(\frac{k_s}{k_e}\right)^6-\frac{43}{8}\left(\frac{k_s}{k_e}\right)^8-\frac{9}{8}\cos\left(2\left(\frac{k_e}{k_s}\right)\right)\Bigg]\nonumber\\
         &=&{\bf F}_{2,{\bf USR}}(k_e,k_s),\\ 
         {\bf F}_{4,{\bf USR}}(k_e,k_s)&=&\Bigg[\frac{1}{12}+\frac{9}{20}\left(\frac{k_s}{k_e}\right)^2+\frac{9}{8}\left(\frac{k_s}{k_e}\right)^4+\frac{9}{12}\left(\frac{k_s}{k_e}\right)^6-\frac{299}{230}\left(\frac{k_s}{k_e}\right)^{12}\Bigg].\eea
Additionally, it is important to mention the explicit expression for the coupling parameters $Z_e$ and $Z_s$ at the end of USR and SRI to USR transition scales $\tau=\tau_e$ and $\tau=\tau_s$ are given by the following expression:
         \bea && Z_e\equiv Z(\tau_e)= \frac{H(\tau_e)\dot{\bar{\phi}}_0(\tau_e)}{\Lambda^3},\\
         && Z_s\equiv Z(\tau_s)= \frac{H(\tau_s)\dot{\bar{\phi}}_0(\tau_s)}{\Lambda^3},\eea
         where $\dot{\bar{\phi}}_0(\tau_e)$ and $\dot{\bar{\phi}}_0(\tau_s)$ can be expressed as:
         \bea &&\dot{\bar{\phi}}_0(\tau_e)=\frac{\Lambda^3}{12H(\tau_e)}\frac{c_2}{c_3}\Bigg[-1+\sqrt{1+\frac{8c_3}{c^2_2}\frac{\lambda^3}{\Lambda^3}}\Bigg], \\ &&\dot{\bar{\phi}}_0(\tau_s)=\frac{\Lambda^3}{12H(\tau_s)}\frac{c_2}{c_3}\Bigg[-1+\sqrt{1+\frac{8c_3}{c^2_2}\frac{\lambda^3}{\Lambda^3}}\Bigg].\eea
         
\subsubsection{Result for the Region III (SRII)}
\label{s5c3}

The one-loop contribution to the primordial power spectrum computed from the comoving curvature perturbation in the SRII region is computed as:
 \bea \label{OLP3} \Bigg[\Delta^{2}_{\zeta,{\bf One-loop}}(p)\Bigg]_{\bf SRII}&=&\Bigg[\Delta^{2}_{\zeta,{\bf Tree}}(p)\Bigg]^2_{\bf SRI}\Bigg\{c_{\bf SRII}+\frac{1}{8{\cal A}^2_*\pi^4}\sum^{4}_{i=1}\widetilde{\cal G}_{i,{\bf SRII}} {\bf F}_{i,{\bf SRII}}(k_{\rm end},k_e)\Bigg\}\Theta(p-k_e)\nonumber\\
 &=&\Bigg[\Delta^{2}_{\zeta,{\bf Tree}}(p)\Bigg]^2_{\bf SRI}\Bigg\{c_{\bf SRII}+\frac{c^4_s}{8{\cal B}^2_*\pi^6}\sum^{4}_{i=1}\widetilde{\cal G}_{i,{\bf SRII}} {\bf F}_{i,{\bf SRII}}(k_{\rm end},k_e)\Bigg\}\Theta(p-k_e).\eea 
where the power spectrum in the SRI region is already defined earlier.Here $c_{\bf SRII}$ represents the regularization scheme dependent parameter for the present computational purpose in the SRII phase.
Also the coefficients in terms CGEFT couplings can be expressed within the interval $\tau_{e}\leq \tau \leq \tau_{\rm end}$ by the following expressions:

\bea \widetilde{\cal G}_{1,{\bf SRII}}&=&\Bigg(\frac{{\cal G}^2_1(\tau_{\rm end})}{c^3_s}\left(\frac{k_e}{k_s}\right)^6-\frac{{\cal G}^2_1(\tau_e)}{c^3_s}\Bigg)\nonumber\\
&=& \Bigg\{\frac{4H^2(\tau_{\rm end})\dot{\bar{\phi}}^6_0(\tau_{\rm end})}{\Lambda^6c^3_s}\Bigg(c_3+9c_4Z_{\rm end}+30c_5Z^2_{\rm end}\Bigg)^2\left(\frac{k_e}{k_s}\right)^6-\frac{4H^2(\tau_e)\dot{\bar{\phi}}^6_0(\tau_e)}{\Lambda^6c^3_s}\Bigg(c_3+9c_4Z_e+30c_5Z^2_e\Bigg)^2\Bigg\},\quad\quad\quad\\
    \widetilde{\cal G}_{2,{\bf SRII}}&=&\Bigg(\frac{{\cal G}^2_2(\tau_{\rm end})H^2(\tau_{\rm end})}{c^5_s}\left(\frac{k_e}{k_s}\right)^4-\frac{{\cal G}^2_2(\tau_e)H^2(\tau_e)}{c^5_s}\Bigg) \nonumber\\
    &=&\Bigg\{\frac{4H^2(\tau_{\rm end})\dot{\bar{\phi}}^6_0(\tau_{\rm end})}{\Lambda^6c^5_s}\Bigg(c_3+6c_4Z_{\rm end}+18c_5Z^2_{\rm end}\Bigg)^2\left(\frac{k_e}{k_s}\right)^4-\frac{4H^2(\tau_e)\dot{\bar{\phi}}^6_0(\tau_e)}{\Lambda^6c^5_s}\Bigg(c_3+6c_4Z_e+18c_5Z^2_e\Bigg)^2\Bigg\},\\
    \widetilde{\cal G}_{3,{\bf SRII}}&=&\Bigg(\frac{{\cal G}^2_3(\tau_{\rm end})}{c^6_s}\left(\frac{k_e}{k_s}\right)^3-\frac{{\cal G}^2_3(\tau_e)}{c^6_s}\Bigg)\nonumber\\
    &=&\Bigg(\frac{2H(\tau_{\rm end})\dot{\bar{\phi}}^3_0(\tau_{\rm end})}{\Lambda^3c^3_s}\Bigg(c_3+7c_4Z_{\rm end}+18c_5Z^2_{\rm end}\Bigg)+\frac{2\dot{\bar{\phi}}^3_0(\tau_{\rm end})H(\tau_{\rm end})\eta(\tau_{\rm end})}{\Lambda^3c^3_s}\Bigg(c_3+6c_4Z_{\rm end}+18c_5Z^2_{\rm end}\Bigg)\Bigg)^2\left(\frac{k_e}{k_s}\right)^3\nonumber\\
    && \quad\quad-\Bigg(\frac{2H(\tau_e)\dot{\bar{\phi}}^3_0(\tau_e)}{\Lambda^3c^3_s}\Bigg(c_3+7c_4Z_e+18c_5Z^2_e\Bigg)+\frac{2\dot{\bar{\phi}}^3_0(\tau_e)H(\tau_e)\eta(\tau_e)}{\Lambda^3c^3_s}\Bigg(c_3+6c_4Z_e+18c_5Z^2_e\Bigg)\Bigg)^2,
    \\
    \widetilde{\cal G}_{4,{\bf SRII}}&=&\Bigg(\frac{{\cal G}^2_2(\tau_{\rm end})H^2(\tau_{\rm end})}{c^7_s}\left(\frac{k_e}{k_s}\right)^2-\frac{{\cal G}^2_2(\tau_e)H^2(\tau_e)}{c^7_s}\Bigg)\nonumber\\
    &=&\Bigg(\frac{\dot{\bar{\phi}}^3_0(\tau_{\rm end})}{\Lambda^3c^{7/2}_s}\bigg\{c_3+3c_4Z_{\rm end}+6c_5\bigg[Z^2_{\rm end}+\frac{\dot{H}(\tau_{\rm end})\dot{\bar{\phi}}^2_0(\tau_{\rm end})}{\Lambda^6}\bigg]\bigg\}-\frac{3\dot{\bar{\phi}}^4_0(\tau_{\rm end})H(\tau_{\rm end})\eta(\tau_{\rm end})}{\Lambda^6c^{7/2}_s}\bigg\{c_4+4c_5Z_{\rm end}\bigg\}\Bigg)^2\left(\frac{k_e}{k_s}\right)^2\nonumber\\
    &&\quad\quad-\Bigg(\frac{\dot{\bar{\phi}}^3_0(\tau_e)}{\Lambda^3c^{7/2}_s}\bigg\{c_3+3c_4Z_e+6c_5\bigg[Z^2_e+\frac{\dot{H}(\tau_e)\dot{\bar{\phi}}^2_0(\tau_e)}{\Lambda^6}\bigg]\bigg\}-\frac{3\dot{\bar{\phi}}^4_0(\tau_e)H(\tau_e)\eta(\tau_e)}{\Lambda^6c^{7/2}_s}\bigg\{c_4+4c_5Z_e\bigg\}\Bigg)^2.\eea

    Also, the momentum and effective sound speed-dependent functions are described by the following expressions:
    \bea {\bf F}_{1,{\bf SRII}}(k_{\rm end},k_e)&=&\Bigg[\frac{81}{64}\left(1-\left(\frac{k_e}{k_{\rm end}}\right)^4\right)+\frac{81}{40}\left(1-\left(\frac{k_e}{k_{\rm end}}\right)^5\right)\left(1+\left(\frac{k_s}{k_e}\right)\right)-\frac{9}{8}\left(1-\left(\frac{k_e}{k_{\rm end}}\right)^2\right)\left(1+\left(\frac{k_s}{k_e}\right)^2\right)\nonumber\\
    &&+\frac{1}{6}\left(1-\left(\frac{k_e}{k_{\rm end}}\right)^6\right)\left(\left(1+\left(\frac{k_s}{k_e}\right)\right)^2+2\left(\frac{k_s}{k_e}\right)\right)+\frac{2}{7}\left(1-\left(\frac{k_e}{k_{\rm end}}\right)^7\right)\left(\frac{k_s}{k_e}\right)\left(1+\left(\frac{k_s}{k_e}\right)\right)\nonumber\eea\bea
    &&+\frac{1}{8}\left(1-\left(\frac{k_e}{k_{\rm end}}\right)^8\right)\left(\frac{k_s}{k_e}\right)^2+\frac{27}{8}\left(1-\left(\frac{k_e}{k_{\rm end}}\right)^2\right)\left(1+\left(\frac{k_s}{k_e}\right)^6\right)-\frac{3}{4}\left(\frac{k_e}{k_{\rm end}}\right)\cos\left(2\left(\frac{k_{\rm end}}{k_e}\right)\right)\nonumber\\
    &&+4\left(\frac{k_e}{k_{\rm end}}\right)^2\ln\left(\frac{k_e}{k_{\rm end}}\right)+6\left(1-\left(\frac{k_e}{k_{\rm end}}\right)^8\right)-\frac{3}{4}\sin\left(1+\left(\frac{k_{\rm end}}{k_s}\right)\right)\sin\left(1-\left(\frac{k_{\rm end}}{k_s}\right)\right)\nonumber\\
    &&-\frac{9}{4}\frac{\displaystyle\left(\frac{k_s}{k_e}\right)}{\displaystyle\left(1-\left(\frac{k_s}{k_e}\right)\right)^2}\Bigg\{\cos\left(2\left(\left(\frac{k_{\rm end}}{k_e}\right)-\left(\frac{k_{\rm end}}{k_s}\right)\right)\right)-\cos\left(2\left(1-\left(\frac{k_e}{k_s}\right)\right)\right)\Bigg\}\nonumber\\
    &&-\frac{9}{4}\frac{\displaystyle 1}{\displaystyle\left(1-\left(\frac{k_s}{k_e}\right)\right)}\Bigg\{\left(\frac{k_s}{k_{\rm end}}\right)\sin\left(2\left(\left(\frac{k_{\rm end}}{k_e}\right)-\left(\frac{k_{\rm end}}{k_s}\right)\right)\right)-\left(\frac{k_s}{k_{e}}\right)\sin\left(2\left(1-\left(\frac{k_e}{k_s}\right)\right)\right)\Bigg\}\nonumber\\
    &&-9\frac{\displaystyle\left(1+\left(\frac{k_s}{k_e}\right)\right)^2}{\displaystyle \left(1-\left(\frac{k_s}{k_e}\right)\right)}\Bigg\{\sin\left(2\left(\left(\frac{k_{\rm end}}{k_e}\right)-\left(\frac{k_{\rm end}}{k_s}\right)\right)\right)-\sin\left(2\left(1-\left(\frac{k_e}{k_s}\right)\right)\right)\Bigg\}\nonumber\\
    &&+\frac{9}{8}\Bigg\{8-18\left(\frac{k_s}{k_e}\right)^4\left(1+\left(\frac{k_s}{k_e}\right)\right)\Bigg\}\Bigg\{\cos\left(2\left(\left(\frac{k_{\rm end}}{k_e}\right)-\left(\frac{k_{\rm end}}{k_s}\right)\right)\right)-\left(\frac{k_e}{k_{\rm end}}\right)\cos\left(2\left(1-\left(\frac{k_e}{k_s}\right)\right)\right)\Bigg\}\nonumber\\
    &&-\frac{9}{16}\left(\frac{k_s}{k_e}\right)^2\left(9+44\left(\frac{k_s}{k_e}\right)^2\right)\ln\left(\frac{k_e}{k_{\rm end}}\right)+9\left(\frac{k_s}{k_e}\right)^4\left(1+\left(\frac{k_s}{k_e}\right)^2\right)\left(1-\left(\frac{k_e}{k_{\rm end}}\right)^2\right)\nonumber\\
    &&+\frac{9}{7}\left(\frac{k_s}{k_e}\right)^6\left(1-\left(\frac{k_e}{k_{\rm end}}\right)^7\right)+22\left(\frac{k_s}{k_e}\right)^2\left(1-\left(\frac{k_e}{k_{\rm end}}\right)\right)+\frac{81}{128}\left(\frac{k_e}{k_{\rm end}}\right)^6\left(\frac{k_s}{k_{\rm end}}\right)^6\left(1-\left(\frac{k_e}{k_{\rm end}}\right)^8\right)\nonumber\\
    &&+\frac{27}{16}\left(\frac{k_e}{k_{\rm end}}\right)^6\left(1-\left(\frac{k_e}{k_{\rm end}}\right)^6\right) \left(1+\left(\frac{k_s}{k_e}\right)^2\right)-\frac{9}{16}\left(8+9\left(\frac{k_s}{k_e}\right)^2\right)\left(\frac{k_e}{k_{\rm end}}\right)^4\ln\left(\frac{k_e}{k_{\rm end}}\right)\nonumber\\
    &&+\frac{81}{64}\left(\frac{k_e}{k_{\rm end}}\right)^6\left(\frac{k_s}{k_{\rm end}}\right)^2\left(1-\left(\frac{k_e}{k_{\rm end}}\right)^4\right)\left(\left(1+\left(\frac{k_s}{k_e}\right)\right)^2+2\left(\frac{k_s}{k_e}\right)\right)\nonumber\\
    &&+\frac{9}{32}\left(1-\left(\frac{k_e}{k_{\rm end}}\right)^2\right)\left(4+18\left(\frac{k_s}{k_e}\right)^2+18\left(\frac{k_s}{k_e}\right)^4\right)\left(\frac{k_e}{k_{\rm end}}\right)^6-2\left(1-\left(\frac{k_e}{k_{\rm end}}\right)^2\right)\nonumber\\
    &&-\frac{27}{16}\left(\frac{k_s}{k_e}\right)^4\left(\frac{k_e}{k_{\rm end}}\right)\Bigg\{\sin\left(2\left(\frac{k_{\rm end}}{k_{e}}\right)\right)-\left(\frac{k_e}{k_{\rm end}}\right)\cos\left(2\left(\frac{k_{\rm end}}{k_{e}}\right)\right)\Bigg\}-\frac{3}{4}\cos\left(2\left(\frac{k_{\rm end}}{k_{e}}\right)\right)\Bigg],\quad\quad\quad\\
         {\bf F}_{2,{\bf SRII}}(k_{\rm end},k_e)&=&\Bigg[\frac{81}{16}\ln\left(\frac{k_e}{k_{\rm end}}\right)+\frac{81}{8}\left(1-\left(\frac{k_e}{k_{\rm end}}\right)\right)\left(1+\left(\frac{k_s}{k_e}\right)\right)+5\left(\frac{k_s}{k_e}\right)^6\ln\left(\frac{k_e}{k_{\rm end}}\right)\nonumber\\
    &&+\frac{1}{2}\left(1-\left(\frac{k_e}{k_{\rm end}}\right)^2\right)\left(\left(1+\left(\frac{k_s}{k_e}\right)\right)^2+2\left(\frac{k_s}{k_e}\right)\right)+\frac{2}{3}\left(1-\left(\frac{k_e}{k_{\rm end}}\right)^3\right)\left(\frac{k_s}{k_e}\right)\left(1+\left(\frac{k_s}{k_e}\right)\right)\nonumber\\
    &&+\frac{1}{4}\left(1-\left(\frac{k_e}{k_{\rm end}}\right)^4\right)\left(\frac{k_s}{k_e}\right)^2+\frac{1}{4}\left(1-\left(\frac{k_e}{k_{\rm end}}\right)^8\right)+\frac{27}{64}\left(1-\left(\frac{k_e}{k_{\rm end}}\right)^4\right)\nonumber\\
    &&-\frac{27}{8}\left(1-\left(\frac{k_e}{k_{\rm end}}\right)^5\right)\left(1+\left(\frac{k_s}{k_e}\right)^6\right)+\frac{3}{8}\left(1-\left(\frac{k_e}{k_{\rm end}}\right)^5\right)\left(1+\left(\frac{k_s}{k_e}\right)^2\right)\nonumber\\
    &&-\frac{9}{16}\frac{\displaystyle\left(\frac{k_s}{k_e}\right)^3}{\displaystyle\left(1-\left(\frac{k_s}{k_e}\right)\right)^2}\Bigg\{\cos\left(2\left(\left(\frac{k_{\rm end}}{k_e}\right)-\left(\frac{k_{\rm end}}{k_s}\right)\right)\right)-\cos\left(2\left(1-\left(\frac{k_e}{k_s}\right)\right)\right)\Bigg\}\nonumber\eea\bea
    &&-\frac{27}{4}\frac{\displaystyle 1}{\displaystyle\left(1-\left(\frac{k_s}{k_e}\right)\right)}\Bigg\{\left(\frac{k_s}{k_{\rm end}}\right)\sin\left(2\left(\left(\frac{k_{\rm end}}{k_e}\right)-\left(\frac{k_{\rm end}}{k_s}\right)\right)\right)-\left(\frac{k_s}{k_{e}}\right)\sin\left(2\left(1-\left(\frac{k_e}{k_s}\right)\right)\right)\Bigg\}\nonumber\\
    &&+\frac{81}{16}\frac{\displaystyle\left(1+\left(\frac{k_s}{k_e}\right)^2\right)}{\displaystyle \left(1-\left(\frac{k_s}{k_e}\right)\right)}\Bigg\{\cos\left(2\left(\left(\frac{k_{\rm end}}{k_e}\right)-\left(\frac{k_{\rm end}}{k_s}\right)\right)\right)-\cos\left(2\left(1-\left(\frac{k_e}{k_s}\right)\right)\right)\Bigg\}\nonumber\\
    &&-\frac{9}{16}\frac{\displaystyle\left(\frac{k_s}{k_e}\right)^3}{\displaystyle\left(1-\left(\frac{k_s}{k_e}\right)\right)}\Bigg\{\sin\left(2\left(\left(\frac{k_{\rm end}}{k_e}\right)-\left(\frac{k_{\rm end}}{k_s}\right)\right)\right)-\sin\left(2\left(1-\left(\frac{k_e}{k_s}\right)\right)\right)\Bigg\}\nonumber\\
    &&-\frac{9}{16}\left(1+\left(\frac{k_e}{k_s}\right)\right)\Bigg\{\left(\frac{k_e}{k_{\rm end}}\right)\cos\left(2\left(\left(\frac{k_{\rm end}}{k_e}\right)-\left(\frac{k_{\rm end}}{k_s}\right)\right)\right)-\cos\left(2\left(1-\left(\frac{k_e}{k_s}\right)\right)\right)\Bigg\}\nonumber\\
    &&-\frac{9}{32}\left(\frac{k_s}{k_e}\right)^2\left(9+8\left(\frac{k_s}{k_e}\right)^2+36\left(\frac{k_s}{k_e}\right)^4\right)\left(1-\left(\frac{k_e}{k_{\rm end}}\right)\right)+\frac{27}{16}\left(\frac{k_s}{k_e}\right)^6\left(1-\left(\frac{k_e}{k_{\rm end}}\right)^3\right)\nonumber\\
    &&+\frac{81}{64}\left(\frac{k_s}{k_e}\right)^6\left(1-\left(\frac{k_e}{k_{\rm end}}\right)^4\right)-\frac{81}{16}\left(\frac{k_s}{k_e}\right)^4\left(1+\left(\frac{k_s}{k_e}\right)^2\right)\nonumber\\
    &&+\left(1+4\left(\frac{k_s}{k_e}\right)^2+\left(\frac{k_s}{k_e}\right)^4\right)\ln\left(\frac{k_e}{k_{\rm end}}\right)-\frac{3}{4}\left(\frac{k_e}{k_{\rm end}}\right)\cos\left(2\left(\frac{k_{\rm end}}{k_e}\right)\right)\Bigg],\\
         {\bf F}_{3,{\bf SRII}}(k_{\rm end},k_e)&=&{\bf F}_{2,{\bf SRII}}(k_{\rm end},k_e),\\
         {\bf F}_{4,{\bf SRII}}(k_{\rm end},k_e)&=&\Bigg[\frac{81}{16}\left(1-\left(\frac{k_e}{k_{\rm end}}\right)^4\right)+\frac{27}{8}\left(1-\left(\frac{k_e}{k_{\rm end}}\right)^3\right)\left(1+\left(\frac{k_s}{k_e}\right)\right)+\frac{81}{16}\left(\frac{k_s}{k_e}\right)^6\ln\left(\frac{k_e}{k_{\rm end}}\right)\nonumber\\
         &&+\frac{81}{32}\left(1-\left(\frac{k_e}{k_{\rm end}}\right)^2\right)\left(\left(1+\left(\frac{k_s}{k_e}\right)\right)^2+2\left(\frac{k_s}{k_e}\right)\right)+10\left(1-\left(\frac{k_e}{k_{\rm end}}\right)\right)\left(\frac{k_s}{k_e}\right)\left(1+\left(\frac{k_s}{k_e}\right)\right)\nonumber\\
         &&+\frac{1}{12}\left(1-\left(\frac{k_e}{k_{\rm end}}\right)^{12}\right)+\frac{9}{40}\left(1-\left(\frac{k_e}{k_{\rm end}}\right)^{10}\right)\left(1+\left(\frac{k_s}{k_e}\right)^2\right)+\frac{27}{64}\left(1-\left(\frac{k_e}{k_{\rm end}}\right)^{4}\right)\left(1+\left(\frac{k_s}{k_e}\right)^6\right)\nonumber\\
         &&+\frac{81}{128}\left(1-\left(\frac{k_e}{k_{\rm end}}\right)^{8}\right)+6\left(\frac{k_s}{k_e}\right)^2\ln\left(\frac{k_e}{k_{\rm end}}\right)\nonumber\\
         &&-\frac{9}{16}\frac{\displaystyle\left(1+\left(\frac{k_s}{k_e}\right)\right)^2}{\displaystyle\left(1-\left(\frac{k_s}{k_e}\right)\right)^2}\Bigg\{\cos\left(2\left(\left(\frac{k_{\rm end}}{k_e}\right)-\left(\frac{k_{\rm end}}{k_s}\right)\right)\right)-\cos\left(2\left(1-\left(\frac{k_e}{k_s}\right)\right)\right)\Bigg\}\nonumber\\
    &&-\frac{27}{8}\frac{\displaystyle \displaystyle\left(1+\left(\frac{k_s}{k_e}\right)\right)^2}{\displaystyle\left(1-\left(\frac{k_s}{k_e}\right)\right)^3}\Bigg\{\sin\left(2\left(\left(\frac{k_{\rm end}}{k_e}\right)-\left(\frac{k_{\rm end}}{k_s}\right)\right)\right)-\sin\left(2\left(1-\left(\frac{k_e}{k_s}\right)\right)\right)\Bigg\}\nonumber\\
    &&-\frac{27}{4}\frac{\displaystyle \displaystyle\left(1+\left(\frac{k_s}{k_e}\right)\right)^2}{\displaystyle\left(1-\left(\frac{k_s}{k_e}\right)\right)}\Bigg\{\left(\frac{k_e}{k_{\rm end}}\right)^2\sin\left(2\left(\left(\frac{k_{\rm end}}{k_e}\right)-\left(\frac{k_{\rm end}}{k_s}\right)\right)\right)+\sin\left(2\left(1-\left(\frac{k_e}{k_s}\right)\right)\right)\Bigg\}\nonumber\\
    &&+\frac{9}{80}\left(\frac{k_s}{k_e}\right)^2\left(9+44\left(\frac{k_s}{k_e}\right)^2\right)\left(1-\left(\frac{k_e}{k_{\rm end}}\right)^{5}\right)+\frac{27}{8}\left(\frac{k_s}{k_e}\right)^4\left(1+\left(\frac{k_s}{k_e}\right)^2\right)\left(1-\left(\frac{k_e}{k_{\rm end}}\right)^{3}\right)\nonumber\\
    &&+\frac{81}{16}\left(\frac{k_s}{k_e}\right)^6\left(1-\left(\frac{k_e}{k_{\rm end}}\right)\right)+\frac{99}{56}\left(1-\left(\frac{k_e}{k_{\rm end}}\right)^7\right)-\frac{81}{16}\left(\frac{k_e}{k_{\rm end}}\right)^6\left(\frac{k_s}{k_{\rm end}}\right)^6\ln\left(\frac{k_e}{k_{\rm end}}\right)\nonumber\eea\bea
    &&+\frac{81}{16}\left(\frac{k_e}{k_{\rm end}}\right)^6\left(1-\left(\frac{k_e}{k_{\rm end}}\right)^2\right)\left(1+\left(\frac{k_s}{k_e}\right)^2\right)\nonumber\\
    &&+\frac{3}{32}\left(\frac{k_e}{k_{\rm end}}\right)^6\left(1-\left(\frac{k_e}{k_{\rm end}}\right)^6\right)\left(4+18\left(\frac{k_s}{k_e}\right)^4+18\left(\frac{k_s}{k_e}\right)^2\right)\nonumber\\
    &&+\frac{81}{64}\left(\frac{k_e}{k_{\rm end}}\right)^6\left(\frac{k_s}{k_{\rm end}}\right)^2\left(1-\left(\frac{k_e}{k_{\rm end}}\right)^4\right)\left(1+4\left(\frac{k_s}{k_e}\right)^2+\left(\frac{k_s}{k_e}\right)^4\right)\nonumber\\
    &&+\frac{9}{128}\left(\frac{k_e}{k_{\rm end}}\right)^4\left(1-\left(\frac{k_e}{k_{\rm end}}\right)^8\right)\left(4+18\left(\frac{k_s}{k_e}\right)^2\right)+\frac{1}{4}\left(\frac{k_e}{k_{\rm end}}\right)^2\left(1-\left(\frac{k_e}{k_{\rm end}}\right)^9\right)\nonumber\\
    &&+\frac{81}{32}\left(\frac{k_s}{k_e}\right)^6\left(\frac{k_e}{k_{\rm end}}\right)\Bigg\{\sin\left(2\left(\frac{k_{\rm end}}{k_{e}}\right)\right)-\left(\frac{k_e}{k_{\rm end}}\right)\cos\left(2\left(\frac{k_{\rm end}}{k_{e}}\right)\right)\Bigg\}\Bigg].\eea
Additionally, it is important to mention the explicit expression for the coupling parameters $Z_{\rm end}$ and $Z_e$ at the end of SRII and USR scales $\tau=\tau_{\rm end}$ and $\tau=\tau_e$ are given by the following expression:
         \bea && Z_{\rm end}\equiv Z(\tau_{\rm end})= \frac{H(\tau_{\rm end})\dot{\bar{\phi}}_0(\tau_{\rm end})}{\Lambda^3},\quad\quad 
         Z_e\equiv Z(\tau_e)= \frac{H(\tau_e)\dot{\bar{\phi}}_0(\tau_e)}{\Lambda^3},\eea
         where $\dot{\bar{\phi}}_0(\tau_{\rm end})$ and $\dot{\bar{\phi}}_0(\tau_e)$ can be expressed as:
         \bea && \dot{\bar{\phi}}_0(\tau_{\rm end})=\frac{\Lambda^3}{12H(\tau_{\rm end})}\frac{c_2}{c_3}\Bigg[-1+\sqrt{1+\frac{8c_3}{c^2_2}\frac{\lambda^3}{\Lambda^3}}\Bigg], \\
         &&\dot{\bar{\phi}}_0(\tau_e)=\frac{\Lambda^3}{12H(\tau_e)}\frac{c_2}{c_3}\Bigg[-1+\sqrt{1+\frac{8c_3}{c^2_2}\frac{\lambda^3}{\Lambda^3}}\Bigg].\eea

\subsection{Final result for the cut-off regularized one-loop corrected total power spectrum}
\label{s5d}

In this subsection we are now going to provide the total cut-off regularized one-loop corrected primordial power spectrum computed from comoving curvature perturbation, where we haven taken the individual contributions from SRI, USR and SRII very carefully. After summing over all the one-loop contributions to the tree level contribution we get the following result:
\bea \Bigg[\Delta^{2}_{\zeta}(p)\Bigg]_{\bf Total}
&=&\Bigg[\Delta^{2}_{\zeta,{\bf Tree}}(p)\Bigg]_{\bf SRI}+\Bigg[\Delta^{2}_{\zeta,{\bf Tree}}(p)\Bigg]_{\bf USR}\Theta(p-k_s)+\Bigg[\Delta^{2}_{\zeta,{\bf Tree}}(k)\Bigg]_{\bf SRII}\Theta(p-k_e)\nonumber\\
&&\quad\quad\quad+ \Bigg[\Delta^{2}_{\zeta,{\bf One-loop}}(p)\Bigg]_{\bf SRI}+\Bigg[\Delta^{2}_{\zeta,{\bf One-loop}}(p)\Bigg]_{\bf USR}\Theta(p-k_s)+\Bigg[\Delta^{2}_{\zeta,{\bf One-loop}}(p)\Bigg]_{\bf SRII}\Theta(p-k_e)\nonumber\\
&=& \Bigg(\Bigg[\Delta^{2}_{\zeta,{\bf Tree}}(p)\Bigg]_{\bf SRI}+ \Bigg[\Delta^{2}_{\zeta,{\bf One-loop}}(p)\Bigg]_{\bf SRI}\Bigg)\nonumber\\
&&\quad\quad\quad\quad\quad+\Bigg(\Bigg[\Delta^{2}_{\zeta,{\bf Tree}}(p)\Bigg]_{\bf USR}+ \Bigg[\Delta^{2}_{\zeta,{\bf One-loop}}(p)\Bigg]_{\bf USR}\Bigg)\Theta(p-k_s)\nonumber\\
&&\nonumber\\
&&\quad\quad\quad\quad\quad\quad\quad\quad\quad+\Bigg(\Bigg[\Delta^{2}_{\zeta,{\bf Tree}}(p)\Bigg]_{\bf SRII}+ \Bigg[\Delta^{2}_{\zeta,{\bf One-loop}}(p)\Bigg]_{\bf SRII}\Bigg)\Theta(p-k_e)\nonumber\\
&\approx&\Bigg[\Delta^{2}_{\zeta,{\bf Tree}}(p)\Bigg]_{\bf SRI}\Bigg\{1+\left(\frac{k_e}{k_s }\right)^{6}\Bigg[\left|\alpha^{(2)}_{\bf k}-\beta^{(2)}_{\bf k}\right|^2\Theta(p-k_s)+\left|\alpha^{(3)}_{\bf k}-\beta^{(3)}_{\bf k}\right|^2\Theta(k-k_e)\Bigg]\nonumber\\
&&\quad\quad\quad\quad\quad\quad\quad\quad\quad\quad\quad+\Bigg[\Delta^{2}_{\zeta,{\bf Tree}}(p)\Bigg]_{\bf SRI}\Bigg\{c_{\bf SRI}-\frac{1}{8{\cal A}^2_*\pi^4}\sum^{4}_{i=1}\widetilde{\cal G}_{i,{\bf SRI}} {\bf F}_{i,{\bf SRI}}(k_s,k_*)\Bigg\}\nonumber\eea\bea
&&\quad\quad\quad\quad\quad\quad\quad\quad\quad\quad\quad+\Bigg[\Delta^{2}_{\zeta,{\bf Tree}}(p)\Bigg]_{\bf SRI}\Bigg\{c_{\bf USR}+\frac{1}{8{\cal A}^2_*\pi^4}\sum^{4}_{i=1}\widetilde{\cal G}_{i,{\bf USR}} {\bf F}_{i,{\bf USR}}(k_e,k_s)\Bigg\}\Theta(p-k_s)\nonumber\\
&&\quad\quad\quad\quad\quad\quad\quad\quad\quad\quad\quad+\Bigg[\Delta^{2}_{\zeta,{\bf Tree}}(p)\Bigg]_{\bf SRI}\Bigg\{c_{\bf SRII}+\frac{1}{8{\cal A}^2_*\pi^4}\sum^{4}_{i=1}\widetilde{\cal G}_{i,{\bf SRII}} {\bf F}_{i,{\bf SRII}}(k_{\rm end},k_e)\Bigg\}\Theta(p-k_e)\Bigg\}\nonumber\\
&\approx& \Bigg[\Delta^{2}_{\zeta,{\bf Tree}}(p)\Bigg]_{\bf SRI}\Bigg\{1+\left(\frac{k_e}{k_s }\right)^{6}\Bigg[\left|\alpha^{(2)}_{\bf k}-\beta^{(2)}_{\bf k}\right|^2\Theta(p-k_s)+\left|\alpha^{(3)}_{\bf k}-\beta^{(3)}_{\bf k}\right|^2\Theta(k-k_e)\Bigg]\nonumber\\
&&\quad\quad\quad\quad\quad\quad+\Bigg[\Delta^{2}_{\zeta,{\bf Tree}}(p)\Bigg]_{\bf SRI}\Bigg\{\bigg(c_{\bf SRI}+c_{\bf USR}\Theta(p-k_s)+c_{\bf SRII}\Theta(p-k_e)\bigg)\nonumber\\
&&\quad\quad\quad\quad\quad\quad\quad-\frac{1}{8{\cal A}^2_*\pi^4}\sum^{4}_{i=1}\bigg(\widetilde{\cal G}_{i,{\bf SRI}} {\bf F}_{i,{\bf SRI}}(k_s,k_*)-\widetilde{\cal G}_{i,{\bf USR}} {\bf F}_{i,{\bf USR}}(k_e,k_s)\Theta(p-k_s)\nonumber\\
&&\quad\quad\quad\quad\quad\quad\quad\quad\quad\quad\quad\quad\quad\quad\quad-\widetilde{\cal G}_{i,{\bf SRII}} {\bf F}_{i,{\bf SRII}}(k_{\rm end},k_e)\Theta(p-k_e)\Bigg)\Bigg\}
\nonumber\\
&=&\Bigg[\Delta^{2}_{\zeta,{\bf Tree}}(p)\Bigg]_{\bf SRI}\Bigg\{1+\left(\frac{k_e}{k_s }\right)^{6}\Bigg[\left|\alpha^{(2)}_{\bf k}-\beta^{(2)}_{\bf k}\right|^2\Theta(p-k_s)+\left|\alpha^{(3)}_{\bf k}-\beta^{(3)}_{\bf k}\right|^2\Theta(k-k_e)\Bigg]\nonumber\eea\bea
&&\quad\quad\quad\quad\quad\quad+\Bigg[\Delta^{2}_{\zeta,{\bf Tree}}(p)\Bigg]_{\bf SRI}\Bigg\{\bigg(c_{\bf SRI}+c_{\bf USR}\Theta(p-k_s)+c_{\bf SRII}\Theta(p-k_e)\bigg)\nonumber\\
&&\quad\quad\quad\quad\quad\quad\quad-\frac{c^4_s}{8{\cal B}^2_*\pi^6}\sum^{4}_{i=1}\bigg(\widetilde{\cal G}_{i,{\bf SRI}} {\bf F}_{i,{\bf SRI}}(k_s,k_*)-\widetilde{\cal G}_{i,{\bf USR}} {\bf F}_{i,{\bf USR}}(k_e,k_s)\Theta(p-k_s)\nonumber\\
&&\quad\quad\quad\quad\quad\quad\quad\quad\quad\quad\quad\quad\quad\quad\quad-\widetilde{\cal G}_{i,{\bf SRII}} {\bf F}_{i,{\bf SRII}}(k_{\rm end},k_e)\Theta(p-k_e)\Bigg)\Bigg\},\eea
where the power spectrum in the SRI region can be recast in the following simplified form:
\bea \Bigg[\Delta^{2}_{\zeta,{\bf Tree}}(p)\Bigg]_{\bf SRI}&=&\left(\frac{H^{4}}{8\pi^{2}{\cal A} c^3_s}\right)_*\Bigg\{1+\Bigg(\frac{p}{k_s}\Bigg)^2\Bigg\}=\left(\frac{H^{4}}{8\pi^{2}{\cal B} c_s}\right)_*\Bigg\{1+\Bigg(\frac{p}{k_s}\Bigg)^2\Bigg\}.\eea
In the present context, all the momentum-dependent, time-dependent fixed functions and coupling parameters and the Bogoliubov coefficients in the USR and SRII phases have already been mentioned before explicitly.

Now it is important to note that, there are issues with the requirement for a nontrivial potential. Potential terms must be viewed as irrelevant physical operators that produce minor adjustments because they are not Galilean symmetry invariant. But if they exist, large renormalizations would usually be applied to operators with mass dimensions less than four, which would render the model of limited relevance. In reality, there is no way to get around this problem because of the presence of an effective potential at least made up of quadratic order contribution is needed to characterize a smooth end to the inflationary period. Surprisingly, the two contributions, $\phi$ (linear) and $\phi^2$ (quadratic) are shielded by a non-renormalization theorem and can thus be reliably regarded as the completely irrelevant deformations of the Galilean symmetry applicable to the underlying theoretical setup. This can be easily proved by one particle irreducible  effective action approach within the framework of path integral, which was explicitly shown in the ref. \cite{Burrage:2010cu}. Such a strong non-renormalization theorem does not allow any further operators in the underlying theory which
violate the symmetry at the quantum mechanical level. This strong non-renormalization theorem established at the level of scalar field is also propagated to the quantum fluctuations induced by the terms appearing in the second and third order perturbed action as written before for the comoving curvature perturbation variable. For this reason it is automatically expected that in presence of Galilean symmetry or in presence of its very mild breaking at the level of cosmological perturbation all the self interactions are protected by the any sorts of radiative quantum corrections. On top of this due to having strong strong non-renormalization theorem the in all the correlation functions computed from the cosmological perturbations generated from covariant Galileon has to be non-renormalizable but stable under radiative quantum corrections. Consequently, in the present theoretical framework the underlying concept of renormalization and resummation is not at all applicable. For this reason the cut off regularized total one-loop corrected primordial power spectrum computed from comving scalar curvature perturbation will fully suffice the purpose and with the help of this result we are going to establish our final conclusion. From the derived structure of the one-loop corrections as obtained from the SRI, USR and SRII phases it is completely evident that it is free from quadratic and logarithmic divergences in the detailed structure of the momentum dependent factors associated with the computation. From the underlying structure of the CGEFT theory and due to not absence of the term $\eta^{'}\zeta^{'}\zeta^{2}$ in the third order action to very mildly break the Galilean symmetry such mentioned divergences are completely absent in the primordial power spectrum, which is obviously a remarkable fact from the present computation. The other factors appearing in the one-loop correction are extremely power law suppressed and some of them are highly oscillatory with having restricted amplitude of oscillation. These contributions are not at all harmful for the present computational purpose and will give a very small correction to the tree level amplitude of the primordial power spectrum. For this reason the cosmological perturbation will be valid throughout the SRI, USR and SRII phases and all the perturbative approximations holds good perfectly during the computation of the one-loop correction. We can clearly expect that due to having extremely suppressed one-loop correction the tree level amplitude will sufficient enough for the enhancement of the spectrum from ${\cal O}(10^{-9})$ (SRI) to ${\cal O}(10^{-2})$ (USR) which is one of the necessary conditions to form PBH. Now for the proper enhancement of the tree level power spectrum and to maintain the perturbative approximations the ratio, $k_e/k_s$ has to be strictly fixed at ${\cal O}(10)$, which basically allows only $2$ e-foldings in the USR regime of the computation. That means that prolonged USR phase is not all allowed for the present computation. On the other hand to produce large mass PBHs position of the SRI to USR transition scale has to be fixed at $k_s\leq 10^{6}$, which automatically implies to restrict $k_e\leq 10^{7}$, so that $k_e/k_s\sim {\cal O}(10)$ always maintained. This helps us to produce solar mass PBHs or more than that. In the next sections we are going to investigate the details  of the above findings and its outcomes in the formation of PBHs. Most importantly, we are going to study the constraints in detail to form a large mass PBH within the framework of single field inflation induced by covarinatized Galileon.
 
\section{Numerical results: Constraints on the PBH mass and evaporation time scale}
\label{s6}

    \begin{figure*}[htb!]
    	\centering
{
      	\includegraphics[width=18cm,height=14cm] {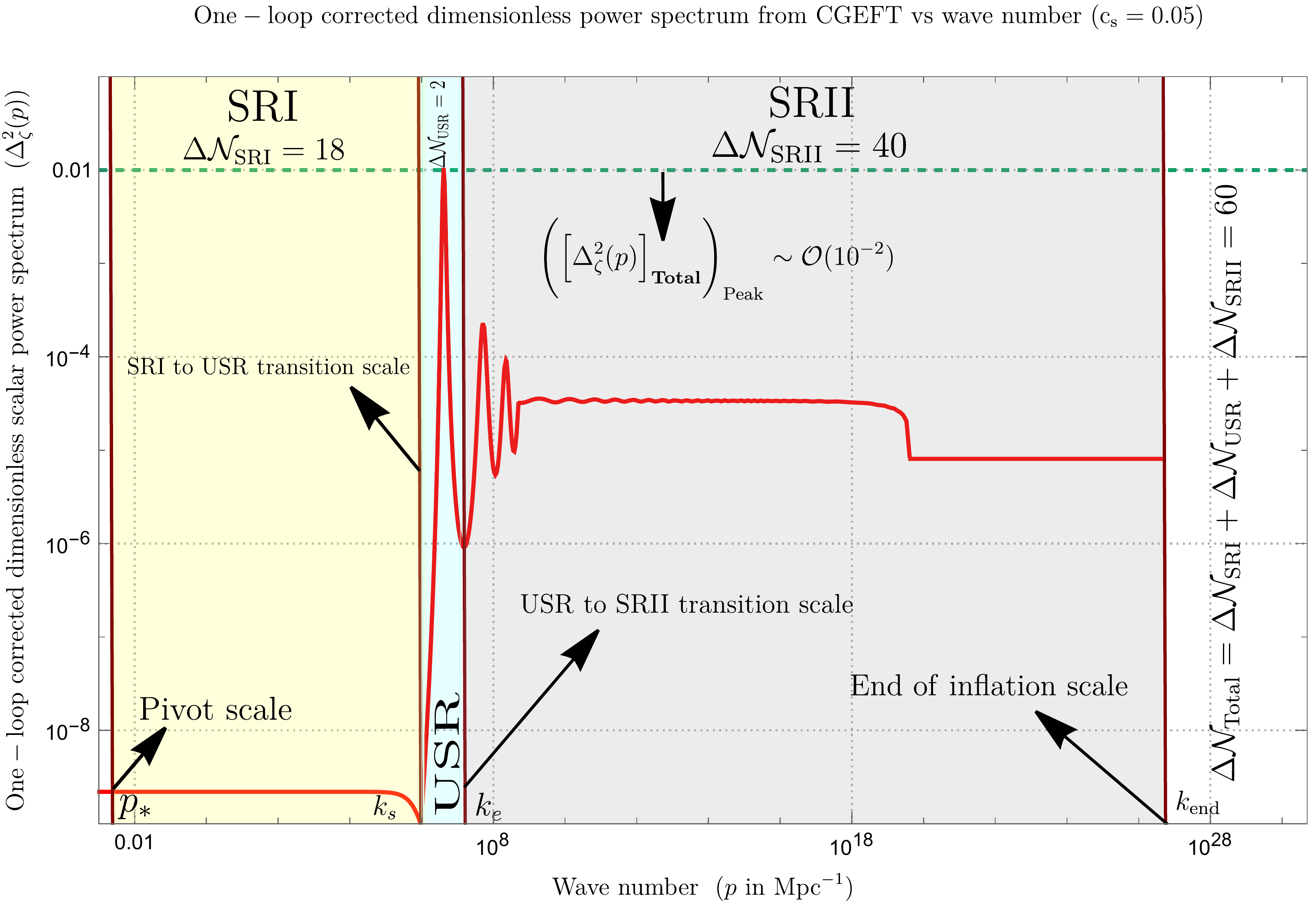}
    }
    	\caption[Optional caption for list of figures]{Behaviour of the dimensionless primordial power spectrum for scalar modes with respect to the wave number. Here we fix the effective sound speed at $c_s=0.05$. We also fix the pivot scale at $p_*=0.02\;{\rm Mpc}^{-1}$, SRI to USR transition scale at $k_s=10^{6}\;{\rm Mpc}^{-1}$, end of USR scale at $k_e=10^{7}\;{\rm Mpc}^{-1}$, end of inflation scale at $k_{\rm end}=10^{27}\;{\rm Mpc}^{-1}$, the regularization parameters, $c_{\bf SRI}=0$, $c_{\bf USR}=0$ and $c_{\bf SRII}=0$. Also we use, $\Delta\eta(\tau_e)=1$ and $\Delta\eta(\tau_s)=-6$. In this plot we have found that, $k_{\rm UV}/k_{\rm IR}=k_e/k_s\approx{\cal O}(10)$ and $k_{\rm end}/k_e\approx{\cal O}(10^{20})$. Peak amplitude ${\cal O}(10^{-2})$ of the corresponding spectrum is achieved at the scale $5k_s\sim 5\times 10^{6}{\rm Mpc}^{-1}$. Total number of e-foldings $60$ is achieved in this analysis, which is sufficient enough for inflation.  } 
    	\label{GL}
    \end{figure*}

    \begin{figure*}[htb!]
    	\centering
{
      	\includegraphics[width=18cm,height=14cm] {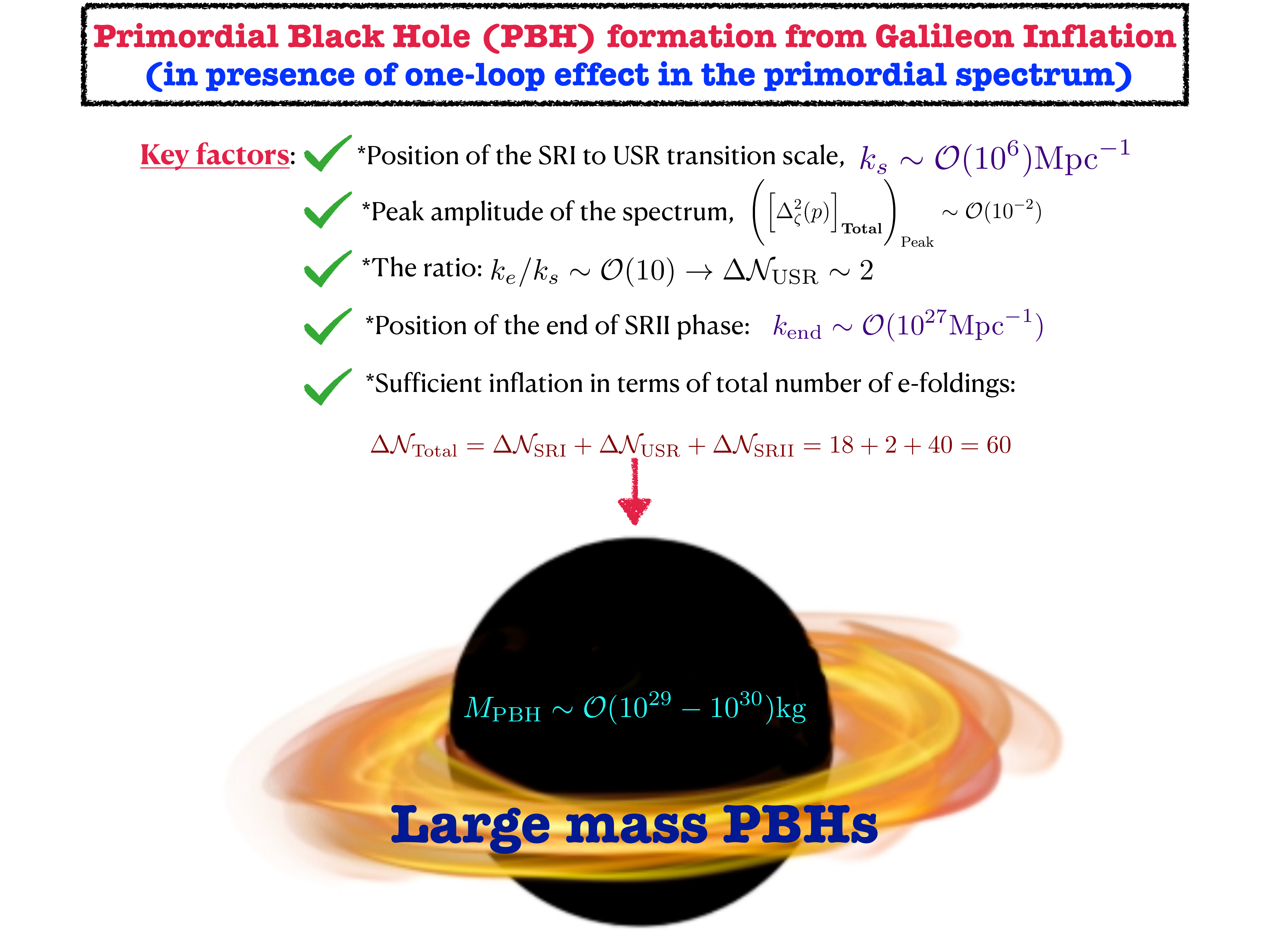}
    }
    	\caption[Optional caption for list of figures]{Representative flow chart of the underlying process of Primordial Black Hole (PBH) formation in presence of one-loop correction in the primordial power spectrum for scalar modes within the framework of single field Galileon inflation. } 
    	\label{PBHGali}
    \end{figure*}
   In figure(\ref{GL}), we have explicitly shown the behaviour of the dimensionless power spectrum for comoving curvature perturbation with respect to the wave number for fixed value of the effective sound speed $c_s=0.05$ for the Galileon. Since renormalization and resummation is not needed for Galileon due to having strong non-renormalization theorem we have only shown the behaviour of the one-loop corrected cut off regularized power spectrum where the cumulative effects from the phases SRI, USR and SRII are encoded. We found just before the the SRI to USR transition the power spectrum falls to a minimum value and then at the corresponding transition point spectrum sharply increases to a peak value, ${\cal O}(10^{-2})$ at $5\times 10^{6}{\rm Mpc}^{-1}$ momentum scale, which is appearing in the mid point in the USR region. The numb er of e-foldings corresponding to the peak value of the amplitude of the power spectrum is $\Delta{\cal N}_{\rm Peak}\sim 20$. After achieving the maximum enhancement in the corresponding power spectrum the amplitude starts falling sharply and attain a value ${\cal O}(10^{-6})$ at the end of the USR phase in the present set up. Further due to having sinusoidal contributions in the power spectrum the the spectrum start oscillating with a very high frequency with having decaying overall amplitude. Within a very short span it will saturate to a certain value with having very small oscillating envelop which persist for longer span and then there is is sharp fall is observed which again saturates to the amplitude ${\cal O}(10^{-5})$ which persist up to the end of SRII phase where the inflation ends as well. In figure(\ref{PBHGali}) we have also depicted the representative flow chart of the underlying process of Primordial Black Hole (PBH) formation in presence of one-loop correction in the primordial power spectrum for scalar modes within the framework of single field Galileon inflation for the better understanding purpose of the overall outcome of the work.
   
   For the numerical purpose we fix, $k_s=10^{6}\;{\rm Mpc}^{-1}$ (where we fix the IR cut-off) and the end of USR at $k_e=10^{7}\;{\rm Mpc}^{-1}$ (where we fix the UV cut-off), end of inflation scale at $k_{\rm end}=10^{27}\;{\rm Mpc}^{-1}$, the regularization parameters, 
   $c_{\bf SRI}=0$, $c_{\bf USR}=0$, $c_{\bf SRII}=0$, $\Delta\eta(\tau_e)=1$ and $\Delta\eta(\tau_s)=-6$. To maintain the perturbative approximations holds good perfectly and to generate sufficient number of e-foldings to achieve inflation we fix, $k_{\rm UV}/k_{\rm IR}=k_e/k_s\approx{\cal O}(10)$ and $k_{\rm end}/k_e\approx{\cal O}(10^{20})$. Though we have shown the figure for the sound speed $c_s=0.05$, our analysis is valid within the regime $0.024<c_s<1$, which means for Galileon causality is strictly maintained within the framework of EFT. 

   From this plot we found that the number of e-foldings allowed in the USR phase is given by:
\bea \Delta {\cal N}_{\rm USR}=\ln(k_e/k_s)\approx\ln(10^{7}/10^{6})=\ln(10)\approx 2,\eea 
which is an important information for the PBH formation in the present Galileon inflationary paradigm. Also, the allowed number of e-foldings for the SRI and SRII periods are given by:
\bea &&\Delta {\cal N}_{\rm SRI}=\ln(k_s/p_*)\approx \ln(10^{6}/0.02)\sim 18,\\
&&\Delta {\cal N}_{\rm SRII}=\ln(k_{\rm end}/k_e)\approx \ln(10^{27}/10^{7})\approx 40.\eea 
As a consequence, the total number of e-foldings allowed by Galileon inflation is given by:
\bea \Delta {\cal N}_{\rm Total}=\Delta {\cal N}_{\rm SRI}+\Delta {\cal N}_{\rm USR}+\Delta {\cal N}_{\rm SRII}\sim 18+2+40=60.\eea 
 This possibility will lead a large mass PBH formation having sufficient number of e-folds for inflation in the single field framework with Galileon.

We found that for Galileon inflation the prolonged USR period is strictly not allowed for PBH formation. In this region the PBH mass can be estimated in terms of the effective sound speed $c_s$ for the underlying CGEFT:
\bea \label{PBH1}M_{\rm PBH}&=&1.13\times 10^{15}\times\bigg(\frac{\gamma}{0.2}\bigg)\bigg(\frac{g_*}{106.75}\bigg)^{-1/6}\bigg(\frac{k_s}{p_*}\bigg)^{-2}M_{\odot}\times c^{2}_s\nonumber\\
&=&0.46\times\bigg(\frac{\gamma}{0.2}\bigg)\bigg(\frac{g_*}{106.75}\bigg)^{-1/6}M_{\odot}\times c^{2}_s
\nonumber\\
&\approx& {\cal O}(10^{29}-10^{30}){\rm kg}\nonumber\\
&\approx& {\cal O}(M_{\odot}),\eea 
where $M_{\odot}\sim 2\times 10^{30}{\rm kg}$ is the solar mass, 
SRI to USR transition scale, $k_s=10^{6}\;{\rm Mpc}^{-1}$ and pivot scale, $p_*=0.02\;{\rm Mpc}^{-1}$, effective sound speed $0.024<c_s<1$, $\gamma\sim 0.2$ and relativistic d.o.f. $g_*\sim 106.75$ for SM and $g_*\sim 226$ for SUSY d.o.f. 

 Finally, the evaporation time scale of the formed large mass PBHs from the Galileon inflationary paradigm can be computed as:
\bea t^{\rm evap}_{\rm PBH}&=&{10}^{64}\bigg(\frac{M_{\rm PBH}}{M_{\odot}}\bigg)^{3}{\rm years}\approx {10}^{64}{\rm years}.\eea 
  Though the span for PBHs formation in terms of number of e-foldings is small ($\Delta {\cal N}_{\rm USR}\sim 2$), but the estimated PBH mass is extremely large ($\sim M_{\odot})$ and the corresponding evaporation time scale is also very large ($\sim {10}^{64}{\rm years}$) in the present context.

\section{Summary and conclusions}
\label{s7}

In this paper, we investigated the process of PBHs formation in the framework of single-field Galileon inflation.
We have demonstrated that the no-go theorem imposed due to having a large one-loop effect on the primordial power spectrum computed from scalar curvature perturbation in all possible classes of single field inflationary frameworks (such as $P(X,\phi)$ inflation, EFT of inflation, which covers various canonical and non-canonical models) to explain the generation of large mass PBHs formation can be completely evaded within the framework of Galileon inflation. The outcomes of our findings are explicitly shown in figure (\ref{GL}) and figure (\ref{PBHGali}). Let us highlight the following points to justify our findings: (1) In the scenario under consideration, the one-loop quantum effects are sub-dominant over the tree-level amplitude of the primordial power spectrum, and perturbation theory holds good perfectly. The prime reason for this fact lies in the very mild breaking of Galilean symmetry in the comoving curvature perturbation, $\zeta\rightarrow\zeta-{H}/{\dot{\bar{\phi}}_0}\left(b\cdot \delta x\right)$ generated from the underlying EFT framework. Due to this small breaking of Galilean symmetry, the cubic interaction term $\eta^{'}\zeta^{'}\zeta^{2}$ is absent in the third order action for the comoving curvature perturbation, as this contribution can be recast as a total derivative term at the boundary, which gives a trivial contribution and can be absorbed in the field redefinition. This aspect was highlighted in the technical part of the discussion. Since the other terms appearing in the third-order action do not have any terms that contain the time derivatives of the second slow-roll parameter $\eta$, the effect of the sharp transition from SRI to USR is not harmful at all for the present computational purpose. We have found that, due to the absence of the specific cubic self-interaction term, no quadratic and logarithmic divergences are appearing in the one-loop contributions computed from the SRI, USR, and SRII regions, respectively. Other contributions that are appearing in the one-loop contributions are either highly power law suppressed or made up of sinusoidally highly oscillating terms with restricted small amplitudes. For this reason, we found from our analysis that the one-loop contribution is subdominant over the result obtained from the tree-level counterpart. The technical details in support of such a strong outcome are presented in the previous sections of this paper for reference.  (2) In this case, there are no strong constraints from renormalization and ressumation on the one-loop corrected power spectrum, as these effects are not important in the context of Galileon EFT due to its strong non-renormalization theorem.
(3) Interestingly, there is no problem in fixing the position of the SRI to USR transition scale at $k_s\sim 10^{6}{\rm Mpc}^{-1}$ necessary for the generation of large fluctuations, which are ${\cal O}(10^{-2})$  required for PBH formation. (4) Consequently, large mass PBHs, $ {\cal O}(M_{\odot})$ whose evaporation time scale is ${10}^{64}{\rm years}$.
(5)  In the absence of any strict restriction on shifting the position of the SRI to USR transition scale by shifting it to a smaller value, one can be able to generate PBHs with extremely large masses $M_{\rm PBH}\gg M_{\odot}$ having a huge evaporation time scale. On the other hand, extremely small mass PBHs $M_{\rm PBH}\ll M_{\odot}$ can also be generated by shifting the position of the SRI to the USR transition scale to a very large value.
The aforesaid implies that {\it PBHs with arbitrary masses can be generated using Galileon inflation, where the no-go result due to quantum loop corrections on the primordial power spectrum is completely evaded}. 

For the immediate future prospect, one can study the density of PBH formation to the tail of the probability distribution function (PDF), and thus it has a direct impact on the non-Gaussinaity. In most of the studies during such studies of PBH formation, it is assumed that the PDF of the comoving curvature perturbation variable is Gaussian. However, in a strict sense, this is not true always. It is expected from the corresponding setup of the underlying problem that PBHs have to be formed in the tail of the mentioned PDF of curvature perturbation. This will be going to change predictions from primordial non-Gaussianity which is responsible for the enhancement of the appearance of the occasional events which directly put an impact on the PBH formation in this context of the discussion. Such non-Gaussian effects on the corresponding PDF of comoving curvature perturbation sometimes reduce the amount of power required to produce such PBHs. For more details on this issue, details see refs. \cite{Gow:2022jfb,Kitajima:2021fpq,Young:2013oia,Biagetti:2021eep}. Recently the issue of the non-Gaussianity in the presence of the USR phase has been addressed in ref. \cite{Choudhury:2023kdb} in detail at the level of the three-point function and the associated bispectrum computed from curvature perturbation. However, from the perspective of PDF, we have not studied anything yet, which we are planning to address in near future studies on the related setup.

{\bf Acknowledgements:} SC would like to thank the work-friendly environment of The Thanu Padmanabhan Centre For Cosmology and Science Popularization (CCSP), Shree Guru Gobind Singh Tricentenary (SGT) University, Gurugram, Delhi-NCR for providing tremendous support in
research and offer the Assistant Professor (Senior Grade) position. SC would like to thank Mayukh Raj Gangopadhyay, Yogesh, Mohit Kumar Sharma for useful discussions. SC would like to specially thanks Soumitra SenGupta for inviting at IACS, Kolkata during the work. SC also thanks all
the members of our newly formed virtual international non-profit consortium Quantum Aspects of the SpaceTime \& Matter (QASTM) for elaborative discussions. The work of MS is supported by Science and Engineering Research Board (SERB), DST, Government of India under the Grant Agreement number CRG/2022/004120 (Core Research Grant). MS is also partially supported by the Ministry of Education and Science of the Republic of Kazakhstan, Grant
No. 0118RK00935 and CAS President's International Fellowship Initiative (PIFI). Last but not
least, we would like to acknowledge our debt to the people belonging to the various parts
of the world for their generous and steady support for research in natural sciences.

\newpage

\bibliography{references2}
\bibliographystyle{utphys}

\end{document}